\documentclass[aip,graphicx]{revtex4-2}

\usepackage{nicefrac} 
\usepackage{xcolor}
\usepackage{amsmath, amsthm, amssymb} 
\usepackage{graphicx}
\usepackage{dcolumn}
\usepackage{bm}
\usepackage{hyperref} 

\usepackage[utf8]{inputenc}
\usepackage[T1]{fontenc}
\usepackage{mathptmx}
\usepackage{etoolbox}
\usepackage{placeins} 

\begin{document}


\title{On the response of neutrally stable flows to oscillatory forcing with application to liquid sheets} 



\author{Colin M. Huber}
\affiliation{School of Mathematical Sciences, 
		Rochester Institute of Technology, 
        Rochester, 
        NY,
        14623, 
        USA}
\author{Nathaniel S. Barlow}
\affiliation{School of Mathematical Sciences, 
		Rochester Institute of Technology, 
        Rochester, 
        NY,
        14623, 
        USA}
\author{Steven J. Weinstein}
\affiliation{School of Mathematical Sciences, 
		Rochester Institute of Technology, 
        Rochester, 
        NY,
        14623, 
        USA}
\affiliation{Department of Chemical Engineering, 
			Rochester Institute of Technology, 
            Rochester, 
            NY,
			14623,            
            USA}


\date{\today}

\begin{abstract}
Industrial coating processes create thin liquid films with tight thickness tolerances, and thus models that predict the response to inevitable disturbances are essential. The mathematical modeling complexities are reduced through linearization as even small thickness variations in films can render a product unsalable.  The signaling problem, considered in this paper, is perhaps the simplest model that incorporates the effects of repetitive (oscillatory) disturbances that are initiated, for example, by room vibrations and pump drives. In prior work, Gordillo and P\'erez (Phys. Fluids 14, 2002) examined the structure of the signaling response for linear operators that admit exponentially growing or damped solutions, i.e., the medium is classified as unstable or stable via classical stability analysis. The signaling problem admits two portions of the solution, the transient behavior due to the start-up of the disturbance and the long-time behavior that is continually forced; the superposition reveals how the forced solution evolves through the passage of a transient.  In this paper, we examine signaling for the linear operator examined by King et al. (King et al. 2016, Phys. Rev. Fluids 1(7)) that governs varicose waves in a thin liquid sheet and that can admit solutions having algebraic growth although the underlying medium is classified as being neutrally stable.  Long-time asymptotic methods are used to extract critical velocities that partition the response into distinct regions having markedly different location-dependent responses, and the amplitudes of oscillatory responses in these regions are determined.  Together, these characterize the magnitude and breadth of the solution response.  Results indicate that the signaling response in neutrally stable flows (that admit algebraic growth) is significantly different from that in exponentially unstable systems.
\end{abstract}


\maketitle 



%
%

%


\tableofcontents

\section{Introduction}
	\label{sig:sec:Intro}

Many liquid film coating processes have tight product specifications and require precise control over process conditions. Such processes are used, for example, to produce printed electronics, liquid display screens, high quality papers used for printing, and more~\cite{cohen1992,kistler1997}. 
Industrial processes are inevitably subject to ambient disturbances, some impulsive such as those from the sudden repositioning of equipment, and some repetitive from vibrations due to pumps and fans. 
Perhaps the simplest model for the response of fluid flows to repetitive disturbances is the so-called ``signaling'' problem, defined in one dimension (1D) as
\begin{equation}
\label{sig:eq:SigProbEx} 
Lh=A_f e^{i\omega_f t} \delta(x),
\end{equation}
where $L$ is a linear governing operator, $t$ is time, $A_f$ is the forcing amplitude, $\omega_f$ is the forcing frequency, $\delta(x)$ is the Dirac delta function, and $x\in(-\infty,\infty)$. Equation~(\ref{sig:eq:SigProbEx}) is subject to constraints to make the problem well-posed; the number of constraints is determined from the number of spatial and time derivatives in $L$. The signaling problem admits two components of the solution, the transient behavior due to the start-up of the disturbance and the long-time signaling behavior due to the oscillatory forcing. While most real-world disturbances occur over multiple frequencies and locations,~(\ref{sig:eq:SigProbEx}) provides a simple model from which the responses to arbitrary forcing functions can be constructed via superposition and the method of Green's functions; these are afforded by the linearity of the operator.
The motivation of this paper lies in the need to determine how the response to signaling from inevitable disturbances in manufacturing flows will propagate; this includes a characterization of both the breadth (how much of the product will be affected) and  the amplitude of the response (how badly will the product be affected).

In order to characterize the response of~(\ref{sig:eq:SigProbEx}), a first step is to examine the stability of the unforced medium itself by assessing response behavior of the homogeneous form $Lh=0$; this approach, known as classical stability analysis, is used extensively in the hydrodynamic literature~\cite{Chandrasekhar,
huerre2000,
schmid2001,
drazin2004}.
The solution to the homogeneous equation is expressed as an infinite sum of complex modes,
\begin{equation}
\label{sig:eq:ModalForm}
h(x,t)=\sum h_k(x,t)=\sum A_k e^{i(k_rx-\omega t)}
\end{equation}
\noindent
with complex amplitude, $A_k$, real wave number, $k_r$, and complex frequency, $\omega=\omega_r+i\omega_i$. The form~(\ref{sig:eq:ModalForm}) is substituted into the homogeneous operator without constraints, and nontrivial solutions are obtained that enable $A_k$ to be arbitrary.  The result is a dispersion relation of the form $\omega=\omega(k_r)$.
Notably, the assumed modal form, $h_k$, can be rewritten as an exponential term multiplied by an oscillatory term,
\begin{equation}
\label{sig:eq:ExponentialForm}
h_k=A_k e^{(\omega_it)} e^{i(k_rx-\omega_r)t}.
\end{equation}
\noindent
In the summation~(\ref{sig:eq:ModalForm}), the mode (or modes) that grows the fastest or decays the slowest will dominate the overall response as time goes to infinity. Therefore, we can express the maximum magnitude of $h_\mathrm{max}$ for $\omega_{i,\mathrm{max}}\equiv $ max$($Im$[\omega])$ and with the associated $A_k$ value $A_\mathrm{max}$ as
\begin{equation}
\label{sig:eq:MaxModalForm}
|h_{\mathrm{max}}|
\sim 
\big|
A_\mathrm{max}e^{\omega_{i\mathrm{max}}t}
\big|
\quad
\text{ as }
\quad
t\to\infty.
\end{equation}
\noindent
In~(\ref{sig:eq:MaxModalForm}), the sign of $\omega_{i,\mathrm{max}}$ is  used to determine the classical stability for a given set of conditions as follows~\cite{Chandrasekhar, HuerreRossi, huerre2000}:
\begin{subequations}
\label{sig:eq:general:ClassicalStabilityClassification}
\begin{align}
\label{sig:eq:general:Stability}
\omega_{i,\mathrm{max}}<0:&\quad\text{ The system is stable: exponential decay results},\\  
\label{sig:eq:general:NeutralStability}
\omega_{i,\mathrm{max}}=0:&\quad\text{ The system is neutrally stable: no growth or decay results},\\
\label{sig:eq:general:Instability}
\omega_{i,\mathrm{max}}>0:&\quad\text{ The system is unstable: exponential growth results}.
\end{align}
\end{subequations}
However, the method fails to properly classify neutrally stable~(\ref{sig:eq:general:NeutralStability}) cases  where the maximum value (or all values) of $\omega_i$ is zero. Such cases can lead to emergent algebraic behavior, as seen in~\cite{king2016,Huber2020} (and the references therein), where the solution magnitude goes as $t$ raised to a fractional power, such as $t^{\nicefrac{1}{2}}$ or $t^{\nicefrac{-1}{2}}$. Such behavior is fascinating because it arises from the summation of modes and cannot be deduced from any individual exponential mode or finite set of modes as in classically stable or unstable systems.

As shown in previous literature~\cite{Gordillo,barlow2017,barlow2012}, the spatio-temporal stability of the underlying medium itself (deduced from the homogeneous governing equation~(\ref{sig:eq:SigProbEx})) is central to the relevance of signaling.  In fact, transients are carried by the homogeneous solution, and these are initiated at $t=0$ in any system through initial conditions.  Previous work on signaling in spatio-temporal systems has focused on systems having exponential growth or decay in time according to~(\ref{sig:eq:MaxModalForm}) and~(\ref{sig:eq:general:ClassicalStabilityClassification}).  
If a flow is temporally stable, then the effects of transients damp exponentially with time and the response to the forcing in~(\ref{sig:eq:SigProbEx}) persists as the dominant solution.
Alternatively, if a flow is temporally unstable but convective, transients propagate away from the source even as they grow exponentially with time, and once again, the response to forcing in~(\ref{sig:eq:SigProbEx}) remains in any finite domain after the transient passes through. In this case, then, the processes can be controlled by minimizing disturbances and adjusting the rate at which exponential growth occurs through the choice of process conditions\footnote{In fluid systems, this can include viscosity, flow rate, etc.}, so that the response magnitude remains sufficiently small over a given finite domain.
If a flow is absolutely unstable, however, transients remain in the domain, never dissipate as they grow exponentially with time, and overwhelm the effect of forcing in~(\ref{sig:eq:SigProbEx}); therefore, the forcing solution in this case is subdominant to the transients at large times and thus is not of practical importance. Absolutely unstable processes cannot be controlled by minimizing process disturbances, as responses remain in the domain as they grow. As a result, it is impossible to maintain a high-precision process in absolutely unstable systems as products are pushed out of specifications.

Thus far, the solution response governed by linear operators that admit algebraic growth has only been studied for impulsive forcings~\cite{king2016,Huber2020}. In this work, we elucidate the structure of the response of a neutrally stable flow to a localized oscillatory forcing (i.e., the signaling problem~(\ref{sig:eq:SigProbEx})).  
\begin{figure}[ht!]
\centering
\begin{minipage}{.5\textwidth}
\includegraphics[keepaspectratio,width=3in]{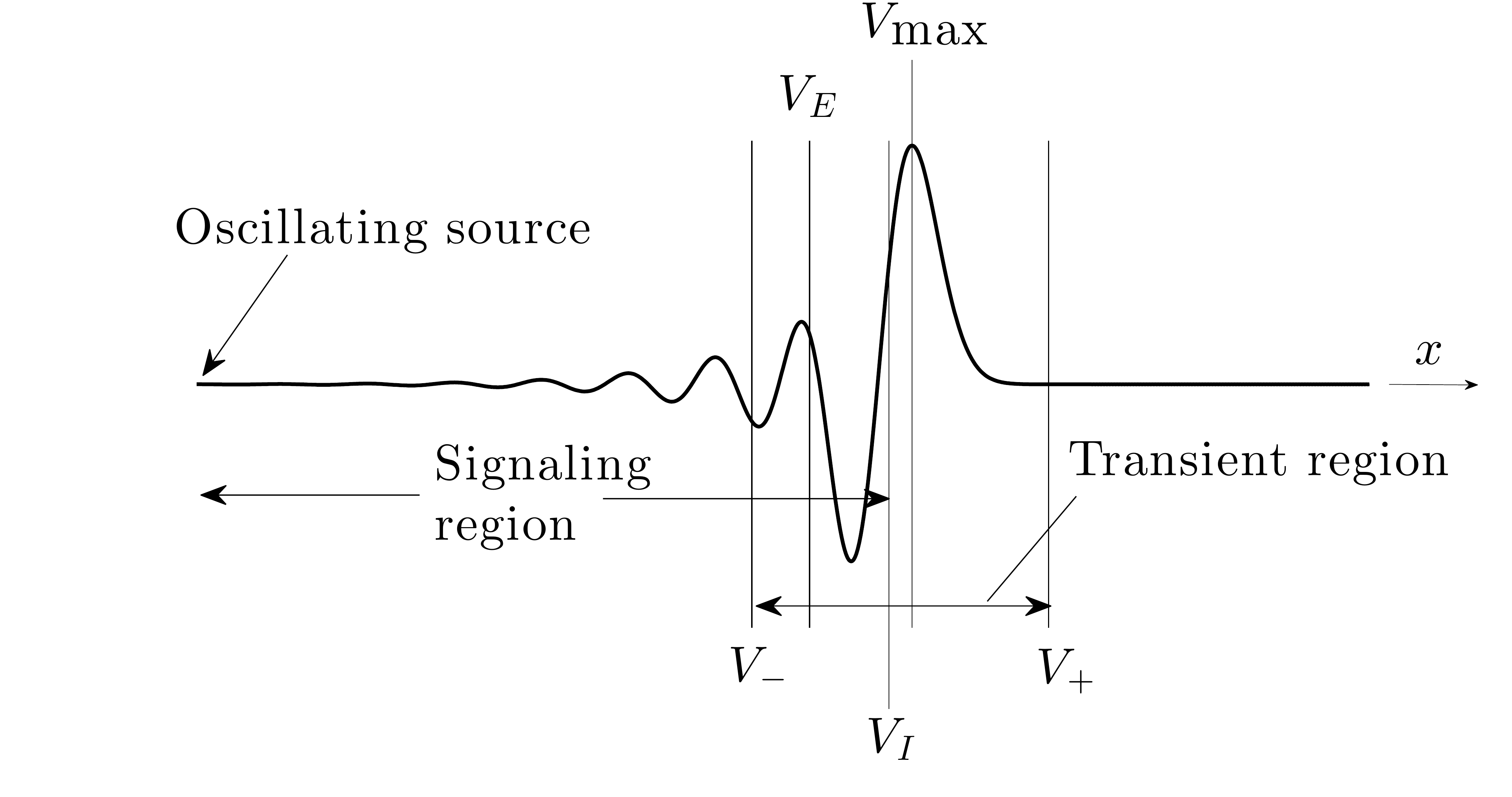}
\centering
\ref{sig:fig:Exp}a
\end{minipage}\hfill
\begin{minipage}{.5\textwidth}
\includegraphics[keepaspectratio,width=3in]{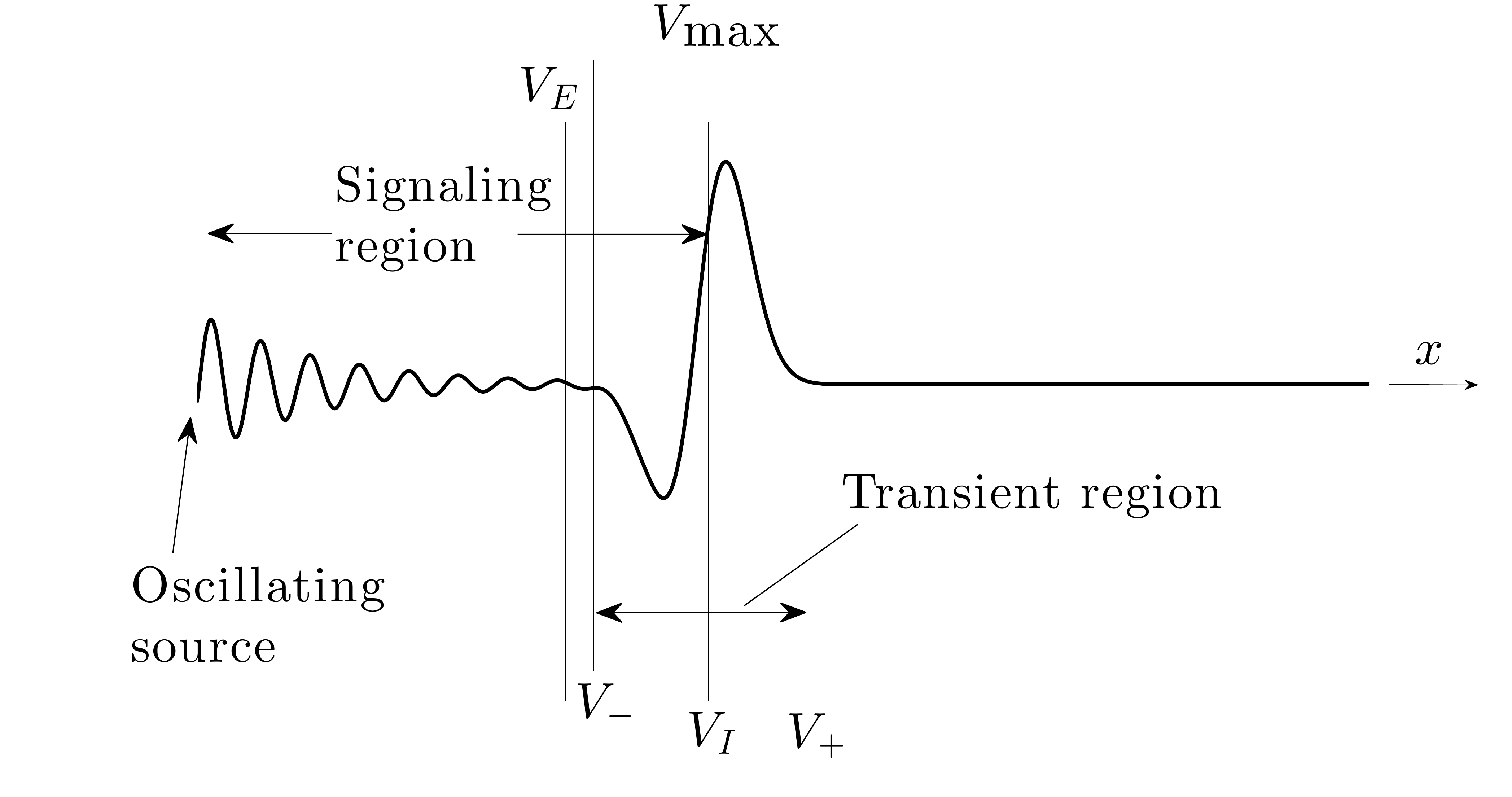}
\centering
\ref{sig:fig:Exp}b
\end{minipage}
\caption{Schematics of exponentially unstable signaling problems studied by Barlow et al.~\cite{barlow2017} replicated here with permission. Fig.~\ref{sig:fig:Exp}a is spatially unstable and Fig.~\ref{sig:fig:Exp}b is spatially stable. The velocities $V_+$ and $V_-$ are the bounds of the transient growth envelope around the peak $V_{\mathrm{max}}$, the velocities $V_I$ are the farthest downstream locations where the oscillating source affects the solution (the signaling region), and the velocity $V_E$ marks the point where the growth rate of waves generated by the oscillatory forcing is the same as those of the transient solution.}
\label{sig:fig:Exp}
\end{figure}
Relevant background is the closely-related signaling problem for operators that admit exponential growth and decay; see Barlow et al.~\cite{barlow2017} and references contained therein for an extensive review.
Fig.~\ref{sig:fig:Exp} provides a schematic of the signaling response in such cases.  As indicated, the structure of the solution is delineated by well-defined velocities; these velocities are extracted naturally from the long-time behavior using the method of steepest descent~\cite{bender1999}.  In particular, one can readily extract the bounding envelope of the transient ($V_+$,$V_-$) from the asymptotic expressions where the growth rate of the transient is zero, the farthest extent of the so-called signaling region ($V_I$) from pole inclusion in the relevant contour integrations, and the velocity ($V_E$) from the equality of the transient and forced oscillation response amplitudes. 
We demonstrate herein that the structure of the signaling response in a system that admits algebraic growth also has critical velocities, but the structure is markedly different. The neutrally stable system studied in this paper only has two types of velocities relevant to the overall structure: Those that correspond to the peak of the transient (of the type $V_\mathrm{max}$ in Fig.~\ref{sig:fig:Exp}) and those due to the inclusion of poles (of the type $V_I$ in Fig.~\ref{sig:fig:Exp}). While our approach is analogous to the studies that map the regions of interaction between exponentially growing waves (in space and time) in a classically unstable signaling problem~\cite{Gordillo,
barlow2017,
barlow2012}, the signaling problem examined here has a distinctly different structure that is explored through asymptotic analysis.

The paper is organized as follows:  In Section~\ref{sig:sec:ProblemStatement}, we pose the signaling problem that governs varicose waves in an inviscid planar liquid sheet at neutral stability. The governing linear operator here admits algebraic growth of waves when initiated by an impulse disturbance, as shown by King et al.~\cite{king2016}. 
In the subsequent sections, the signaling response is fully characterized through Fourier analysis and asymptotic methods. This examination begins in Section~\ref{sig:sec:FourierIntegral}, where the Fourier integral solution is provided.
In Section~\ref{sig:sec:CriticalVelocities}, critical velocities are determined that delineate regions of the signaling response --- this parallels the structure found in responses that exhibit exponential growth or decay as discussed above. Section~\ref{sig:sec:BoundingAmplitudes} describes the method to extract the bounding amplitudes of the oscillatory signaling response via analytical means. In Section~\ref{sig:sec:Results}, results showing the structure of the signaling response are provided, and Section~\ref{sig:app:CriticalVelocities} provides additional discussion on the relationship between critical velocities and the breadth of the response. The results are discussed in Section~\ref{sig:sec:Discussion}; this section also includes discussion which contrasts signaling in systems that admit algebraic and exponential growth. Concluding remarks are provided in Section~\ref{sig:sec:Conclusions}.

\section{Formulation}
	\label{sig:sec:ProblemStatement}

To uncover key features of signaling in a neutrally stable flow, we introduce localized oscillatory forcing to the model of King et al.~\cite{king2016} as,
\begin{subequations}\label{sig:eq:H_PDE}
\begin{equation}
\frac{\partial^2 H}{\partial t^2}+
c^2\frac{\partial^2H}{\partial x^2} +
2c \frac{\partial^2H}{\partial x \partial t}+
B^2
\frac{\partial^{4}H}{\partial x^{4}}=
A_f e^{(i\omega_f t)}\delta (x),
\end{equation}
\begin{equation}
H(x,0)=H_0\delta (x),\qquad
\frac{\partial H}{\partial t}(x,0)=U_0\delta (x),\qquad
H\to 0 \quad as \quad x\to\pm\infty,
\end{equation}

\end{subequations}
\noindent
where $H$ is the system response (given physical context in the discussion below), $H_0$ is the real amplitude of the impulsive perturbation in initial height, $U_0$ is the real amplitude of the impulsive perturbation in initial surface velocity, and the parameters $B$ and $c$ are real-valued coefficients. In order to study different phases of sinusoidal forcing, $A_f$ is taken to be complex. Note that in King et al., $\delta(t)$ is used instead of $e^{i\omega_f t}$ in the right-hand side of~(\ref{sig:eq:H_PDE}a), as the focus of that work is the response of the operator to impulsive disturbances~\cite{king2016}.   

\begin{figure}[ht!]
\centering
\includegraphics[keepaspectratio,width=6in]{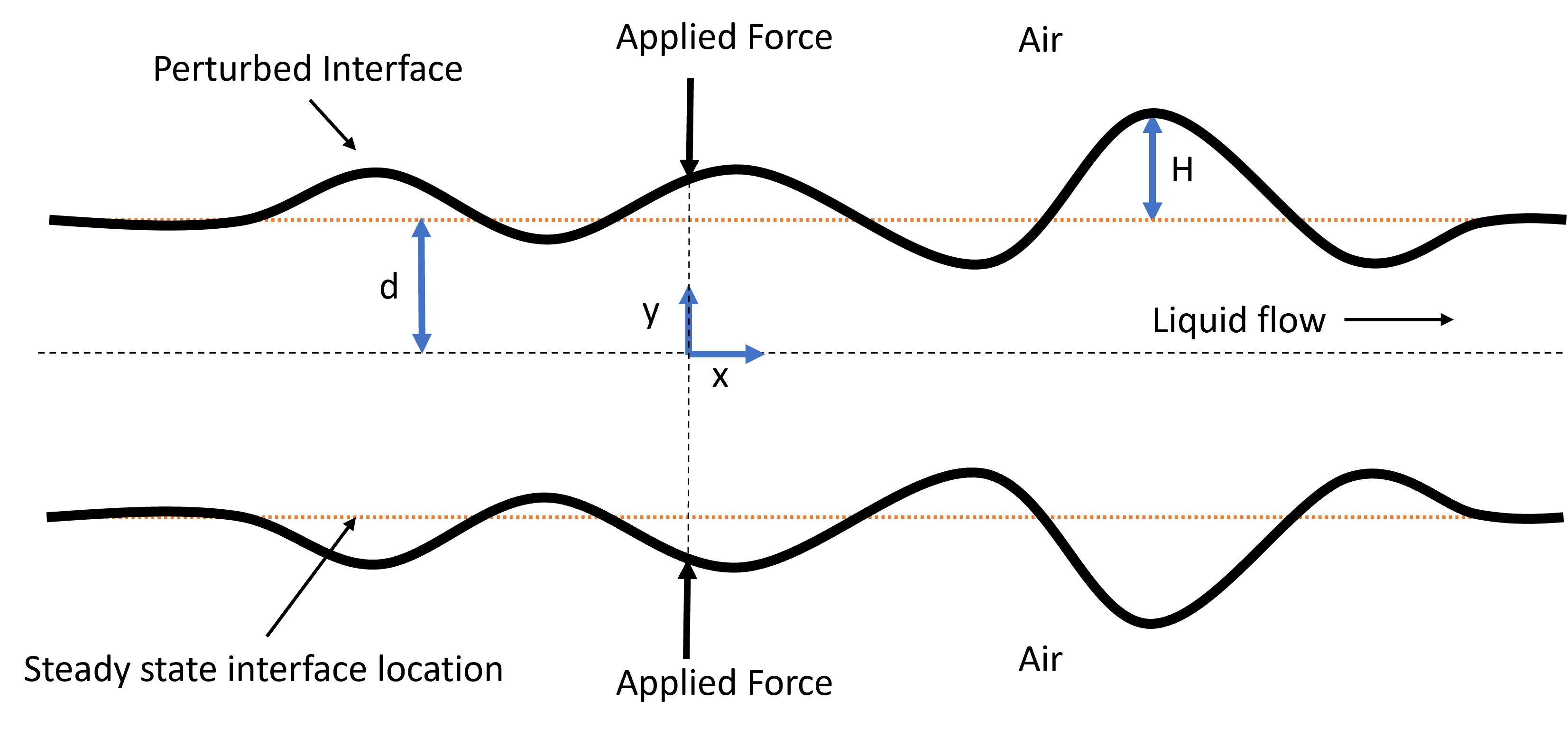}
\caption{Schematic of varicose waves in a planar liquid sheet. The flow is invariant in the $z$ direction, which is oriented into and out of the page. The non-dimensional steady state half-sheet thickness is $d$, and the perturbations from it are given by height $H$. Note that the indicated response $H$ has been exaggerated for clarity, $H$ must be much smaller than $d$ to be consistent with the linearity assumption.
}
\label{sig:fig:VaricoseSchematic}
\end{figure}

Physically, the linear operator in~(\ref{sig:eq:H_PDE}) governs the propagation of long varicose waves in an inviscid planar liquid sheet in the absence of ambient gas once the identifications $B=We^{-1/2}$ and $c=1$ are made~\cite{lin2003}. In this context, $We=\frac{\rho_l u^2 h_0}{\sigma}$ is the dimensionless Weber number, where $\rho_l$ is the liquid density, $u$ is the liquid velocity in the sheet, $h_0$ is the half-sheet thickness, and $\sigma$ is the surface tension; a schematic is provided in Fig.~\ref{sig:fig:VaricoseSchematic}. This flow has long been known to be neutrally stable~\cite{lin2003,deluca1997} in accordance with~(\ref{sig:eq:general:ClassicalStabilityClassification}) and exhibits algebraic instability~\cite{king2016,barlow2011} when $U_0$ is nonzero. In the context of varicose waves in liquid sheets, the oscillatory forcing introduced here has only been studied for responses, taken in \textit{the presence of ambient gas}, that admit exponential growth \cite{barlow2012} and are governed by a different operator than~(\ref{sig:eq:H_PDE}). In the following sections, we uncover, for the first time, the structure that arises when signaling is imposed on a neutrally stable flow.

Because the partial differential equation (PDE) in~(\ref{sig:eq:H_PDE}) is linear, the dependent variable, $H$, can be expressed by superposition as follows:
\[ H(x,t)=g(x,t)+h(x,t), \]
\noindent
where $g$ and $h$ are expressed as
\begin{subequations}\label{sig:eq:g_PDE}
\begin{equation}
\frac{\partial^2 g}{\partial t^2}+
c^2\frac{\partial^2g}{\partial x^2} +
2c \frac{\partial^2g}{\partial x \partial t}+
B^2
\frac{\partial^{4}g}{\partial x^{4}}=0,
\end{equation}
\begin{equation}
g(x,0)=H_0\delta (x),\qquad
\frac{\partial g}{\partial t}(x,0)=U_0\delta (x),\qquad
g\to 0 \quad as \quad x\to\pm\infty,
\end{equation}
\end{subequations}
\noindent
and
\begin{subequations}\label{sig:eq:h_PDE}
\begin{equation}
\frac{\partial^2 h}{\partial t^2}+
c^2\frac{\partial^2h}{\partial x^2} +
2c \frac{\partial^2h}{\partial x \partial t}+
B^2
\frac{\partial^{4}h}{\partial x^{4}}=
A_f e^{(i\omega_f t)}\delta (x),
\end{equation}
\begin{equation}
h(x,0)=0,\qquad
\frac{\partial h}{\partial t}(x,0)=0,\qquad
h\to 0 \quad as \quad x\to\pm\infty.
\end{equation}
\end{subequations}

\noindent
The split between~(\ref{sig:eq:g_PDE}) and~(\ref{sig:eq:h_PDE}) allows us to study the effect of the forcing in~(\ref{sig:eq:h_PDE}) without repeating the work done by King et al.~\cite{king2016}. Note that equation~(\ref{sig:eq:g_PDE}) omits the impulsive forcing ($A\delta(x)\delta(t)$) present in King et al. as it can be shown that it is equivalent to a disturbance in the initial surface velocity with $A=U_0$ in~(\ref{sig:eq:g_PDE}). 
King et al. found that, for nonzero $U_0$, the solution propagates and grows as $t^{\nicefrac{1}{2}}$ along the single velocity $\nicefrac{x}{t}=c$; however, a nonzero displacement $H_0$ in~(\ref{sig:eq:g_PDE}b) yields a damped solution. The superposition above allows us to focus specifically on the response to the oscillatory forcing in~(\ref{sig:eq:h_PDE}), and how the long-time response to the forcing (i.e., signaling) evolves past the transients invoked by the start-up.

\section{Fourier Integral Solution}
	\label{sig:sec:FourierIntegral}

The solution to~(\ref{sig:eq:h_PDE}) is found by taking the Fourier transform in $x$, solving the resulting ordinary differential equation in time, and setting up the Fourier inversion integral (see Appendix~\ref{sig:app:FourierSolution} for additional details). The inversion integral is broken into two integrals, $\mathbb{J}$ and $\mathbb{L}$ as
\begin{subequations}
\label{sig:eq:InversionIntegral}
\begin{equation}
h=\frac{\mathbb{J}+\mathbb{L}}{2\pi},
\end{equation}
\begin{equation}
\label{sig:eq:ITilde}
\mathbb{J}=
\int\limits_{-\infty}^{\infty}
\frac{
-\frac{A_fe^{i(\omega_f)t}}{B^2}e^{i(kV)t}
\left(
e^{iB(k-k_2)(k-k_3)t}-1
\right)}
{(k-k_0)(k-k_1)(k-k_2)(k-k_3)}dk,
\end{equation}
\begin{equation}
\label{sig:eq:IHat}
\mathbb{L}=
\int\limits_{-\infty}^{\infty}
\frac{iA_f\sin(Bk^2t)e^{i(kV-kc)t}}
{k^2B^2(k-k_0)(k-k_1)}dk,
\end{equation}
\begin{equation*}
k_0=\frac{-c-\sqrt{c^2-4B\omega_f}}{2B}
,\qquad
k_1=\frac{-c+\sqrt{c^2-4B\omega_f}}{2B},
\end{equation*}
\begin{equation}
\label{sig:eq:PoleLocs}
k_2=\frac{c-\sqrt{c^2+4B\omega_f}}{2B}
,\qquad
k_3=\frac{c+\sqrt{c^2+4B\omega_f}}{2B},
\end{equation}

\end{subequations}

\noindent
where $V\equiv\nicefrac{x}{t}$ is taken to be a fixed velocity at which we evaluate the integral.  Note that the introduction of $V$ as a fixed parameter naturally arises through analysis of the long-time asymptotics of the integral.  Furthermore, from prior work discussed in Section~\ref{sig:sec:Intro}, structural changes in the solution are naturally characterized by different values of $V$. 

A key feature of the integrals in~(\ref{sig:eq:InversionIntegral}) are the poles $k_0$ through $k_3$ which arise from the forcing function. When we evaluate~(\ref{sig:eq:ITilde}) and~(\ref{sig:eq:IHat}) through complex contour integration, the location of the poles determines whether they have an effect on the solution; contours are examined in Supplemental Material Sections~\ref{sig:sup:BowTieContour} and~\ref{sig:sup:HalfPlaneContour}.
A significant change occurs in the structure based on the sign of square root discriminants in~(\ref{sig:eq:PoleLocs}). The critical value of forcing frequency, denoted as $\omega_c$, arises when the discriminants of $k_0$ and $k_1$ are zero, i.e.
\begin{equation}\label{sig:eq:CriticalFrequency}
\omega_c=\frac{c^2}{4B}.
\end{equation}

\begin{figure}[ht!]
\centering
\begin{minipage}{.45\textwidth}
\centering
\includegraphics[keepaspectratio,width=3in]{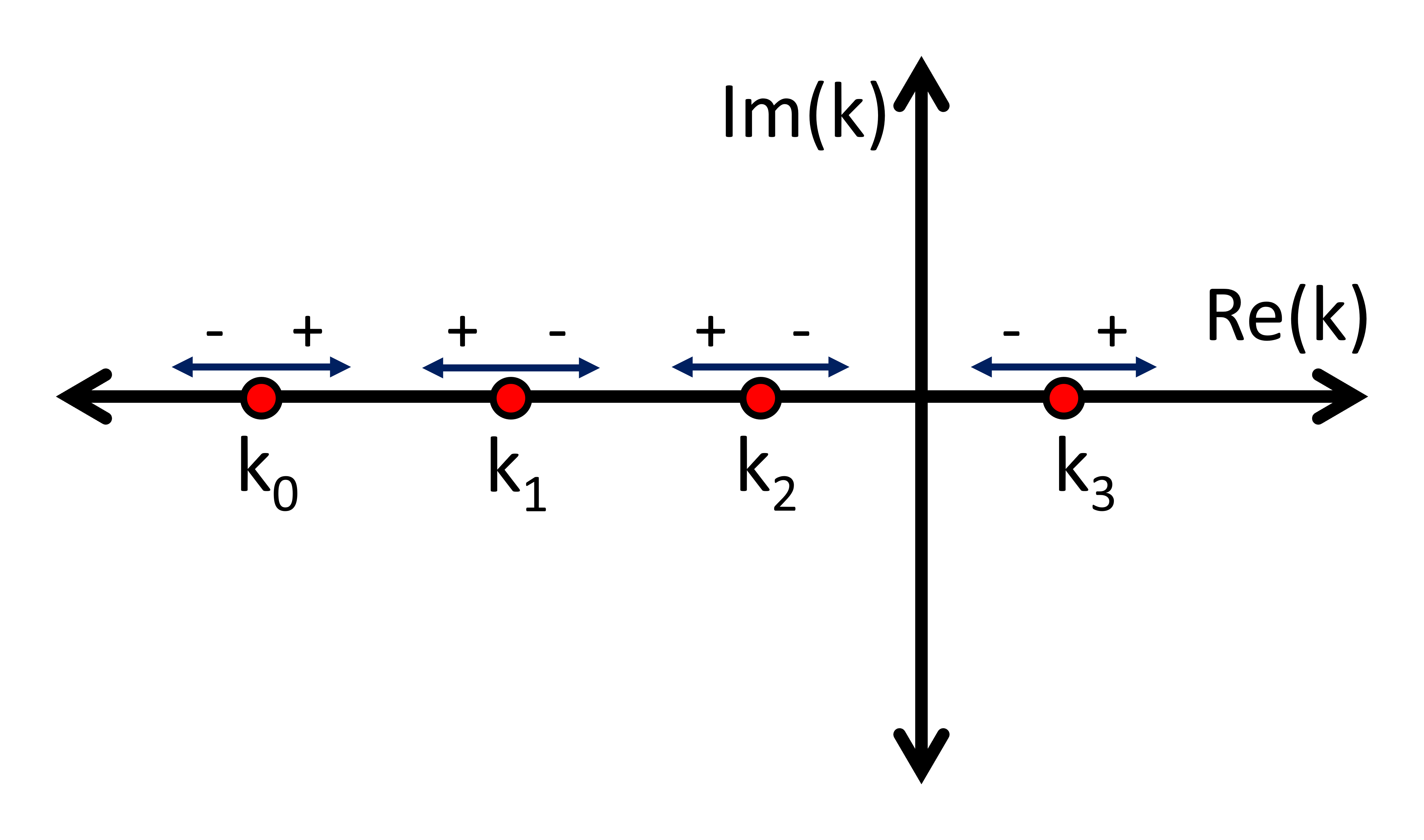}
\caption{$\omega_f<\omega_c$: General schematic showing the movement of the poles with small increases ($+$) and decreases ($-$) in the forcing frequency, $\omega_f$, for $\omega_f<\omega_c$. Note that all four poles lie on the real $k$ axis.}
\label{sig:fig:LTCPoles}
\end{minipage}\hfill
\begin{minipage}{.45\textwidth}
\centering
\includegraphics[keepaspectratio,width=2.5in]{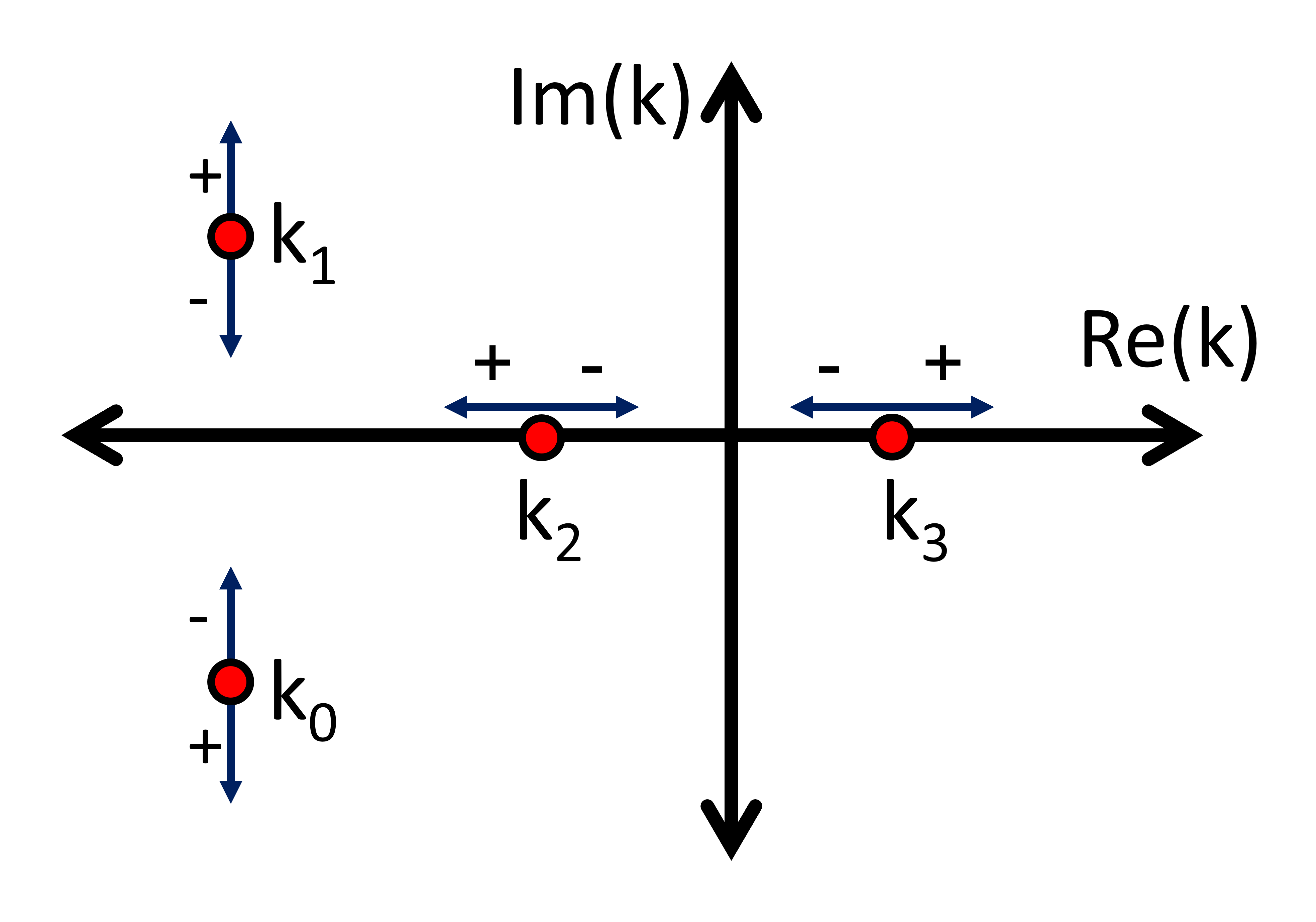}
\caption{$\omega_f>\omega_c$: General schematic showing the movement of the poles with small increases ($+$) and decreases ($-$) in the forcing frequency, $\omega_f$, for $\omega_f>\omega_c$. Note that $k_0$ and $k_1$ lie off of the real axis, and $k_2$ and $k_3$ lie on the real axis.}
\label{sig:fig:GTCPoles}
\end{minipage}
\end{figure}

\noindent
When the system is forced at a frequency lower than the critical value ($\omega_f<\omega_c$), all four poles lie on the real $k$ axis, as shown in Fig.~\ref{sig:fig:LTCPoles}.
When the system is forced at a frequency higher than the critical value ($\omega_f>\omega_c$), poles $k_0$ and $k_1$~(\ref{sig:eq:PoleLocs}) lie off of the real $k$ axis and poles $k_2$ and $k_3$~(\ref{sig:eq:PoleLocs}) lie on the real $k$ axis, as shown in Fig.~\ref{sig:fig:GTCPoles}. In what follows, we will explore the structure arising from values of $\omega_f<\omega_c$ and $\omega_f>\omega_c$. The case of $\omega_f=\omega_c$  is examined using the Fourier Series Solution (FSS) (Appendix~\ref{sig:app:FSS}) of~(\ref{sig:eq:h_PDE})  and addressed at the end of Section~\ref{sig:app:CriticalVelocities}.  
As shall be seen in Appendix~\ref{sig:app:FourierSolution}, the location of the poles in Figs.~\ref{sig:fig:LTCPoles} and~\ref{sig:fig:GTCPoles} is crucial to the asymptotic evaluation of the Fourier integral solution~(\ref{sig:eq:InversionIntegral}) in the limit as $t\to\infty$. 
A key detail of the analysis is that contour integration is used to evaluate the integrals in~(\ref{sig:eq:InversionIntegral}), and the position of the contour in the complex plane is a function of the velocity $V$.
This means that, for a fixed set of parameters, the velocity at which the integral is being evaluated determines the poles that are enclosed, and this, in turn, determines the structure of the solution. The details of the asymptotic evaluation of the inversion integrals is provided in Appendix~\ref{sig:app:FourierSolution}.

\subsection{Approach to determine critical velocities}
\label{sig:sec:CriticalVelocities}

Critical velocities examined in spatio-temporal problems have been used to delineate structural changes in solution responses having exponential growth or decay~\cite{huerre2000} as shown in Fig.~\ref{sig:fig:Exp}. However, many of the critical velocities in such systems are based on transitions between exponentially growing and decaying portions of the solution response~\cite{Gordillo,barlow2017,barlow2012}.  Thus, many of the ``bounding velocities'' and associated structural changes are different when an operator that admits algebraic growth is considered.

Here, we define \textit{critical velocities} as those which --- based on pole, saddle, and contour orientation --- delineate structural changes in the solution. Of these critical velocities, there are two categories, observable and non-observable. \textit{Observable critical velocities} delineate distinct regions of the solution response that persist in the long-time limit, while \textit{non-observable critical velocities} are those where structural changes do not persist in that limit.
The widths of these regions are crucial to understanding how disturbances propagate in a forced neutral flow such as~(\ref{sig:eq:h_PDE}). 
Although observable critical velocities may be approximated by inspection from a numerical or FSS (Appendix~\ref{sig:app:FSS}), analytical expressions for these velocities emerge from the long-time asymptotic solution. All of the critical velocities can be extracted early in the process of the asymptotic analysis, but require more work to definitively distinguish between those that lead to observable changes in solution response in the long time limit (observable critical velocities) and those that do not (non-observable critical velocities). 

Two critical velocities can be identified immediately without additional analysis. The zero velocity, $V=0$, delineates the upstream and downstream directions and is the location of the imposed oscillatory disturbance. The peak velocity, $V=c$, is extracted from the homogeneous form of the equation by King et al.~\cite{king2016} and provides the velocity of the algebraically growing peak. 
In order to determine the additional critical velocities relevant to the signaling problem, we note that the Fourier integrals in~(\ref{sig:eq:InversionIntegral}) are of similar form, and can be written generally as:
\begin{equation}
\label{sig:eq:SampleFourierInversion}
Integral=
\int\limits_{-\infty}^{\infty} \frac{1}{f(k)}e^{\phi(k)t}dk.
\end{equation}
\noindent
In~(\ref{sig:eq:SampleFourierInversion}), there are a finite number of simple poles, $k_i$, where $f(k_i)=0$ and some value of $k_s(V)$, which is an $n^\text{th}$ order saddle point of $\phi(k)$, such that $\frac{d^{n-1}}{dk^{n-1}}(\phi)\big|_{k_s}=0$ and $\frac{d^{n}}{dk^{n}}(\phi)\big|_{k_s}\neq 0$. The behavior of the integral near saddle points is relevant to the long-time asymptotic behavior via the method of steepest descents~\cite{bender1999}. The location of the saddle point is a function of $V$ (i.e., $k_s=k_s(V)$), indicating that any given velocity corresponds to a saddle point. Each saddle point leads to a different arrangement of the complex contours used in the asymptotic analysis (see Appendix~\ref{sig:app:FourierSolution}).
For a given velocity and set of parameters, the poles are either  enclosed, not enclosed, or intersected by the contour. Critical velocities are extracted by finding all the real values of $V$ where a pole of $f(k)$ intersects a contour. 
The simplest critical velocities to determine are those associated with the poles that lie on the locus of saddle points, as it is only a matter of setting $k_s(V)=k_i$ and solving for $V$. The velocities associated with poles that lie off the locus of saddles are more difficult to find, as they require establishing how the contours sweep through the complex $k$ plane as shown in Appendix~\ref{sig:app:FourierSolution} and the Supplemental Material. Note that, for~(\ref{sig:eq:h_PDE}), the locus of saddles lies along the entirety of the real $k$ axis. In what follows, no distinction will be made between the ``locus of saddles'' and ``real $k$ axis''.

Overall, then, every critical velocity (besides $V=0$ and $V=c$) corresponds to either: (i) a change in whether a given pole is enclosed or not within a contour (going from not enclosed to enclosed or vice versa) or (ii) the situation where a saddle point and pole are coincident. 
As defined earlier, these velocities indicate where structural changes in the solution \textit{could} occur at long times.
Once all the critical velocities are found, they can be overlaid on a numerical or Fourier series solution to visually determine those which yield significant changes in the solution (i.e., are observable critical velocities) without having to undertake a full asymptotic analysis. 
If a term in the asymptotic form associated with a critical velocity decays exponentially in time when active, then it will not have any lingering structural effect on the solution response at large times. These velocities correspond to the off-axis pole locations in Fig.~\ref{sig:fig:GTCPoles} and are non-observable. The only observable critical velocities, then, are those associated with the coincidence of a pole and saddle along the real $k$ axis. In fact, it is the difference in the number of on-axis poles that accounts for the differences in the solution structure associated with the cases $\omega_f<\omega_c$ and $\omega_f>\omega_c$ configurations as will be seen. 
Note that observable critical velocities bound upstream and downstream regions of the response, and it is possible to identify portions of the response corresponding to transients and the oscillatory forcing. In light of comments made in the introduction regarding damage to thin film coatings, the breadth of these regions, as well as the amplitude of the response within them, is of practical relevance.

\subsection{Approach to determine bounding amplitudes}
\label{sig:sec:BoundingAmplitudes}

Within each region delineated by the critical velocities of Section~\ref{sig:sec:CriticalVelocities}, there are three potential outcomes; the disturbance grows, decays, or oscillates with a constant amplitude.
At large times, the spatial resolution needed to accurately predict features of the solution is quite large (see Appendix~\ref{sig:app:FSSResolution} for more details).  As the solution has an oscillatory character in each region, a bounding amplitude enables us to determine the maximum damage to the liquid sheet without needing to obtain the full solution to the required resolution. 
The bounding amplitude can certainly be extracted numerically by looking at the solution response, although this must be done individually for every set of parameters. Bounding amplitudes can be extracted analytically using the long-time asymptotic solutions and, as such, this section provides a higher level discussion about how such amplitudes are extracted.

To extract the amplitude of the solution within a single region between two adjacent critical velocities $V_n$ and $V_m$, we note that the asymptotic solution is expressed as a sum of one or more distinct exponentials with different phases (see Sections~\ref{sig:sec:LTCSol} and~\ref{sig:sec:GTCSol}). The general form of the asymptotic solution is written as

\begin{equation}
	\label{sig:eq:SampleExpSum}
	Real[h(x,t)] 
	\sim Real\left[ \sum_{n=1}^N
	\alpha_ne^{i\beta_n(V)t}
	\right]
	\quad V\in[V_n,V_m]
	 \quad \text{as} \quad t\to\infty,
\end{equation}

\noindent
where $\alpha_n$ is some complex constant. The maximum possible amplitude ($\mathcal{A}$) of~(\ref{sig:eq:SampleExpSum}) (see Supplemental Material Section~\ref{sig:app:BoundingAmp}) can be expressed as 
\begin{equation}
	\label{sig:eq:SampleAmp}
	\Big|
	Real[h(x,t)]
	\Big| \lesssim
	\mathcal{A}
	= \sum_{n=1}^N
	\big|
	\alpha_n
	\big|
	\qquad V\in[V_n,V_m]
	\qquad \text{as} \qquad t\to\infty.
\end{equation}

\noindent
Using~(\ref{sig:eq:SampleAmp}), the extracted amplitudes are obtained and reported in Sections~\ref{sig:sec:LTCSol} and~\ref{sig:sec:GTCSol}.
Note that the algebraic growth, which can occur at the velocity of $V=c$, is not considered when determining the bounding amplitudes --- this velocity is handled separately as will be seen.

\subsection{Long-time asymptotic solution for $\omega_f<\omega_c$}
	\label{sig:sec:LTCSol}
	
Following the approach of Barlow et al.~\cite{barlow2010}, a Fourier Series Solution (FSS) of~(\ref{sig:eq:h_PDE}) is used to generate Fig.~\ref{sig:fig:FourierLT}, which provides a typical solution response for a case where $\omega_f<\omega_c$; see Appendix~\ref{sig:app:FSS} for details.
Note that the large number of oscillations in the solution are not caused by numerical noise or aliasing --- the solution indeed responds in this way, and when magnified in any region, is entirely smooth (see Appendix~\ref{sig:app:FSSResolution} for additional details on resolving oscillations). Fig.~\ref{sig:fig:FourierLT} is presented in terms of the velocity $V\equiv\nicefrac{x}{t}$ as this domain provides critical insight into the structure of the solution; we elaborate on this below. Note that at any time $t$, the $V$ domain may be interpreted as being simply a rescaling of $x$; the $V$ domain has the advantage that the structural features are kept stationary in time for a fixed $V$ domain.

\begin{figure}[ht!]
\centering
\includegraphics[keepaspectratio,width=6in]{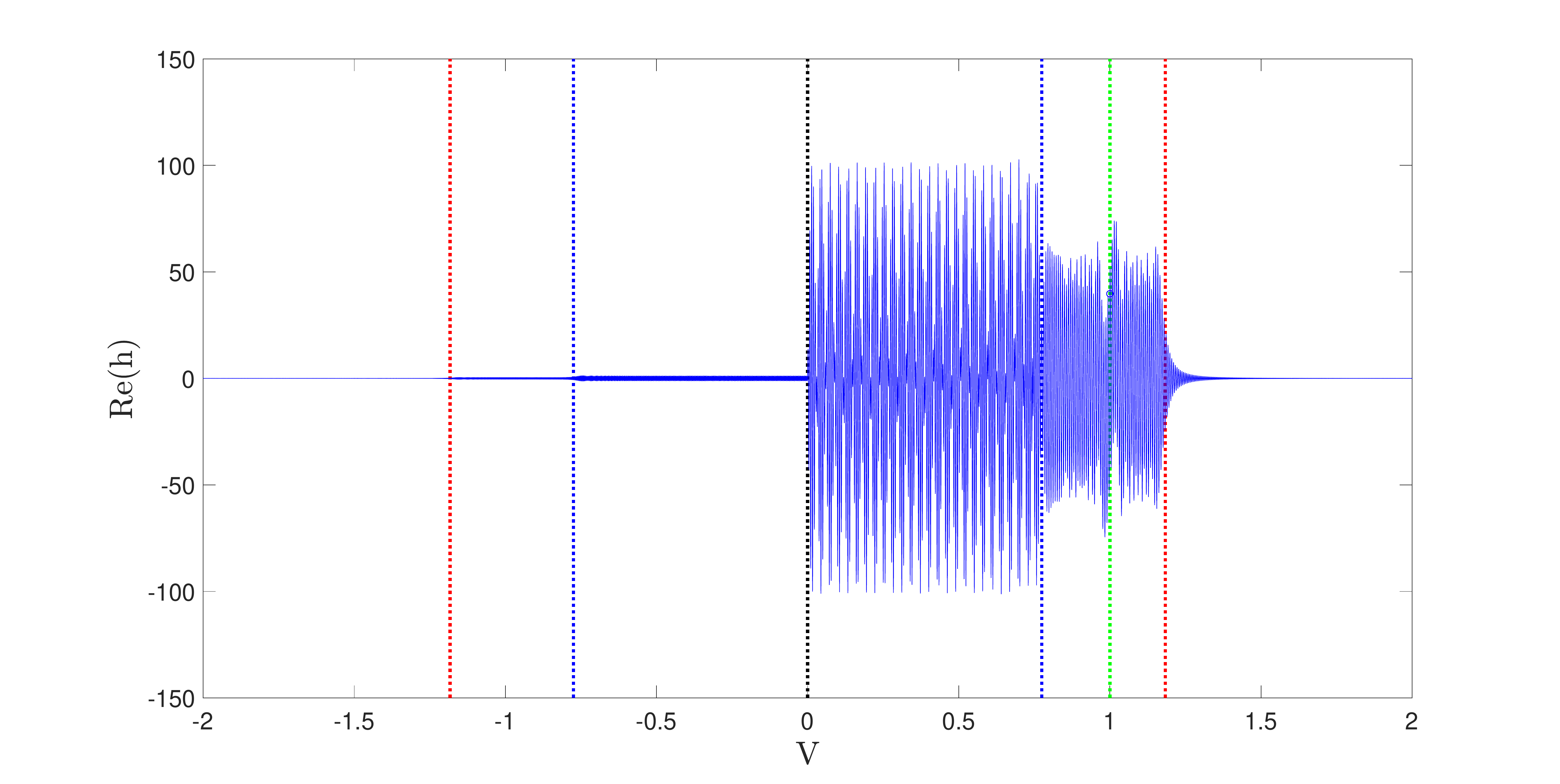}
\caption{Fourier Series Solution~(Appendix~\ref{sig:app:FSS}) of~(\ref{sig:eq:h_PDE}) for $\omega_f<\omega_c$ and $t=10000$ with $A_f=1$, $\omega_f=0.1$, $B=1$, $c=1$, and $L=N=100000$. Observable critical velocities (extracted analytically from the integral solution) are marked with dotted vertical lines (see Section~\ref{sig:sec:CriticalVelocities} for more details). Results are plotted against $V$ in order to capture the critical velocities and separate regions.
}
\label{sig:fig:FourierLT}
\end{figure}

\noindent
When viewing the solution in terms of $V$, the frequencies of the oscillations increase with time; as additional wave cycles are compressed into a fixed range of $V$, the density of the wave cycles increases, leading to solutions which appear to be rectangular blocks (Figs.~\ref{sig:fig:FourierLT} and~\ref{sig:fig:FourierGT}).
It is important to note, however, that the region between any two velocities $V_n$ and $V_m$ at time $t$ covers a spatial distance of $(V_m-V_n)t$ and thus expands with time. 
If we look at a fixed value of $t$, $V$ is a direct scaling of $x$ for the entire domain. 
If we look at a fixed point in $x$, however, the point will move towards decreasingly smaller $V$ values as time progresses. 
For any non-zero $x$ location, small times correspond to large magnitudes of $V$.  As $t$ increases, the magnitude of $V$ decreases until the $x$ location sees a change in structure at an observable critical velocity. As $t$  continues to increase and approaches infinity, the magnitude of $V$ approaches zero for the $x$ location. Therefore, the long-time solution for any finite $x$ value may be deduced based on the behavior of the solution near $V=0$ where there are no further transitions in structure across critical velocities. 

Applying the methodology from Sections~\ref{sig:sec:CriticalVelocities} and~\ref{sig:sec:BoundingAmplitudes} to the long-time asymptotic solution of~(\ref{sig:eq:InversionIntegral}) (see Appendix~\ref{sig:app:LTC}) leads to analytical expressions for the solution response, bounding amplitudes of each region, and the associated critical velocities that separate them.  The results are: 
\begin{equation*}
V_1=-\sqrt{c^2+4B\omega_f}
,\quad
V_2=-\sqrt{c^2-4B\omega_f},
\end{equation*}
\begin{equation}
\label{sig:eq:AmpVelLTC}
V_3=\sqrt{c^2-4B\omega_f}
,\quad
V_4=\sqrt{c^2+4B\omega_f}.
\end{equation}

\noindent
For the $\omega_f<\omega_c$ case, the leading-order behaviors within each of the six regions are as follows:

\begin{subequations}
	\label{sig:eq:LTCRegionSolutions}
\noindent
\underline{$-\infty<V<V_1 \qquad t\to\infty$:}
		\begin{equation*}
		h\sim O\left(t^{-\frac{1}{2}}\right),
		\end{equation*}
		\begin{equation}
		\label{sig:eq:AmpLT:Region1}
		\mathcal{A}_{(\omega_f<\omega_c)}=0.
		\end{equation}

\noindent
\underline{$V_1<V<V_2 \qquad t\to\infty$:}
		\begin{equation*}
		h\sim
		-i\frac{A_f}{B^2}
		\frac{e^{i\omega_ft}e^{ik_3Vt}}
		{(k_3-k_0)(k_3-k_1)(k_3-k_2)},
		\end{equation*}
		\begin{equation}
		\label{sig:eq:AmpLT:Region2}	
		\mathcal{A}_{(\omega_f<\omega_c)}
		=\frac{A_f}{B^2}
		\Bigg|
		\frac{1}{(k_3-k_0)(k_3-k_1)(k_3-k_2)}
		\Bigg|.
		\end{equation}
	
\noindent
\underline{$V_2<V<0 \qquad t\to\infty$:}
		\begin{equation*}
		h\sim
		-i\frac{A_f}{B^2}
		\frac{e^{i\omega_ft}e^{ik_3Vt}}
		{(k_3-k_0)(k_3-k_1)(k_3-k_2)}
		-i\frac{A_f}{B^2}
		\frac{e^{i\omega_ft}e^{ik_0Vt}}
		{(k_0-k_1)(k_0-k_2)(k_0-k_3)},
		\end{equation*}
		\begin{equation}
		\label{sig:eq:AmpLT:Region3}
		\mathcal{A}_{(\omega_f<\omega_c)}
		=\frac{A_f}{B^2}
		\Bigg[
		\bigg|
		\frac{1}{(k_3-k_0)(k_3-k_1)(k_3-k_2)}
		\bigg|
		+
		\bigg|
		\frac{1}{(k_0-k_1)(k_0-k_2)(k_0-k_3)}
		\bigg|
		\Bigg].
		\end{equation}

\noindent
\underline{$0<V<V_3 \qquad t\to\infty$:}
		\begin{equation*}
		h\sim
		i\frac{A_f}{B^2}
		\frac{e^{i\omega_ft}e^{ik_2Vt}}
		{(k_2-k_0)(k_2-k_1)(k_2-k_3)}
		-i\frac{A_f}{B^2}
		\frac{e^{i\omega_ft}e^{ik_1Vt}}
		{(k_1-k_0)(k_1-k_2)(k_1-k_3)},
		\end{equation*}
		\begin{equation}
		\label{sig:eq:AmpLT:Region4}
		\mathcal{A}_{(\omega_f<\omega_c)}
		=\frac{A_f}{B^2}
		\Bigg[
		\bigg|
		\frac{1}{(k_2-k_0)(k_2-k_1)(k_2-k_3)}
		\bigg|
		+ 
		\bigg|
		\frac{1}{(k_1-k_0)(k_1-k_2)(k_1-k_3)}
		\bigg|
		\Bigg].
		\end{equation}

\noindent
\underline{$V_3<V<V_4 \qquad t\to\infty$:}
		\begin{equation*}
		h\sim
		i\frac{A_f}{B^2}
		\frac{e^{i\omega_ft}e^{ik_2Vt}}
		{(k_2-k_0)(k_2-k_1)(k_2-k_3)},
		\end{equation*}
		\begin{equation}
		\label{sig:eq:AmpLT:Region5}	
		\mathcal{A}_{(\omega_f<\omega_c)}
		=\frac{A_f}{B^2}
		\Bigg|
		\frac{1}{(k_2-k_0)(k_2-k_1)(k_2-k_3)}
		\Bigg|.	
		\end{equation}
	
\noindent
\underline{$V_4<V<\infty \qquad t\to\infty$:	}
		\begin{equation*}
		h\sim O\left(t^{-\frac{1}{2}}\right),
		\end{equation*}
		\begin{equation}
		\label{sig:eq:AmpLT:Region6}
		\mathcal{A}_{(\omega_f<\omega_c)}=0.
		\end{equation}

\end{subequations}

\noindent
Higher order asymptotic corrections to~(\ref{sig:eq:LTCRegionSolutions}) can be found in Appendix~\ref{sig:app:LTC}, and the poles $k_0$ through $k_3$ are defined in~(\ref{sig:eq:PoleLocs}). A detailed examination of the precise response in each region (i.e., the solutions for $h$ in~(\ref{sig:eq:LTCRegionSolutions})) is provided in Section~\ref{sig:sec:Results:LTCRegime} to follow.

\begin{figure}[ht!]
\centering
\includegraphics[keepaspectratio,width=6in]{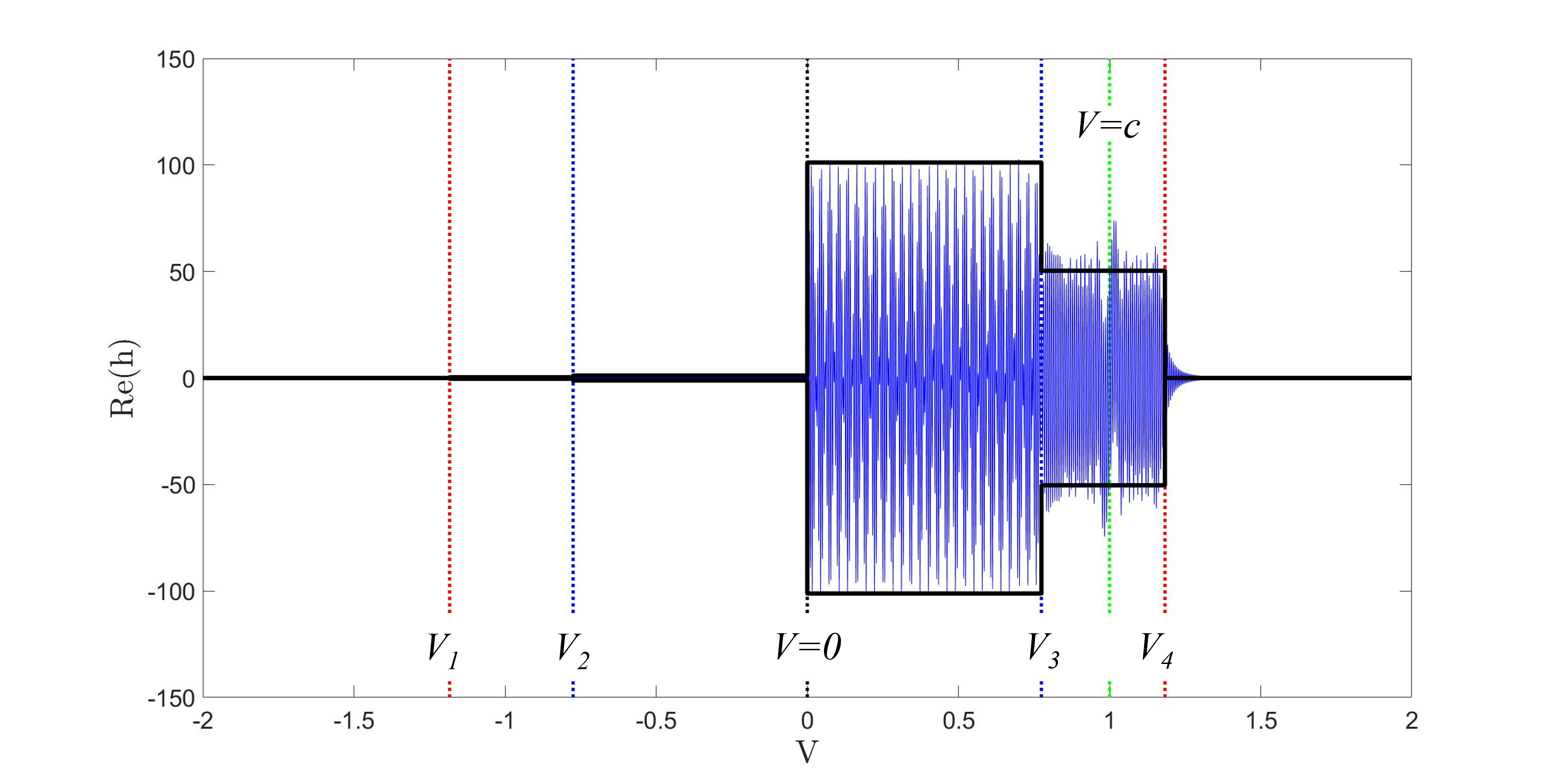}
\caption{Bounding amplitudes (given by~(\ref{sig:eq:LTCRegionSolutions}))  overlaid on the Fourier series solution in Fig.~\ref{sig:fig:FourierLT} for $\omega_f<\omega_c$. Critical velocities (extracted analytically from the integral solution) are marked with dotted vertical lines. The observable critical velocities from~(\ref{sig:eq:AmpVelLTC}) are labeled.
}
\label{sig:fig:FourierBoxLT}
\end{figure}

Fig.~\ref{sig:fig:FourierBoxLT} overlays the amplitude/critical-velocity asymptotic results given by~(\ref{sig:eq:LTCRegionSolutions}) on Fig.~\ref{sig:fig:FourierLT}.  As indicated, the bounding amplitudes and delineated regions agree well with the FSS~(Appendix~\ref{sig:app:FSS})  of~(\ref{sig:eq:h_PDE}) except near $V=c=1$ and near the critical velocities. This discrepancy is due to the data being generated for finite times and nonuniform asymptotic limits as will be discussed in Section~\ref{sig:sec:Results} below. This agreement is seen over a wide range of analogous plots. Thus, we conclude that the amplitude and breadth of each region is well-represented by the asymptotically extracted results.
Note that, if $A_f$ is chosen to be purely real, as it is in Figs.~\ref{sig:fig:FourierLT} and~\ref{sig:fig:FourierBoxLT}, there is no algebraic growth at $V=c$. The algebraic growth occurs only in the imaginary part of $h$, which can only be accessed in the real part of the solution is $A_f$ is complex (see the end of Appendix~\ref{sig:app:LTC}).

\subsection{Long-time asymptotic solution for $\omega_f>\omega_c$}
	\label{sig:sec:GTCSol}

\begin{figure}[ht!]
\centering
\includegraphics[keepaspectratio,width=6in]{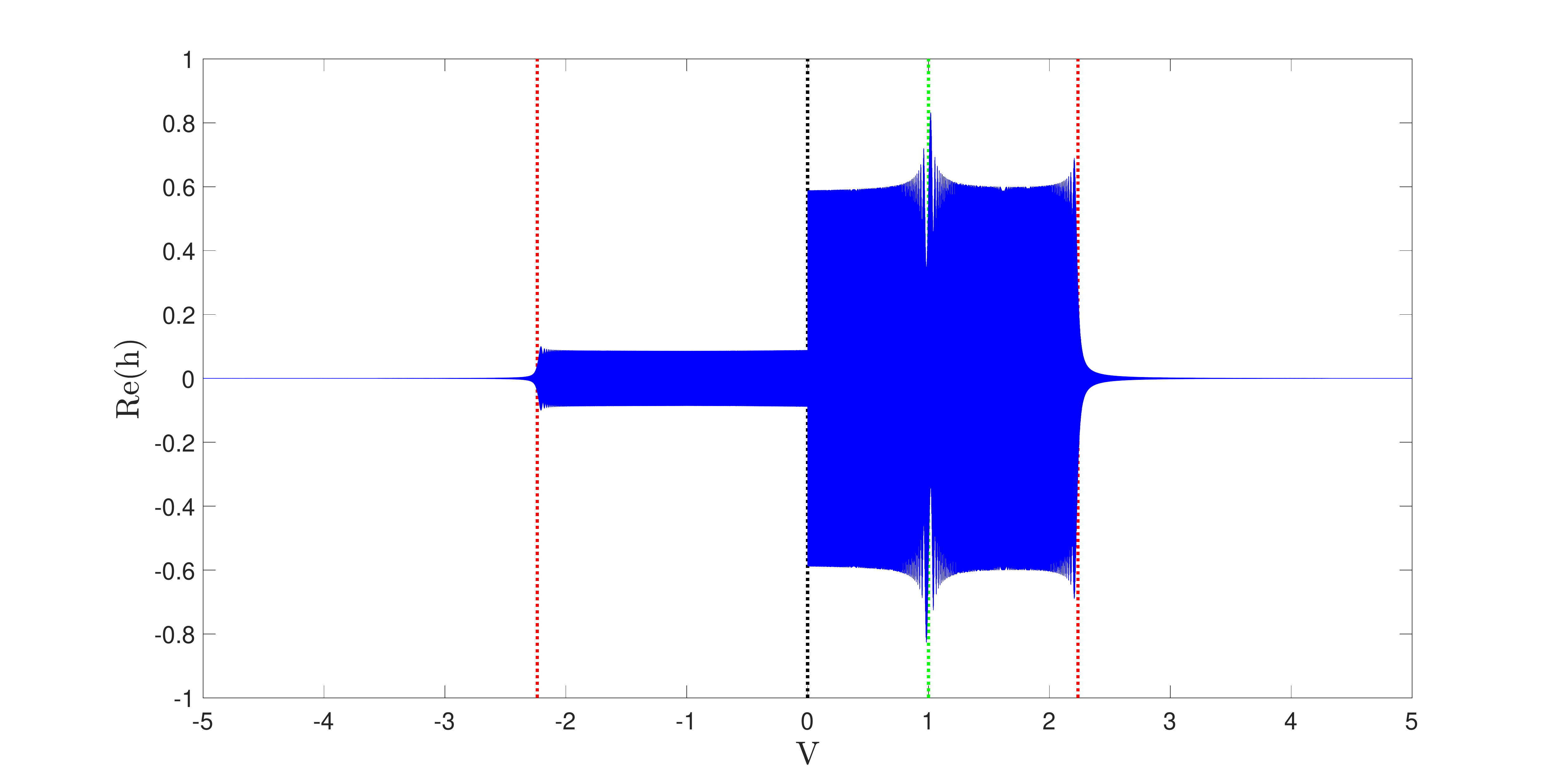}
\caption{ 
Fourier series solution (Appendix~\ref{sig:app:FSS}) of~(\ref{sig:eq:h_PDE}) for $\omega_f>\omega_c$ and $t=10000$ with $A_f=1$, $\omega_f=1$, $B=1$, $c=1$, and $L=N=100000$. Critical velocities (extracted analytically from the integral solution) are marked with dotted vertical lines. Results are plotted against $V$ in order to capture the critical velocities and separate regions.
}
\label{sig:fig:FourierGT}
\end{figure}

The FSS (Appendix~\ref{sig:app:FSS})  of~(\ref{sig:eq:h_PDE}) is used to generate Fig.~\ref{sig:fig:FourierGT} which provides a typical solution response for a case where $\omega_f>\omega_c$.  As in the $\omega_f<\omega_c$ case just considered, there are a large number of oscillations present in the solution. 
At large times, in fact, the solution appears to be a solid block as indicated, but the solution is completely smooth and oscillatory upon magnification.
There are four regions of distinctly different amplitudes, and the observable critical velocities derived using the above-described algorithm delineate discrete regions. As discussed in Section~\ref{sig:sec:LTCSol}, Fig.~\ref{sig:fig:FourierGT} is presented in terms of $V$ instead of $x$ to better capture the long-time behavior.

Applying the methodology from Sections~\ref{sig:sec:CriticalVelocities} and~\ref{sig:sec:BoundingAmplitudes} to the results from Appendix~\ref{sig:app:GTC} leads to a solution for each region of the solution. The regions are separated by the critical velocities
\begin{equation}
\label{sig:eq:AmpVelGTC}
V_1=-\sqrt{c^2+4B\omega_f}
,\quad
V_4=\sqrt{c^2+4B\omega_f},
\end{equation}

\noindent
which are precisely the same as for $V_1$ and $V_4$ in the $\omega_f<\omega_c$ case in~\ref{sig:eq:AmpVelLTC}. The velocities $V_2$ and $V_3$ correspond to the poles $k_0$ and $k_1$ in Fig.~\ref{sig:fig:GTCPoles}, but do not contribute to the long-time behavior --- the off-axis poles add contributions that damp with time and the associated critical velocities are thus non-observable. The poles lying off the real $k$ axis accounts for the change in number of regions between the $\omega_f<\omega_c$ and $\omega_f>\omega_c$ cases.

\FloatBarrier
The leading-order asymptotic behaviors within each of the four regions are as follows:

\begin{subequations}
	\label{sig:eq:GTCRegionSolutions}
\noindent
\underline{$-\infty<V<V_1 \qquad t\to\infty$:}
		\begin{equation*}
		h\sim O\left(t^{-\frac{1}{2}}\right),
		\end{equation*}
		\begin{equation}
		\label{sig:eq:AmpGT:Region1}
		\mathcal{A}_{(\omega_f>\omega_c)}=0.
		\end{equation}

\noindent
\underline{$V_1<V<0 \qquad t\to\infty$:	}
		\begin{equation*}
		h\sim
		-i\frac{A_f}{B^2}
		\frac{e^{i\omega_ft}e^{ik_3Vt}}
		{(k_3-k_0)(k_3-k_1)(k_3-k_2)},
		\end{equation*}
		\begin{equation}
		\label{sig:eq:AmpGT:Region2}
		\mathcal{A}_{(\omega_f>\omega_c)}=
		\frac{A_f}{B^2}
		\Bigg|
		\frac{1}{(k_3-k_0)(k_3-k_1)(k_3-k_2)}
		\Bigg|.
		\end{equation}

\noindent
\underline{$0<V<V_4 \qquad t\to\infty$:}
		\begin{equation*}
		h\sim
		i\frac{A_f}{B^2}
		\frac{e^{i\omega_ft}e^{ik_2Vt}}
		{(k_2-k_0)(k_2-k_1)(k_2-k_3)},
		\end{equation*}
		\begin{equation}
		\label{sig:eq:AmpGT:Region3}
		\mathcal{A}_{(\omega_f>\omega_c)}=
		\frac{A_f}{B^2}
		\Bigg|
		\frac{1}{(k_2-k_0)(k_2-k_1)(k_2-k_3)}
		\Bigg|.
		\end{equation}

\noindent
\underline{$V_4<V<\infty \qquad t\to\infty$:}
		\begin{equation*}
		h\sim O\left(t^{-\frac{1}{2}}\right),
		\end{equation*}
		\begin{equation}
		\label{sig:eq:AmpGT:Region4}
		\mathcal{A}_{(\omega_f>\omega_c)}=0.
		\end{equation}

\end{subequations}

\noindent
Higher order corrections to~(\ref{sig:eq:GTCRegionSolutions}) can be found in Appendix~\ref{sig:app:GTC}, and the poles $k_0$ through $k_3$ are defined in~(\ref{sig:eq:PoleLocs}). A detailed examination of the precise response in each region (i.e., the solutions for $h$ in~(\ref{sig:eq:GTCRegionSolutions})) is given Section~\ref{sig:sec:Results:GTCRegime}.

\begin{figure}[ht!]
\centering
\includegraphics[keepaspectratio,width=6in]{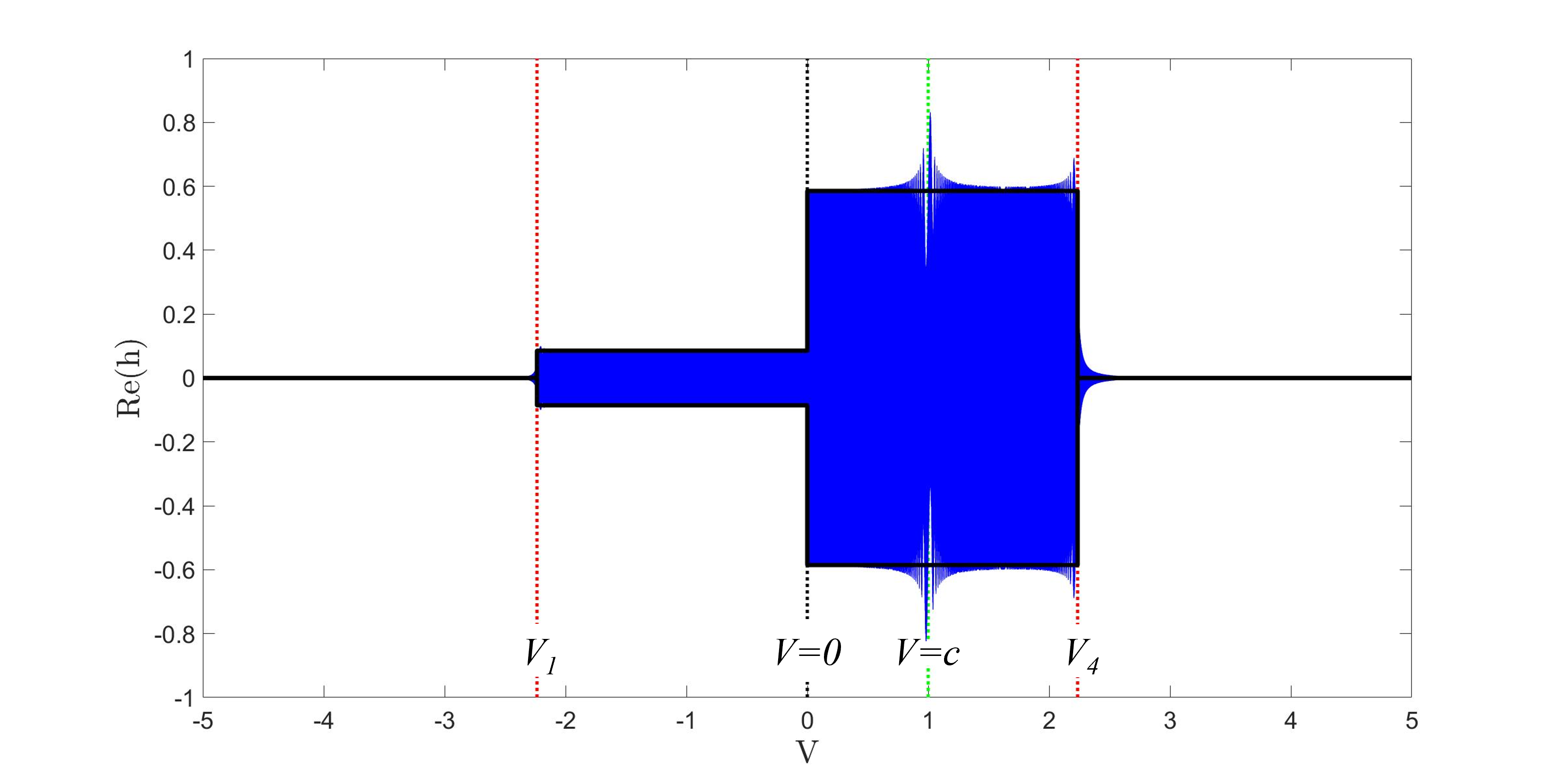}
\caption{ 
Bounding amplitudes (given by~(\ref{sig:eq:GTCRegionSolutions})) overlaid on the Fourier series solution (Appendix~\ref{sig:app:FSS}) in Fig.~\ref{sig:fig:FourierGT} for $\omega_f>\omega_c$. Critical velocities (extracted analytically from the integral solution) are marked with dotted vertical lines. The observable critical velocities from~(\ref{sig:eq:AmpVelGTC}) are labeled.
}
\label{sig:fig:FourierBoxGT}
\end{figure}

Fig.~\ref{sig:fig:FourierBoxGT} overlays the amplitude/critical-velocity asymptotic results given by~(\ref{sig:eq:GTCRegionSolutions}) on Fig.~\ref{sig:fig:FourierGT}.  As with the previous case of $\omega_f<\omega_c$, the bounding amplitudes and delineated regions agree well with the FSS (Appendix~\ref{sig:app:FSS}) results except near $V=c=1$ and near the critical velocities (these discrepancies will be discussed in Section~\ref{sig:sec:Results}). Thus, we conclude that the amplitude and breadth of each region is well-represented by the asymptotically extracted results.
As with the $\omega_f<\omega_c$ case previously considered, if $A_f$ is chosen to be purely real, as it is in Figs.~\ref{sig:fig:FourierGT} and~\ref{sig:fig:FourierBoxGT}, there is no algebraic growth at $V=c$. The algebraic growth occurs only in the imaginary part of $h$ as shown at the end of Appendix~\ref{sig:app:GTC}. A complex $A_f$ is needed to access growth in the real solution.


\section{Results}
\label{sig:sec:Results}

In the what follows, we present the solution response for the $\omega_f<\omega_c$ and $\omega_f>\omega_c$ cases. Just as with Figs.~\ref{sig:fig:FourierBoxLT} and~\ref{sig:fig:FourierBoxGT}, the $V$ domain is used instead of $x$ as discussed in Section~\ref{sig:sec:LTCSol}. 
Note that the qualitative features shown in Figs.~\ref{sig:fig:FourierBoxLT} and~\ref{sig:fig:FourierBoxGT} are observed for all surveyed parameter values (see Supplemental Material Section~\ref{sig:sup:RelativeAmplitudes} for additional details).  Although the exact amplitudes and critical velocities can vary, the amplitude proportions between the regions are maintained.  

\subsection{Details of solution response for $\omega_f<\omega_c$}
	\label{sig:sec:Results:LTCRegime}

\begin{figure}[ht!]
\centering
\includegraphics[keepaspectratio,width=6in]{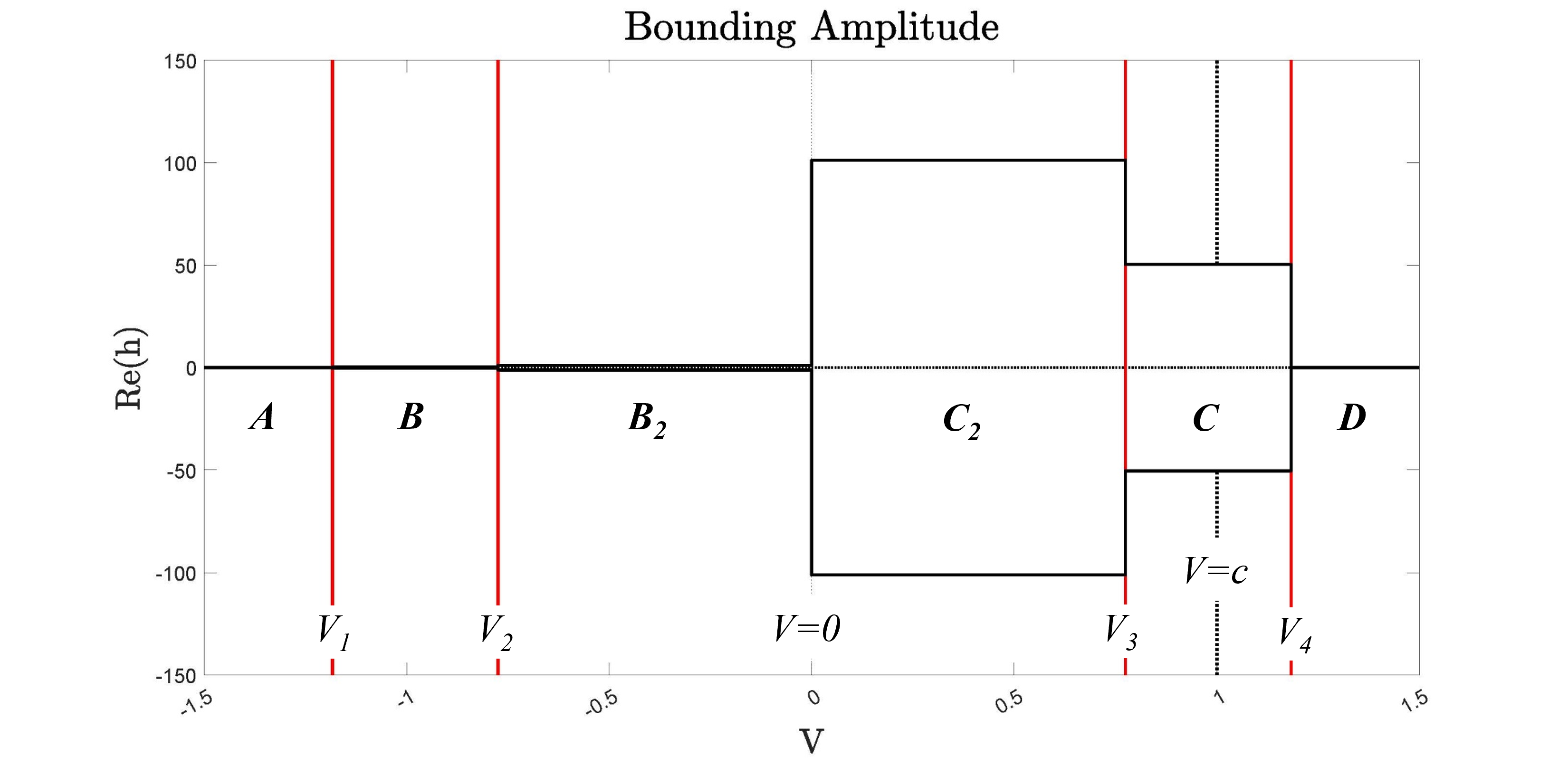}
\caption{Infinite-time bounding amplitude of the solution to~(\ref{sig:eq:h_PDE}), $h(x,t)$, plotted against the velocity for $\omega_f<\omega_c$. The velocities $V_2$ and $V_4$ are the \textit{Leading Edges of the Forced Solution}. The velocities $V_2$ and $V_3$ are the \textit{Leading  Edges of the Additionally Forced Solutions}. Regions $A$ and $D$ are the \textit{Asymptotically Undisturbed Regions}, $B$ and $C$ are the \textit{Forced Regions}, and $B_2$ and $C_2$ are the \textit{Additionally Forced Regions}. The figure was generated using $\omega_f=0.1$, $A_f=1$, $B=1$, and $c=1$. The vertical line at $V=c=1$ represents the location of the algebraically growing peak, if one were to occur. The observable critical velocities, $V_1$ through $V_4$ are extracted from the asymptotic solution and are given by~(\ref{sig:eq:AmpVelLTC}).
Note that the relative amplitudes of the regions are representative of a general parameter set (this is expounded upon in Supplemental Material Section~\ref{sig:sup:RelativeAmplitudes})
}
\label{sig:fig:WireframeLT}
\end{figure}

\noindent
Thus far, we have examined the efficacy of critical velocities and bounding amplitudes in characterizing the solution response provided in Fig.~\ref{sig:fig:FourierBoxLT}.  We now examine the details of the solution response for $\omega_f<\omega_c$ given by~(\ref{sig:eq:CriticalFrequency}), specifically focusing on the waveforms in each delineated region.  To this end, the apparent distinct regions of the response delineated by amplitudes/critical-velocities in Fig.~\ref{sig:fig:FourierBoxLT} are labeled for reference in Fig.~\ref{sig:fig:WireframeLT}. The labeling convention is chosen to aid in the comparison to the $\omega_c>\omega_f$ case later. Upstream ($V<0$) and downstream ($V>0$) of the disturbance location ($V=0$), regions are grouped into pairs which exhibit similar behavior. These groups are the 
\textit{Asymptotically Undisturbed Regions} 
($A$ and $D$), the 
\textit{Forced Regions}
($B$ and $C$), and the 
\textit{Additionally Forced Regions}
($B_2$ and $C_2$).

The results are presented from the outermost regions to the innermost regions. This order is consistent with the physical response as it sweeps through the domain, as it is the order in which any point fixed in $x$ will move through the regions. In particular, for any $x$ at small time, the magnitude of $V$ is large, but as time goes on, the magnitude of $V$ gets smaller.  Note from~(\ref{sig:eq:LTCRegionSolutions}) that the upstream and downstream solutions have similar forms. As such, the distinction between upstream and downstream is largely ignored except for the case of $V=c$ which is addressed in Section~\ref{sig:sec:LTC:VEqC}.

\FloatBarrier
\subsubsection{\textit{Asymptotically Undisturbed Regions} (Regions $A$ and $D$ in Fig.~\ref{sig:fig:WireframeLT})}
	\label{sig:sec:LTC:Unforced}
\begin{figure}[ht!]
\centering
\includegraphics[keepaspectratio,width=6in]{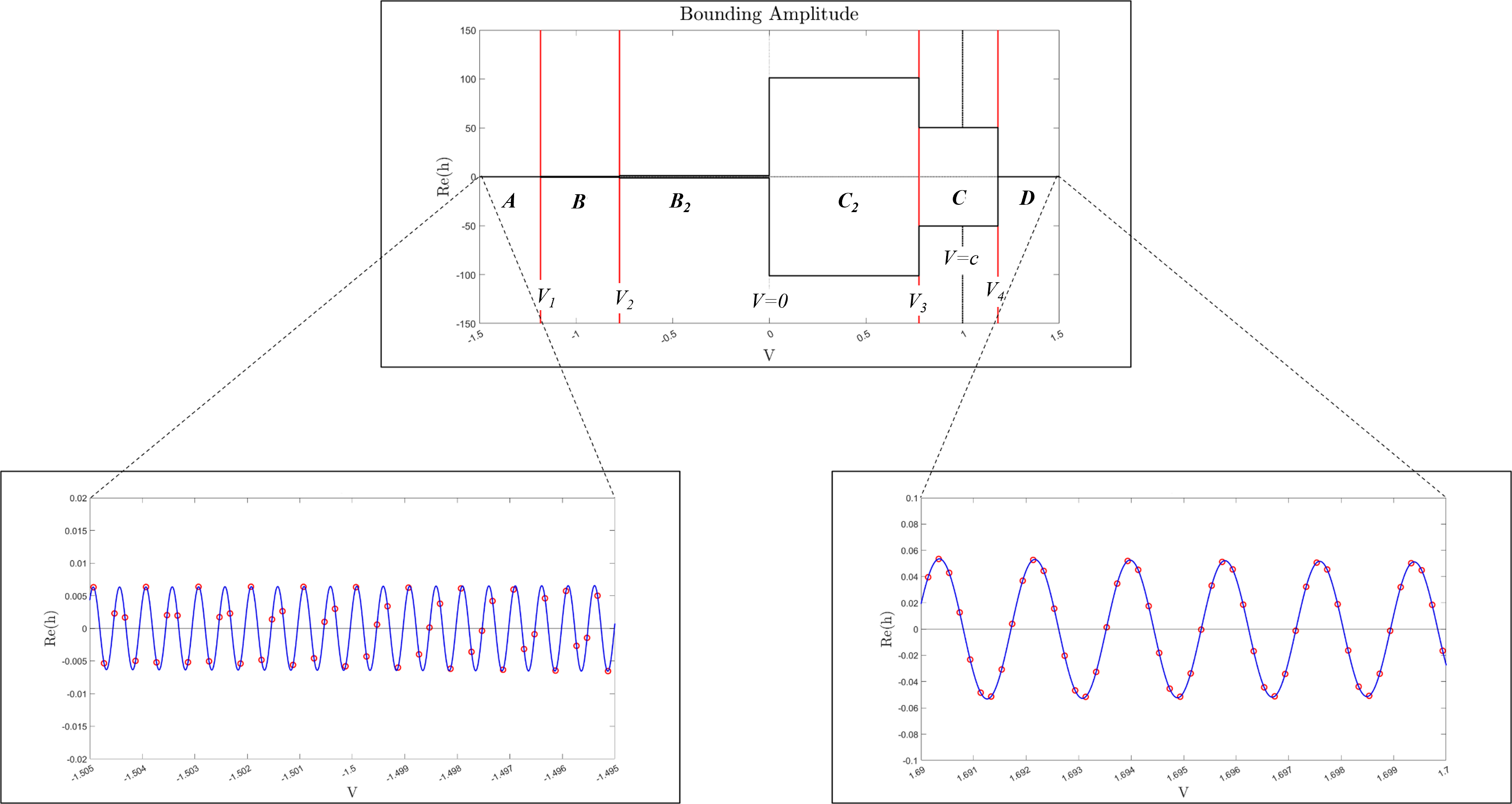}
\caption{Magnified view of the response in the \textit{asymptotically undisturbed regions} $A$ (left) and $D$ (right) of Fig.~\ref{sig:fig:WireframeLT} at $t=10000$ with $\omega_f=0.1$, $B=1$, $c=1$, $L=50000$, and $N=10000$. The FSS (Appendix~\ref{sig:app:FSS}) of~(\ref{sig:eq:h_PDE}) (\textcolor{red}{$\circ$}) agrees with the long-time asymptotic solution (Appendix~\ref{sig:app:LTC}) (blue curve). Bounding amplitudes and leading order asymptotic solutions are given in~(\ref{sig:eq:LTCRegionSolutions})
}
\label{sig:fig:LTC:Unforced}
\end{figure}

The \textit{Asymptotically Undisturbed Regions}, labeled $A$ and $D$ in Fig.~\ref{sig:fig:WireframeLT}, are magnified in Fig.~\ref{sig:fig:LTC:Unforced}. These regions are in advance of any structural features brought about by the forcing which persist at long time.  There are transient responses in the region which decay in time~(\ref{sig:eq:AmpLT:Region1})(\ref{sig:eq:AmpLT:Region6}), but there is no persistent oscillatory behavior induced by the forcing function (none of the poles indicated in Fig.~\ref{sig:fig:LTCPoles} are enclosed in relevant contours --- see Appendix~\ref{sig:app:LTC}).

In Fig.~\ref{sig:fig:LTC:Unforced} it is worth noting that, while the FSS (Appendix~\ref{sig:app:FSS}) (\textcolor{red}{$\circ$}) matches the asymptotic solution (blue lines) of~(\ref{sig:eq:h_PDE}), neither one has reached the long-time amplitude of zero (solid horizontal lines in the sub-plots).  The indicated solutions are shown at a time of $t=10000$; the solution does tend towards zero as time goes off to infinity in accordance with the bounding amplitudes and asymptotic behaviors in~(\ref{sig:eq:LTCRegionSolutions}). 

\FloatBarrier
\subsubsection{\textit{Leading Edges of the Forced Solution} ($V_1$ and $V_4$ in Fig.~\ref{sig:fig:WireframeLT})}

\begin{figure}[ht!]
\centering
\includegraphics[keepaspectratio,width=6in]{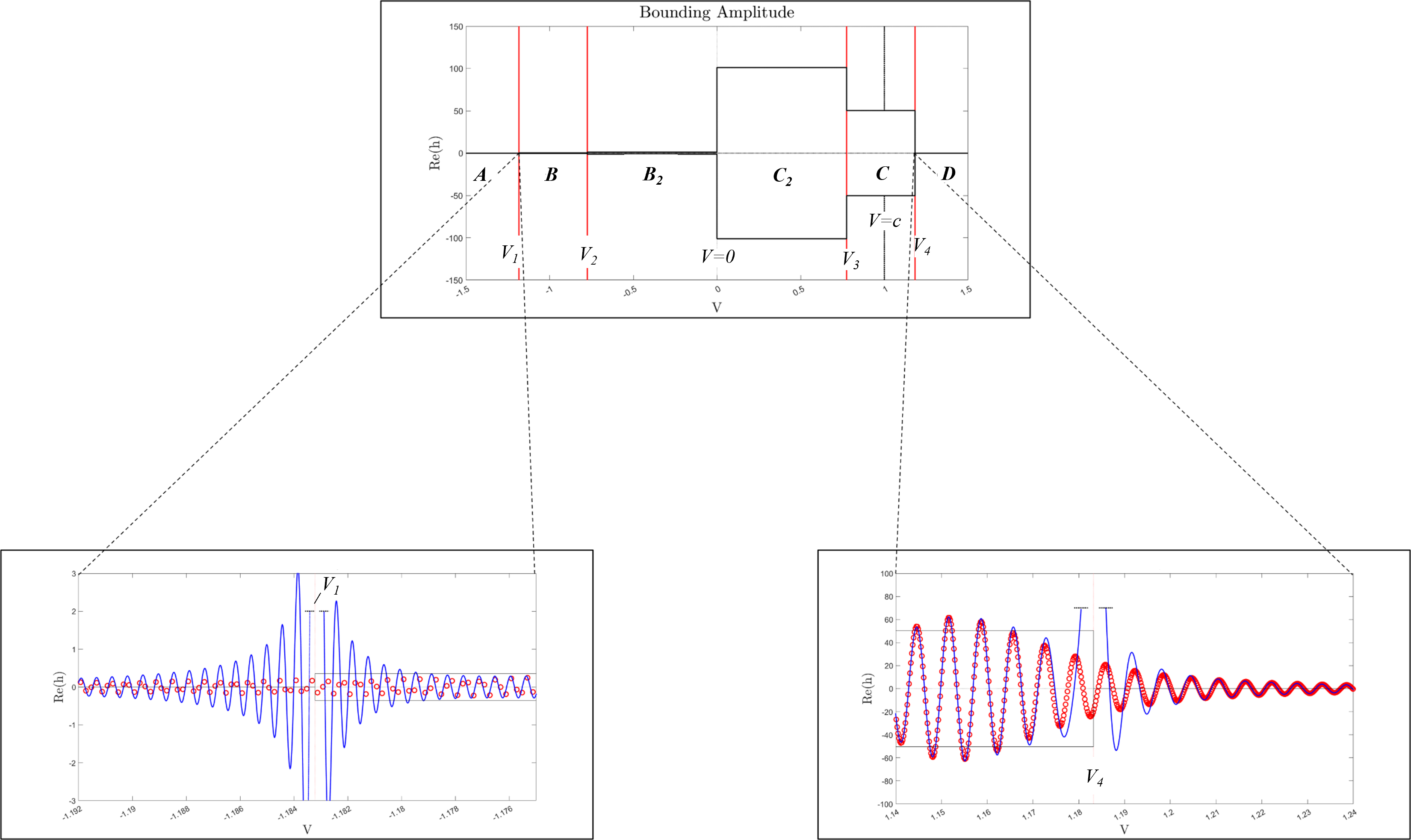}
\caption{
Magnified view of the response near the \textit{Leading Edge of the Forced Solution} of Fig.~\ref{sig:fig:WireframeLT} at $t=10000$ with $\omega_f=0.1$, $B=1$, $c=1$, $L=50000$, and $N=10000$. The FSS (Appendix~\ref{sig:app:FSS}) (\textcolor{red}{$\circ$}) agrees with the long-time asymptotic solution (Appendix~\ref{sig:app:LTC}) (blue curve) away from the discontinuities located at the critical velocities. Note that this data was generated with the full asymptotic form from Appendix~\ref{sig:app:LTC}, not the leading order form given by~(\ref{sig:eq:LTCRegionSolutions}).
Bounding amplitudes and leading order asymptotic solutions are given in~(\ref{sig:eq:LTCRegionSolutions})}
\label{sig:fig:LTC:LeadingEdge} 
\end{figure}

The critical velocities $V_1$ and $V_4$ in Fig.~\ref{sig:fig:WireframeLT} represent the leading edges of the forcing's effect in the upstream and downstream directions at long times, respectively; these regions are magnified in Fig.~\ref{sig:fig:LTC:LeadingEdge}. There is a non-zero bounding amplitude according to Equations~(\ref{sig:eq:AmpLT:Region2}) and~(\ref{sig:eq:AmpLT:Region5}) for $V_1<V<V_3$ and $V_2<V<V_4$. These velocities mark where the poles $k_2$ and $k_3$ (Fig.~\ref{sig:fig:LTCPoles}) are enclosed in the integration contours (see Appendix~\ref{sig:app:LTC}).

\noindent
As indicated in Fig.~\ref{sig:fig:LTC:LeadingEdge}, the FSS (Appendix~\ref{sig:app:FSS}) of~(\ref{sig:eq:h_PDE}) of~(\ref{sig:eq:h_PDE}) is continuous across $V_1$ and $V_4$, but the asymptotic solution diverges. This is consistent with the residual singularities present in the asymptotic results~(\ref{sig:eq:LTCRegionSolutions}). Away from the critical velocities, the long-time asymptotic solution (\ref{sig:eq:LTCRegionSolutions}) agrees well with the FSS~(Appendix~\ref{sig:app:FSS}). The tapered shape of the curve is due to the data being generated at a finite time. As time goes on, the transition becomes more pronounced (not pictured here) as the amplitudes of the \textit{Asymptotically Undisturbed Regions} ($A$ and $D$) go to zero and the amplitudes of the \textit{Forced Regions} ($B$ and $C$) tend to the long-time bounding amplitude (solid horizontal lines in Fig.~\ref{sig:fig:LTC:LeadingEdge}).

\FloatBarrier
\subsubsection{\textit{Forced Regions} (Regions $B$ and $C$ in Fig.~\ref{sig:fig:WireframeLT})}
	\label{sig:sec:LTC:ForcedRegions}
\begin{figure}[ht!]
\centering
\includegraphics[keepaspectratio,width=6in]{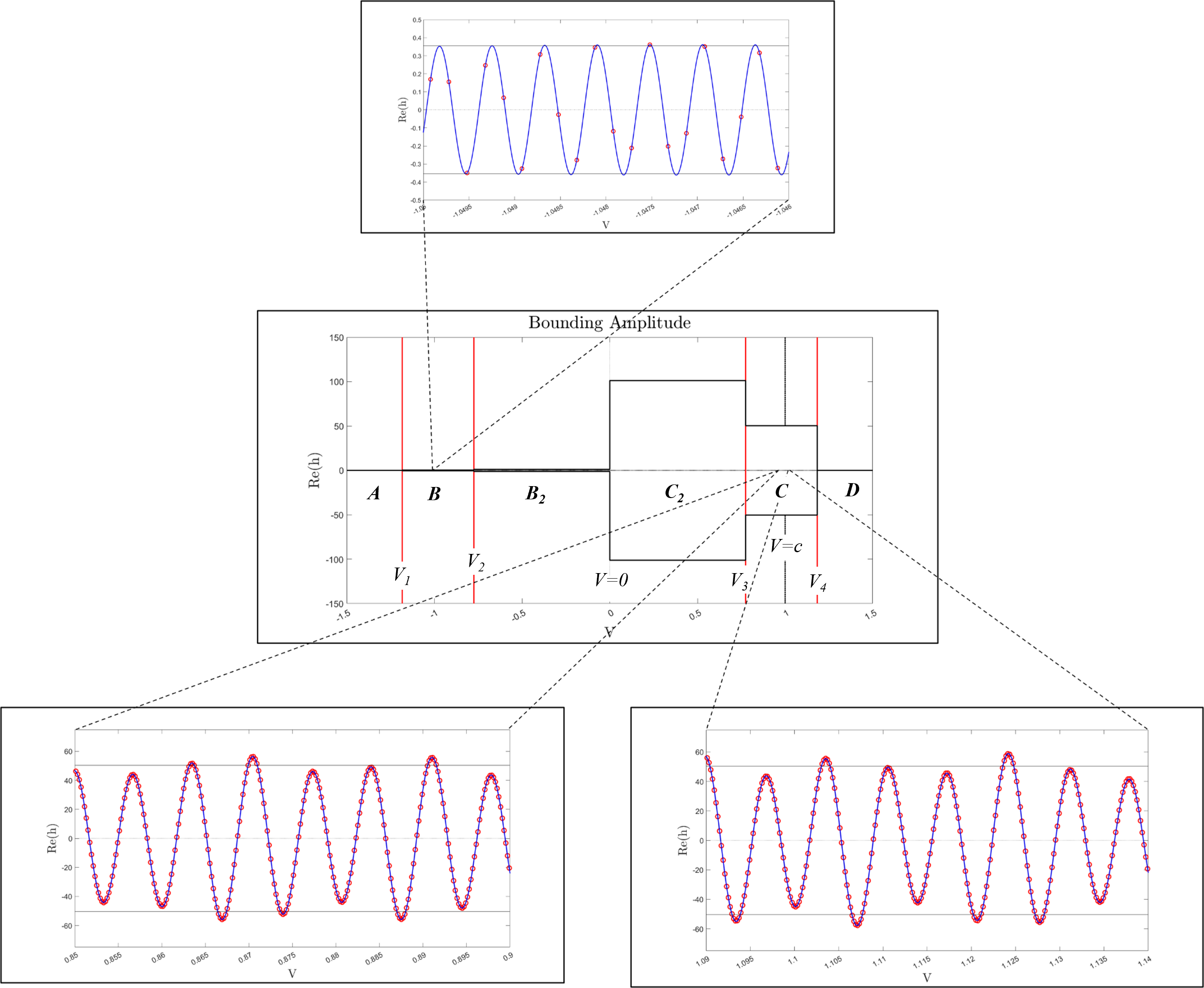}
\caption{
Magnified view of the response in the \textit{Forced Regions} $B$ (top) and $C$ (bottom) of Fig.~\ref{sig:fig:WireframeLT} at $t=10000$ with $\omega_f=0.1$, $B=1$, $c=1$, $L=50000$, and $N=10000$. The FSS (Appendix~\ref{sig:app:FSS}) of~(\ref{sig:eq:h_PDE}) (\textcolor{red}{$\circ$}) agrees with the long-time asymptotic solution (Appendix~\ref{sig:app:LTC}) (blue curve). For the downstream behavior, the two sub-plots show the response on either side of $V=c$, which is the velocity of the growing peak if one were to occur for complex $A_f$ as show in Appendix~\ref{sig:app:LTC}. All the FSS data were generated with the same sampling frequency in $V$; the apparent difference in the plots arises because the upstream sub-plot is magnified further to resolve the waveform.
Bounding amplitudes and leading order asymptotic solutions are given in~(\ref{sig:eq:LTCRegionSolutions}).
}
\label{sig:fig:LTC:TransientRegion}
\end{figure}

Regions $B$ and $C$ in Fig.~\ref{sig:fig:WireframeLT} are influenced by the contributions of a single pole each (see~(\ref{sig:eq:AmpLT:Region2}) and~(\ref{sig:eq:AmpLT:Region5})). Magnifications of these regions are provided in Fig.~\ref{sig:fig:LTC:TransientRegion}. In these regions, any point in the $x$ domain will be reside in either region for only a finite time interval, as $V\equiv\nicefrac{x}{t}$ continues to be reduced in time for a fixed $x$.
The sampling frequency of the FSS appears lower in the upstream sub-plot of Fig.~\ref{sig:fig:LTC:TransientRegion}, but that is because the $V$ axis scale is magnified further than for the other sub-plot in order to resolve the high frequency oscillations.

The two lower sub-plots within Fig.~\ref{sig:fig:LTC:TransientRegion} show the response within the downstream \textit{Forced Region} on either side of and away from the critical velocity $V=c$. That velocity is always contained within region $C$ because $V_3=\sqrt{c^2-4B\omega_f}<c$ and $V_4=\sqrt{c^24B\omega_f}>c$.
Other than near $V=c$, the asymptotic solution~(\ref{sig:eq:LTCRegionSolutions}) matches the FSS (Appendix~\ref{sig:app:FSS}) of~(\ref{sig:eq:h_PDE}). Because the solutions are shown for a time of $t=10000$, their amplitudes do not precisely match the predicted long-time amplitude (the solid horizontal lines in Fig.~\ref{sig:fig:LTC:TransientRegion}) from~(\ref{sig:eq:AmpLT:Region2}); however, they approach the predicted bounding values as time goes to infinity.

\FloatBarrier
\subsubsection{\textit{Leading Edges of the Additionally Forced Regions} ($V_2$ and $V_3$ in Fig.~\ref{sig:fig:WireframeLT})}
	\label{sig:sec:LTC:LeadingEdgeOfAdForced}
\begin{figure}[ht!]
\centering
\includegraphics[keepaspectratio,width=6in]{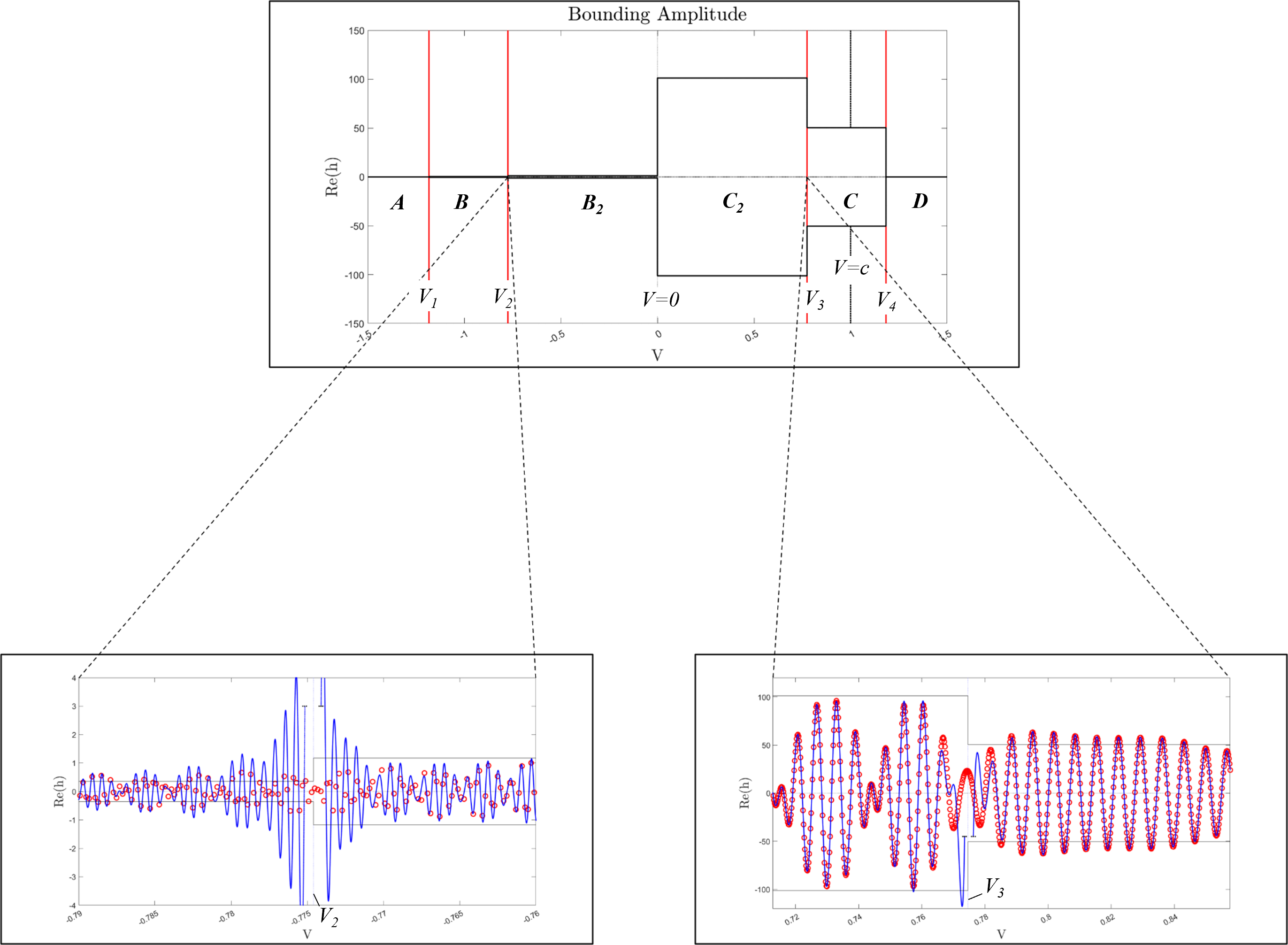}
\caption{
Magnified view of the response near the \textit{Leading Edges of the Additionally Forced Regions} of Fig.~\ref{sig:fig:WireframeLT} at $t=10000$ with $\omega_f=0.1$, $B=1$, $c=1$, $L=50000$, and $N=10000$. The FSS (Appendix~\ref{sig:app:FSS}) of~(\ref{sig:eq:h_PDE}) (\textcolor{red}{$\circ$}) agrees with the long-time asymptotic solution (Appendix~\ref{sig:app:LTC}) (blue curve) away from the discontinuity directly at the critical velocity. Note that this data was generated with the full asymptotic form from Appendix~\ref{sig:app:LTC}, not the leading order form given by~(\ref{sig:eq:LTCRegionSolutions}).
Bounding amplitudes and leading order asymptotic solutions are given in~(\ref{sig:eq:LTCRegionSolutions})
}
\label{sig:fig:LTC:ForcingEdge}
\end{figure}

The critical velocities $V_2$ and $V_3$ in Fig.~\ref{sig:fig:WireframeLT} mark the \textit{Leading Edges of the Additionally Forced Regions}; these transitions are magnified in Fig.~\ref{sig:fig:LTC:ForcingEdge}. This boundary also represents the leading edge of the long-time solution over any finite physical domain (in $x$), that is to say the effects of the other regions have swept through, leaving only oscillations from the forcing.

As mentioned previously, the long-time asymptotic solution diverges near the critical velocities due to residual singularities in terms which decay in time (see Appendix~\ref{sig:app:LTC} for this case and Appendix~\ref{sig:app:GTC} for the $\omega_f>\omega_c$ case). On either side of these discontinuities, the FSS and asymptotic solutions align.  Again, any discrepancies between the bounding amplitudes, indicated on the plots, disappear as $t$ increases. The high frequency of oscillations upstream make the comparison between the long-time asymptotic solution~(\ref{sig:eq:LTCRegionSolutions}) and the FSS (Appendix~\ref{sig:app:FSS}) of~(\ref{sig:eq:h_PDE}) difficult to resolve on a single plot.

\FloatBarrier
\subsubsection{\textit{Additionally Forced Regions} (Regions $B_2$ and $C_2$ in Fig.~\ref{sig:fig:WireframeLT})}
	\label{sig:sec:LTC:AdForced}
\begin{figure}[ht!]
\centering
\includegraphics[keepaspectratio,width=6in]{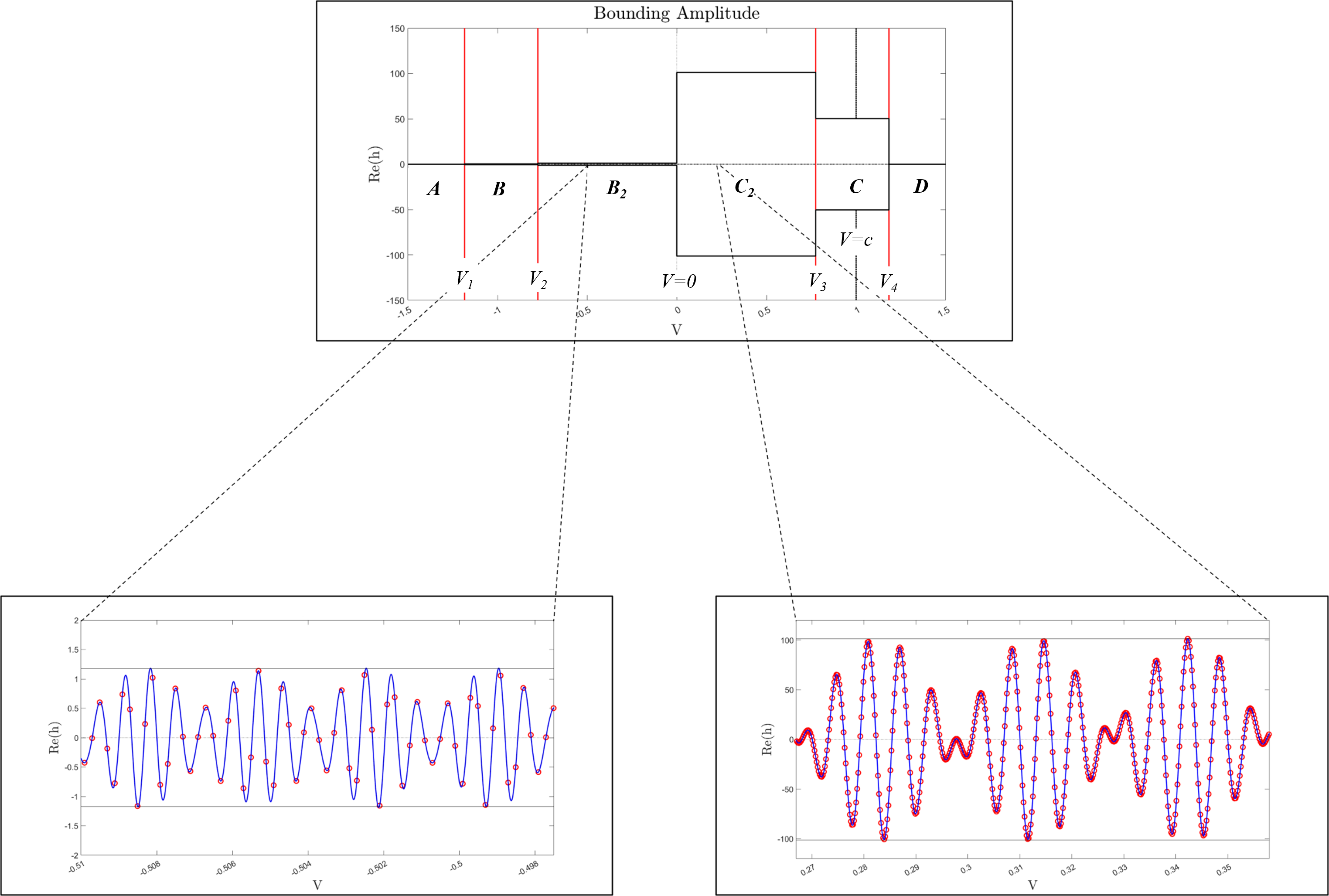}
\caption{Magnified view of the response in the \textit{Long-Time Forced Regions} $B_F$ (left) and $C_f$ (right) of Fig.~\ref{sig:fig:WireframeLT} at $t=10000$ with $\omega_f=0.1$, $B=1$, $c=1$, $L=50000$, and $N=10000$. The FSS (Appendix~\ref{sig:app:FSS}) of~(\ref{sig:eq:h_PDE}) (\textcolor{red}{$\circ$}) agrees with the long-time asymptotic solution (Appendix~\ref{sig:app:LTC})(blue curve). All the FSS data were generated with the same sampling frequency in $V$. Note that the upstream (left) sub-plot is magnified more than the downstream sub-plot in order to resolve the waveform. Bounding amplitudes and leading order asymptotic solutions are given in~(\ref{sig:eq:LTCRegionSolutions})
}
\label{sig:fig:LTC:Forced}
\end{figure}

The regions $B_2$ and $C_2$ in Fig.~\ref{sig:fig:WireframeLT} are the \textit{Additionally Forced Regions} of the solution and represent the long-time response over any finite $x$ domain; given enough time, any point in the $x$ domain will be affected by one of the two, as there are no further structural changes in solution possible as $V$ approaches zero in Fig.~\ref{sig:fig:WireframeLT} (as $t$ goes to infinity for a fixed $x$, $V$ goes to zero). Fig.~\ref{sig:fig:LTC:Forced} shows the agreement between the long-time asymptotic solution~(\ref{sig:eq:LTCRegionSolutions}) and the FSS (Appendix~\ref{sig:app:FSS}) of~(\ref{sig:eq:h_PDE}) away from the bounding velocities $V_2$ and $V_3$ from~(\ref{sig:eq:AmpVelLTC}) discussed in Section~\ref{sig:sec:LTC:LeadingEdgeOfAdForced} and $V=0$ discussed in Section~\ref{sig:sec:LTC:UpDownSep}.

Region $B$ and $B_2$ are both influenced by the pole $k_3$ (see~(\ref{sig:eq:AmpLT:Region2}) and~(\ref{sig:eq:AmpLT:Region3})), but $B_2$ also contains the influence of the pole $k_0$. The same holds true for $C$ and $C_2$ and the poles $k_2$ and $k_1$. The overall structure of the solution in regions $B_2$ and $C_2$ is that of two oscillatory modes with different frequencies which travel at two different speeds.
Within the regions $B_2$ and $C_2$, the asymptotic solution agrees with the FSS. The velocity sampling frequency of both plots are the same (data points per velocity range). Region $B_2$ has higher frequency oscillations than Region $C_2$,and it requires higher magnification to resolve. The vertical scales differ between the sub-plots by two orders of magnitude. Both regions demonstrate a beat-like structure with a lower frequency.

It is important to differentiate here between the long-time solution in $V$ and the long-time solution in $x$. To determine the overall behavior, all values of $V$ are studied as they are relevant to the structure of the solution. However, for any finite $x$ domain, regions $B_2$ and $C_2$ represent the final signaling response as they will persist for all time at any given $x$ (i.e., $V$ approaches zero in Fig.~\ref{sig:fig:WireframeLT}).

\FloatBarrier
\subsubsection{\textit{Upstream/Downstream Division} ($V=0$ in Fig.~\ref{sig:fig:WireframeLT})}
	\label{sig:sec:LTC:UpDownSep}
\begin{figure}[ht!]
\centering
\includegraphics[keepaspectratio,width=6in]{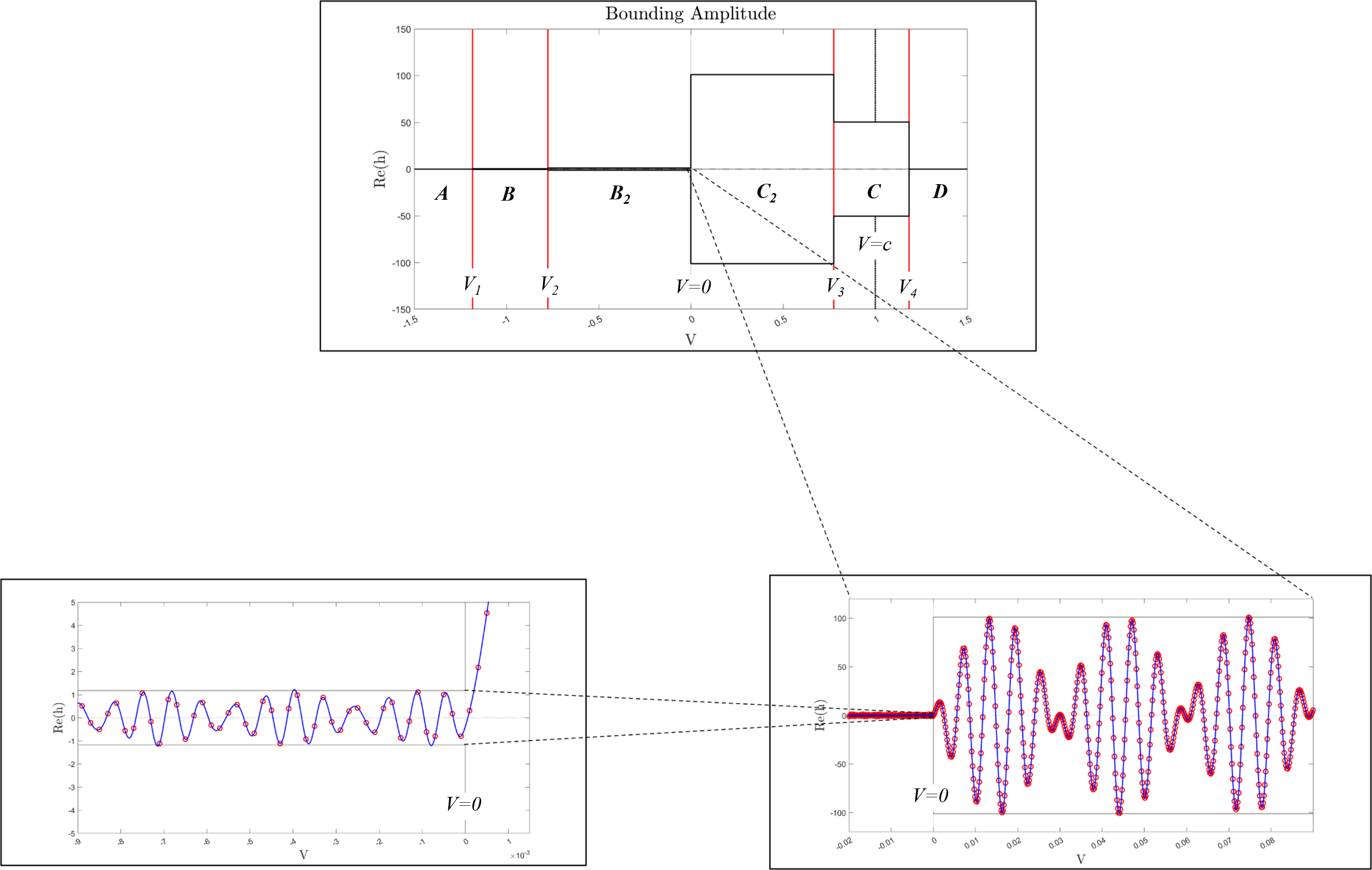}
\caption{
Magnified view of the response at the \textit{Upstream/Downstream Division} of Fig.~\ref{sig:fig:WireframeLT} at $t=10000$ with $\omega_f=0.1$, $B=1$, $c=1$, $L=50000$, and $N=10000$. The FSS (Appendix~\ref{sig:app:FSS}) of~(\ref{sig:eq:h_PDE}) (\textcolor{red}{$\circ$}) agrees with the long-time asymptotic solution (Appendix~\ref{sig:app:LTC}) (blue curve). Note that this data was generated with the full asymptotic form from Appendix~\ref{sig:app:FourierSolution}, not the simplified form given by~(\ref{sig:eq:LTCRegionSolutions}).
All the FSS data were generated with the same sampling frequency in $V$; the apparent difference in the plots (the spread in the circles for FSS data points) arises because the upstream sub-plot is magnified much further to resolve the waveform.
Due to a significant difference in amplitudes, the sub-plot for the upstream response is a further magnification from the downstream sub-plot. Bounding amplitudes and leading order asymptotic solutions are given in~(\ref{sig:eq:LTCRegionSolutions})
}
\label{sig:fig:LTC:UpDownSep}
\end{figure}

The critical velocity $V=0$ separates the upstream and downstream behavior of the response. It is the only critical velocity across which the long-time asymptotic solution (\ref{sig:eq:LTCRegionSolutions}) is continuous. As with all the other transitions in the various regions of Fig.~\ref{sig:fig:WireframeLT}, the FSS (Appendix~\ref{sig:app:FSS}) of~(\ref{sig:eq:h_PDE}) is continuous. The region near this velocity is magnified in Fig.~\ref{sig:fig:LTC:UpDownSep}.
As shown in Section~\ref{sig:sec:LTC:AdForced}, the response in the upstream and downstream forced regions have the same structure, albeit at vastly different scales. The disparities in amplitude and frequency make it impossible to plot the responses of both regions on a single scale. As such, Fig.~\ref{sig:fig:LTC:UpDownSep} includes two levels of magnification, one to show the downstream behavior (right) and another to show the upstream behavior (left).

\FloatBarrier
\subsubsection{Velocity of algebraic growth ($V=c$ in Fig.~\ref{sig:fig:WireframeLT})}
	\label{sig:sec:LTC:VEqC}
	
\begin{figure}[ht!]
\centering
\includegraphics[keepaspectratio,width=6in]{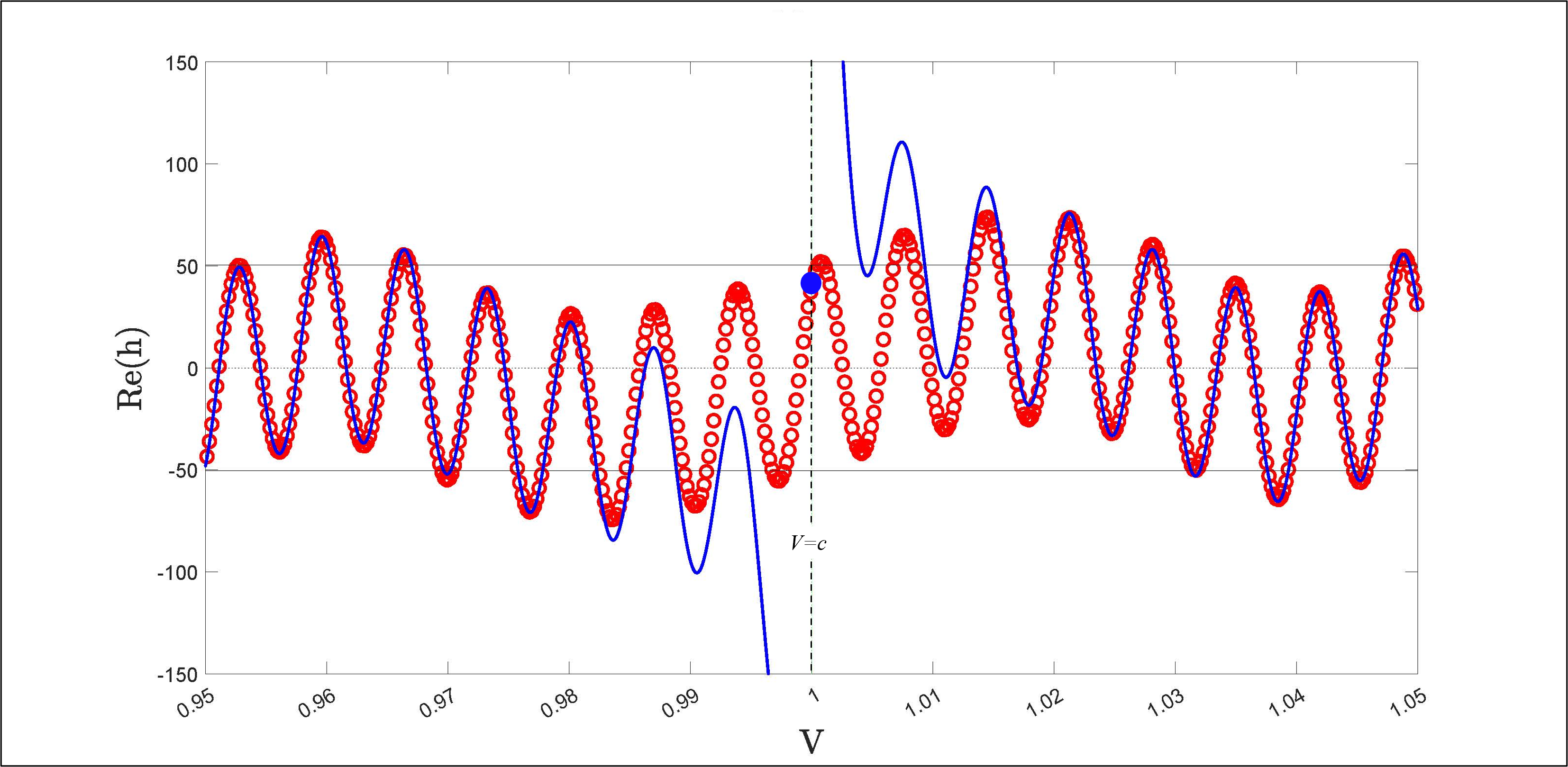}
\caption{ 
Magnified view of the response near $V=c$ at $t=10000$ with $\omega_f=0.1$, $B=1$, $c=1$, $L=50000$, and $N=10000$. The FSS (Appendix~\ref{sig:app:FSS}) of~(\ref{sig:eq:h_PDE}) (\textcolor{red}{$\circ$}) agrees with the long-time asymptotic solution (Appendix~\ref{sig:app:LTC}) (blue curve) away from the discontinuity directly at the critical velocity. Note that this data was generated with the full asymptotic form from Appendix~\ref{sig:app:FourierSolution}, not the simplified form given by~(\ref{sig:eq:LTCRegionSolutions}). Unlike for the other critical velocities, there is an exact solution (\textcolor{blue}{$\bullet$}) for $V=c$. Bounding amplitudes and leading order asymptotic solutions are given in~(\ref{sig:eq:LTCRegionSolutions})
}
\label{sig:fig:LTC:VEqC}
\end{figure}

Near the critical velocity of $V=c$, the long-time asymptotic solution~(\ref{sig:eq:LTCRegionSolutions}) diverges from the FSS (Appendix~\ref{sig:app:FSS}) of~(\ref{sig:eq:h_PDE}). This divergence arises from the non-uniform limit taken in the asymptotic expansion, which is expressed as $(V-c)t$ going to infinity instead of just $t$ going to infinity. As such, increasingly longer times are needed to properly resolve the behavior near $V=c$. The integrals which cause the divergence are solvable in closed form at $V=c$, leading to a non divergent solution, marked in Fig.~\ref{sig:fig:LTC:VEqC} at $V=1$. On either side of the critical velocity, the asymptotic solution agrees with the FSS, as shown in the plots in Fig.~\ref{sig:fig:LTC:TransientRegion}.

The structure of the solution at $V=c$ depends on the phase of the initial disturbance. This dependence is analogous to that found by King et al.~\cite{king2016} where the growth occurs if the initial surface velocity is  perturbed but not when the initial surface height is perturbed. In the forced problem, the phase of the growth is controlled by the phase of the forcing amplitude, $A_f$. Choosing $A_f$ to be real keeps all the growth out of the real part of $h(x,t)$, equivalent to suppressing the initial surface velocity in the unforced problem of King et al.~\cite{king2016} In the unforced problem~(\ref{sig:eq:g_PDE}), the magnitude of the growth integral is controlled by $U_0$, the initial surface velocity. 
Ultimately, the response at $V=c$ is a transient solution created by the start-up of the disturbance. As such, it is separate from the long-time oscillatory solution in each region. The overall long-time asymptotic solution is the superposition of the oscillatory terms and this transiently growing and convecting peak along $V=c$.

\FloatBarrier
\subsection{Details of solution response for $\omega_f>\omega_c$}
	\label{sig:sec:Results:GTCRegime}

\begin{figure}[ht!]
\centering
\includegraphics[keepaspectratio,width=6in]{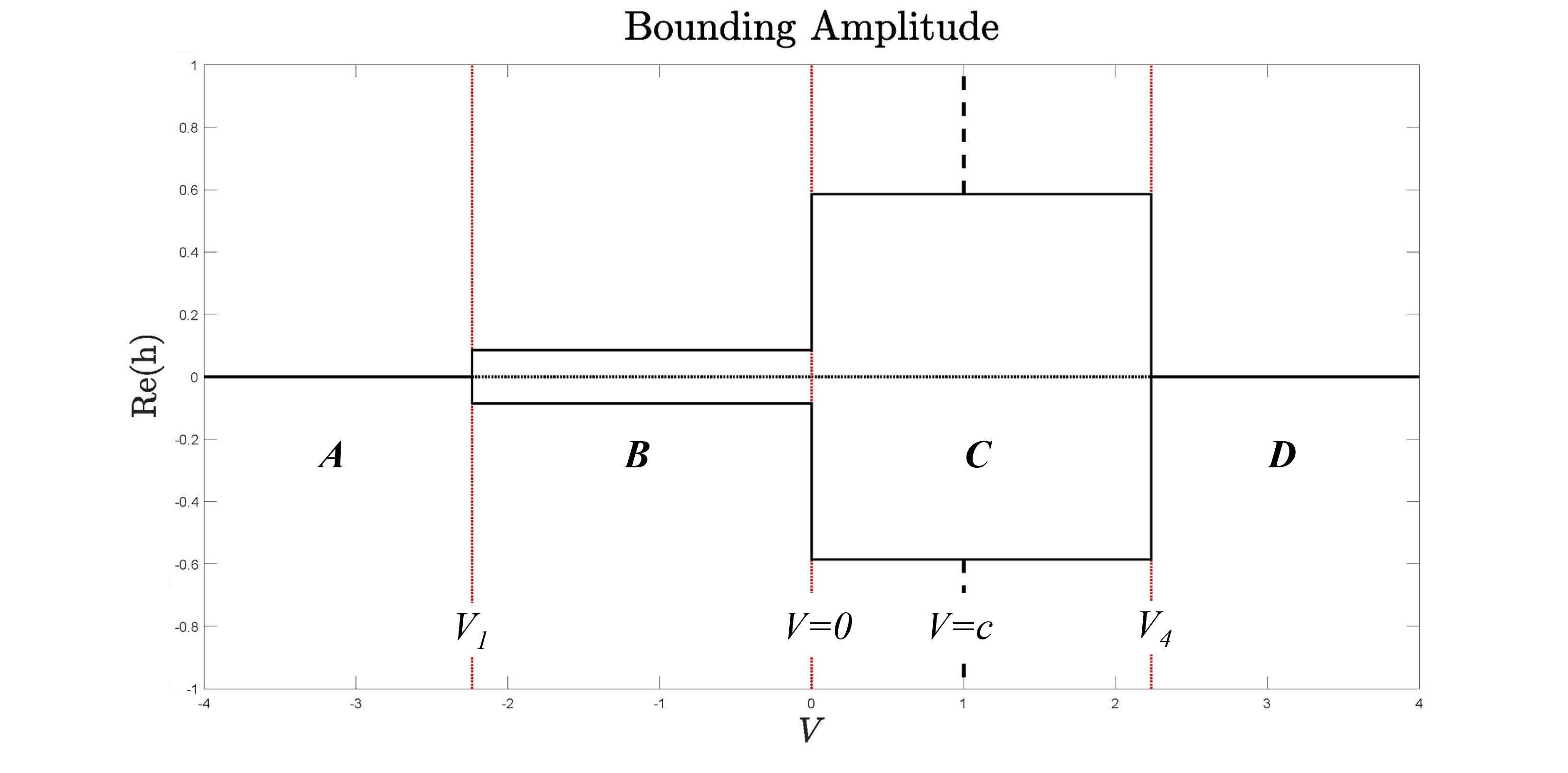}
\caption{Infinite time bounding envelope of the solution, $h(x,t)$ given by~(\ref{sig:eq:GTCRegionSolutions}), plotted against the velocity for $\omega_f>\omega_c$. The velocities $V_1$ and $V_4$ are the \textit{Leading Edge of the Forced Solution}. Regions $A$ and $D$ are the \textit{Asymptotically Undisturbed Regions}, and $B$ and $C$ (excluding $V=c$) are the \textit{Forced Solutions}. The data was generated using $\omega_f=1$, $A_f=1$, $B=1$, and $c=1$. The vertical line at $V=c=1$ represents the location of the growing peak, if one occurs. 
}
\label{sig:fig:WireframeGT}
\end{figure}

Above the critical frequency, $\omega_c$ given in~(\ref{sig:eq:CriticalFrequency}),
the solution has a similar structure as the $\omega_f<\omega_c$ case. As the forcing frequency is increased towards the critical frequency in Fig.~\ref{sig:fig:WireframeLT}, the velocities $V_2$ and $V_3$ move towards $V=0$. Across the critical frequency, the regions $B_2$ and $C_2$ collapse, leaving only the regions $B$ and $C$ as shown in Fig.~\ref{sig:fig:WireframeGT}. Additional details on this transition are included in Section~\ref{sig:app:CriticalVelocities}.

One feature of the solution for $\omega_f>\omega_c$ is that the oscillation frequencies in the response are much higher than those for $\omega_f<\omega_c$, and thus significant magnification is needed to observe individual wave-forms and transitions between regions.  The same qualitative features of the solution are seen across bounding velocities --- the FSS~(Appendix~\ref{sig:app:FSS}) is always continuous and smooth, the asymptotic solution agrees with the FSS away from critical velocities (but not close to them), and discontinuities across critical velocities can be observed in some transitions between regions. In what follows, we discuss the physics of these regions in a more abbreviated form, comparing and contrasting with the $\omega_f<\omega_c$ case discussed in Section~\ref{sig:sec:Results:LTCRegime}.

The long-time \textit{Forced Solution} near $V=0$ corresponds to the regions $B$ and $C$ in Fig.~\ref{sig:fig:WireframeGT}; this differs from the $\omega_f<\omega_c$ case, where regions $B_2$ and $C_2$ formed the long time amplitude. 
As mentioned before, Fig.~\ref{sig:fig:WireframeGT} has fewer distinct regions than Fig.~\ref{sig:fig:WireframeLT}. Due to the collapse of $V_2$ and $V_3$ as $\omega_f$ transitions past $\omega_c$, the $\omega_f>\omega_c$ case has no equivalent regions to $B_2$ and $C_2$ as for the $\omega_f<\omega_c$ case (compare Figs.~\ref{sig:fig:WireframeLT} and~\ref{sig:fig:WireframeGT}). This difference is also seen by inspection of the asymptotic forms and bounding amplitudes for the $\omega_f<\omega_c$ and $\omega_f>\omega_c$ cases given by~(\ref{sig:eq:LTCRegionSolutions}) and~(\ref{sig:eq:GTCRegionSolutions}), respectively; it can be seen that the \textit{Forced Regions} of both cases are the result of the contributions of a single pole each, $k_3$ in the upstream regions and $k_2$ in the downstream regions (see Fig.~\ref{sig:fig:LTCPoles} for pole structure).


As with the case where $\omega_f<\omega_c$, the outer regions $A$ and $D$ in Fig.~\ref{sig:fig:WireframeGT} are the \textit{Asymptotically Undisturbed Regions} that limit to zero amplitude as $t$ approaches infinity.
The critical velocity of $V=0$ divides the upstream and downstream behavior as before. Across this velocity, both the Fourier series and asymptotic solutions are continuous and the solutions have different frequencies and amplitudes on either side as for $\omega_f<\omega_c$.
Similarly to the $\omega_f<\omega_c$ results, the asymptotic solution and FSS disagree near the critical velocity of $V=c$; however, in this case, it is for a different reason.
For $\omega_f>\omega_c$, the FSS and asymptotic solutions agree when $V=c$, and this is true for nearby velocities as well.  In the $\omega_f<\omega_c$ case, however, the asymptotic solution diverges for velocities near $V=c$ due to the nonumiformity of the large time limit (see Section~\ref{sig:sec:LTC:VEqC} above). The reason for this behavior difference is that the real part of the solution does not diverge when $\omega_f>\omega_c$; it is the \textit{imaginary} part which demonstrates similar divergence to the $\omega_f<\omega_c$ case. If $A_f$ is real, the long-time asymptotic solution is continuous across $V=c$, whereas it is not for $\omega_f<\omega_c$.

\section{Effect of forcing frequency on response regions}
	\label{sig:app:CriticalVelocities}
\begin{figure}[ht!]
\includegraphics[keepaspectratio,width=6in]{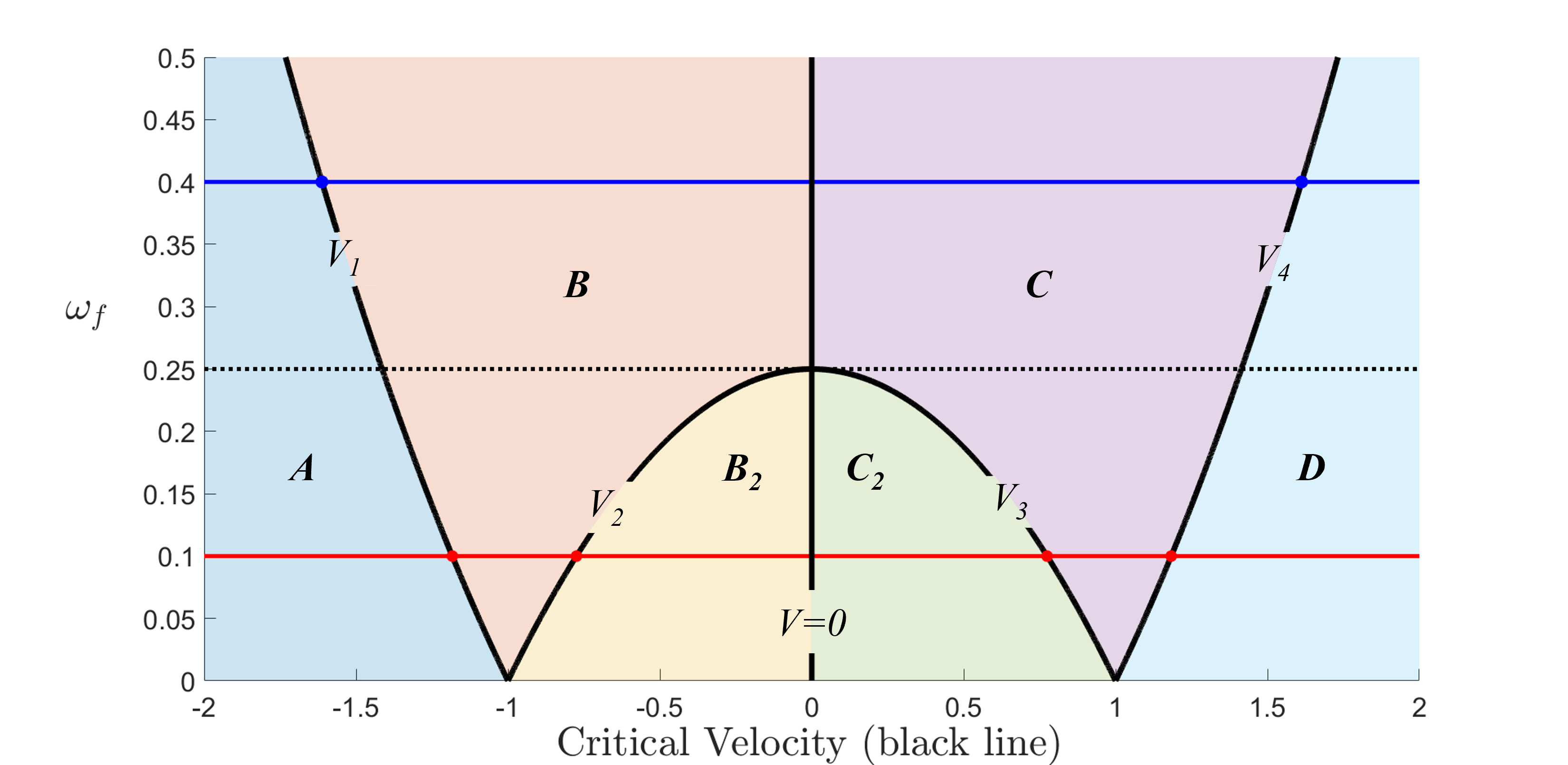}
\caption{Plot tracking critical velocities (black curves) (given in~(\ref{sig:eq:AmpVelLTC}) and~(\ref{sig:eq:AmpVelGTC})) for different values of the forcing frequency. A horizontal line at $\omega=0.25$ marks the critical frequency $\omega_c$.  The plot was generated using $B=1$ and $c=1$. Note that the critical velocity $V=c=1$ is not included on this plot. Two horizontal lines (blue at $\omega_f=0.4$ and red at $\omega_f=0.1$) are included to demonstrate how to extract the critical velocities $V_1$, $V_2$, $V_3$, and $V_4$ for any given value of $\omega_f$.}
\label{sig:fig:CriticalVelocities}
\end{figure}

We now describe how the breadth of the regions described in Sections~\ref{sig:sec:Results:LTCRegime} and~\ref{sig:sec:Results:GTCRegime} change with varying $\omega_f$. Fig.~\ref{sig:fig:CriticalVelocities} shows the loci of observable critical velocities in the solution to~(\ref{sig:eq:h_PDE}). It was generated by plotting the four velocities $V_1$ through $V_4$, given in~(\ref{sig:eq:AmpVelLTC}) and~(\ref{sig:eq:AmpVelGTC}), for a range of $\omega_f$ values holding all other parameter fixed at the same values used in prior figures.  
For a value of $\omega_f$ below the critical frequency (red horizontal line), we can see  that there are four critical velocities, two upstream and two downstream. The outer pair ($V_1$ and $V_4$) are the \textit{Leading Edges of the Forced Solution} and the inner pair ($V_2$ and $V_3$) are the \textit{Leading Edges of the Additionally Forced Regions}. For a value of $\omega_f$ above the critical frequency (blue horizontal line), we can see that there are only two critical velocities ($V_1$ and $V_4$) --- the pair are the \textit{Leading Edges of the Forced Solution}. Note that the two inner velocities converge on $V=0$ as $\omega$ approaches $\omega_c$ from below.

Relating Fig.~\ref{sig:fig:CriticalVelocities} back to the long-time asymptotic forms, we can see that the velocity $V_1$ is the rate at which the contribution of $k_3$ travels away from the forcing location. The same is true for $V_2$ and $k_0$, $V_3$ and $k_1$, and $V_4$ and $k_2$. As $\omega_f$ is increased towards the critical frequency, the contributions from $k_2$ and $k_3$ propagate faster through the domain, and those of $k_0$ and $k_1$ propagate more slowly. Above the critical frequency, the contributions of $k_0$ and $k_1$ decay in time, and are therefore not relevant to the bounding amplitudes. See Figs.~\ref{sig:fig:LTCPoles} and~\ref{sig:fig:GTCPoles} for the pole structure on either side of $\omega_c$.

Through the inspection of~(\ref{sig:eq:AmpLT:Region3}) and~(\ref{sig:eq:AmpLT:Region4}), we see that the bounding amplitudes of regions $B_2$ and $C_2$ go to infinity (there is a $(k_0-k_1)$ in the denominators) and the widths go to zero ($V_2,V_3\to 0$) as $\omega_f\to\omega_c$ from below. 
Due to the change in pole structure ($k_0$ and $k_1$ are coincident when $\omega_f=\omega_c$~(\ref{sig:eq:PoleLocs})) we did not perform the full asymptotic analysis for this threshold case. However, we did simulate results using the FSS (Appendix~\ref{sig:app:FSS}) at the critical frequency. Although not shown here, the result appears to have a single algebraically growing peak  at $V=0$. The behavior predicted in Fig.~\ref{sig:fig:CriticalVelocities} as $\omega_f\to\omega_c$ from below is consistent with that from the FSS. The infinite bounding amplitude corresponds to the peak growing without bound.


\section{Discussion}
\label{sig:sec:Discussion}

We now comment on differences between the signaling response from operators that enable exponential growth (Fig.~\ref{sig:fig:Exp}) and those that only admit algebraically growing solutions (Figs.~\ref{sig:fig:WireframeLT} and~\ref{sig:fig:WireframeGT}).
In an exponentially unstable system, the bounding envelope for the transient corresponds to $V_-$ and $V_+$ in Fig.~\ref{sig:fig:Exp}.  These correspond precisely to the velocities at which the rate of exponential growth is equal to zero, and there is a continuous locus of growing amplitudes for $V\in[V_-,V_+]$.  In the algebraically growing system, there is only one velocity, $V=c$, at which long-time growth can occur (see Figs.~\ref{sig:fig:WireframeLT} and~\ref{sig:fig:WireframeGT}). The interaction between transient and forced responses is significantly altered; instead of a broad region of competition between growing forced and transient responses even at large times, there are only large regions of constant amplitude that expand away from the source with time in accordance with the bounding critical velocities.  Additionally, the speed $V_E$ in Fig.~\ref{sig:fig:Exp} is not present in algebraically growing systems, as this corresponds to the velocity at which the arguments of exponential growth of the transient are equal to that of the transient.  There is no relevant analogous speed in the algebraically growing system.
Note, however, that there is a direct relationship between exponential and algebraic signaling for the velocity $V_I$ in Fig.~\ref{sig:fig:Exp} with those in this paper.  Just as with $V_I$, all the critical velocities ($V_1$ through $V_4$) in the neutrally stable response correspond to locations where the influence of poles (arising from the oscillatory forcing) in the Fourier integral changes.  This corresponds to their inclusion in contours used to evaluate the integrals at large time.

While classical stability analysis~(\ref{sig:eq:general:ClassicalStabilityClassification}) fails to classify the behavior of the convective (i.e., $c\neq 0$) transients in solutions to~(\ref{sig:eq:h_PDE}), it correctly predicts that there is no growth or decay in the long-time response of~(\ref{sig:eq:h_PDE}) for any given $x$ ($V$ approaches 0). This is an intriguing result because the transient behavior arises from the homogeneous solution to which classical stability analysis is applied.
The signaling response of~(\ref{sig:eq:h_PDE}) has both upstream and downstream effects of forcing as time goes to infinity (see the long-time forced regions of Section~\ref{sig:sec:LTC:AdForced}).
The effects neither grow nor decay in space and thus, given enough time, the effects of the forcing will affect the whole domain with constant amplitude oscillations. This behavior is unlike that of an exponential signaling problem~\cite{barlow2017}, where the forced solution either grows or decays in space at large times.  That said, there is an upstream and downstream behavior present in both algebraic and exponentially growing systems induced by the forcing.

\section{Conclusions}
	\label{sig:sec:Conclusions}

The oscillatory forcing of varicose waves in a thin planar inviscid liquid sheet (and in the work of~\cite{Gordillo, barlow2017, barlow2012}) at neutral stability leads to a transient growing peak and a long-time oscillatory response. We find that the behavior of the transient solution is highly dependent on how the disturbance is initiated. Only certain phases of the forcing function lead to an algebraically growing peak in the response. This dependence on initial condition is analogous to the findings of King et al.~\cite{king2016}, where only forcing the initial surface velocity leads to algebraic growth.
The bounding amplitudes and critical velocities, which characterize the oscillatory response, can be extracted from the long-time asymptotic analysis of the Fourier integral solution; specifically, the amplitudes are extracted from the long-time asymptotic solution as the sum of multiple sinusoids and the critical velocities are extracted from the coincidence of a pole from the particular solution and a saddle point from the method of steepest descent. We find that there is a significant change in structure for forcing frequencies above and below a critical forcing frequency. 
The signaling response travels both upstream and downstream from the disturbed location. This behavior, combined with the spatial invariance within the regions, means that the disturbance will affect the entire domain after a sufficiently long time. From a manufacturing perspective, the system is vulnerable to such disturbances, as the damage to the product will not be localized.

\bibliography{neutral2022}

\begin{appendix}

\section{Fourier Series Solution (FSS) to~(\ref{sig:eq:h_PDE})}
	\label{sig:app:FSSTotal}

\subsection{Form of the FSS}
	\label{sig:app:FSS}
In order to validate the asymptotic analysis employed in Appendix~\ref{sig:app:FourierSolution}, the solution to~(\ref{sig:eq:h_PDE}) over the infinite domain $x\in(-\infty,\infty)$ is approximated with a Fourier series solution (FSS) over a finite domain of $x\in[-L,L]$ with periodic boundary conditions $h(-L,t)=h(L,t)$ as done by Barlow et al.~\cite{barlow2010}. The finite length $L$ is taken to be sufficiently large so that the response is unaffected by an increase in $L$ during the simulation time. The FSS is given by
\begin{subequations}\label{sig:eq:app:FSS}
	\begin{equation}
		h_F(x,t)=
		\sum_{n=-N}^N
		\left(\left(
		F_{1_n}e^{r_{1_n}t}+F_{2_n}e^{r_{2_n}t}+F_{3_n}e^{i\omega_ft}
		\right)e^{ik_nx}\right),
	\end{equation}
where
	\begin{equation}
		x\in [-L,L],
		\qquad
		k_n=\frac{n\pi}{L},
	\end{equation}
\begin{equation}
F_{1_n}=\frac{h_0}{2L}-F_{2_n}-F_{3_n},\quad
F_{2_n}=\frac{
\frac{u_0}{2L}-r_{1_n}\frac{h_0}{2L}+(r_{1_n}-i\omega_f)F_{3_n}
}{(r_{2_n}-r_{1_n})},\quad
\end{equation}
\begin{equation}
F_{3_n}=
\frac{\frac{A_f}{2L}}
{(i\omega_f)^2+\alpha_n(i\omega_f)+\beta_n},
\end{equation}
\begin{equation}
r_{1_n}=-i(ck_n-Bk_n^2),\qquad
r_{2_n}=-i(ck_n+Bk_n^2),
\end{equation}
\begin{equation}
\alpha_n = (2ick_n), \qquad
\beta_n = (ick_n)^2+
B^2(k_n)^{4}.
\end{equation}

\end{subequations} 
Notably,
\begin{equation}
	h(x,t)=
	\lim_{N\to\infty}
	\left(
	\lim_{L\to\infty}
	\left(
	h_F(x,t)
	\right)
	\right)
\end{equation}

\subsection{Spatial resolution of solutions}
	\label{sig:app:FSSResolution}
A key issue with the implementation of the long-time asymptotic solutions given in equations~(\ref{sig:eq:LTCRegionSolutions}) and~(\ref{sig:eq:GTCRegionSolutions}) is the resolution of high frequency sinusoids. An important distinction to make first is that the spatial resolution discussed here is entirely separate from how many terms are included in the FSS. Spatial resolution here refers to how many data points are sampled over a given $V$ domain. While the values of the step size in $V$ could be tested manually, it is possible to approximate the required step size by examining the asymptotic solutions.

To explore this issue, we first suppose that the phase of the highest frequency exponential (in $V$) from the long-time asymptotic solution, denoted generally as $f$, has some associated frequency which is a function of time, $\omega_{A}(t)$, and some phase shift which may be a function of time, $\phi(t)$, such that,

\begin{equation}
	\label{sig:eq:SampleForAliasing}
f=e^{i(\omega_{A}(t)V + \phi(t))}.
\end{equation}
\noindent
As an example, for region $C$ in Fig.~\ref{sig:fig:WireframeLT}, $\omega_A(t)=k_2t$ and $\phi(t)=i\omega_ft$ (extracted from the exponent in~(\ref{sig:eq:AmpLT:Region5})).
For a fixed time, it follows that the mode given in (\ref{sig:eq:SampleForAliasing}) has a wavelength of $\nicefrac{2\pi}{\omega_{A}(t)}$ in $V$. We may then choose the number of points in our FSS (Appendix~\ref{sig:app:FSS}) such that the maximum value for the step size in $V$ is, for instance,
\begin{equation}
\Delta V = \frac{\pi}{10\omega_{A}(t)}.
\end{equation}

\noindent
As an example, the $\Delta V$ needed to achieve this level of resolution for region $C$ in Fig.~\ref{sig:fig:WireframeLT} would be
\begin{equation}
\Delta V_{(\omega_f<\omega_c,~\mathrm{Region}~C)} = \frac{\pi}{10 k_2 t}
=
\left(
	\frac{B\pi}
	{5\left(c-\sqrt{c^2+4B\omega_f}\right)}
\right)
\frac{1}{t}.
\end{equation}
\noindent
Note that, as we look at increasingly longer times, a smaller step size in $V$ is required to achieve the same resolution relative to the wave pattern. This is especially problematic as $t\to\infty$ because $\Delta V\to 0$. As long as the same velocity points are sampled, however, the FSS (Appendix~\ref{sig:app:FSS}) and the asymptotic solutions can be compared. However, if $\Delta V$ is too large, oscillations in the solution will not be adequately resolved and the true solution structure will be obscured.

\section{Fourier integral solution to~(\ref{sig:eq:h_PDE})}
	\label{sig:app:FourierSolution}

The solution to the system~(\ref{sig:eq:h_PDE}) is found by taking its Fourier transform, solving the resulting ordinary differential equation in the time domain through a Laplace transform, and then using the Fourier inversion integral, as stated in the main text. After performing those steps, but before rearranging the solution into the form given by~(\ref{sig:eq:InversionIntegral}) in the main text, the following intermediate solution is obtained,

\begin{multline}
\label{sig:eq:app:UnsimplifiedIntegrals}
h(x,t)=
-\frac{A_f}{2\pi B^2}
	\int\limits_{-\infty}^{\infty}
	\Bigg(
		\frac{\cos (Bk^2 t)e^{ik(V-c)t}}
		{(k-k_0)(k-k_1)(k-k_2)(k-k_3)}
	\Bigg)dk 
\\-\frac{iA_fc}{2\pi B^3}
	\int\limits_{-\infty}^{\infty}
	\Bigg(
		\frac{\sin (Bk^2 t)e^{ik(V-c)t}}
		{k(k-k_0)(k-k_1)(k-k_2)(k-k_3)}
	\Bigg)dk
\\-\frac{iA_f\omega_F}{2\pi B^3}
	\int\limits_{-\infty}^{\infty}
	\Bigg(
		\frac{\sin (Bk^2 t)e^{ik(V-c)t}}
		{k^2(k-k_0)(k-k_1)(k-k_2)(k-k_3)}
	\Bigg)dk
\\+\frac{A_f}{2\pi B^2}
	\int\limits_{-\infty}^{\infty}
	\Bigg(
		\frac{e^{i(kV+\omega_f)t}}
		{(k-k_0)(k-k_1)(k-k_2)(k-k_3)}
	\Bigg)dk.
\end{multline}
The definitions of $k_0$ through $k_3$ are given in equation~(\ref{sig:eq:PoleLocs}) and are written here for completeness:
\begin{equation*}
k_0=\frac{-c-\sqrt{c^2-4B\omega_f}}{2B}
,\qquad
k_1=\frac{-c+\sqrt{c^2-4B\omega_f}}{2B},
\end{equation*}
\begin{equation}
k_2=\frac{c-\sqrt{c^2+4B\omega_f}}{2B}
,\qquad
k_3=\frac{c+\sqrt{c^2+4B\omega_f}}{2B}.
\end{equation}

The four integrals in~(\ref{sig:eq:app:UnsimplifiedIntegrals}) are subsequently combined and written in terms of two new integrals, $\mathbb{J}$ and $\mathbb{L}$, as indicated in equations~(\ref{sig:eq:ITilde}) and~(\ref{sig:eq:IHat}) in the main text.

\section{Asymptotic evaluation of integrals $\mathbb{J}$ and $\mathbb{L}$} 
	\label{sig:app:AsympFormsGeneral}
The following sections outline the
long-time asymptotic behavior of the integrals $\mathbb{J}$~(\ref{sig:eq:ITilde})and $\mathbb{L}$~(\ref{sig:eq:IHat}) for an arbitrary value of $\omega_f$. In Appendix~\ref{sig:app:AsympFormsForOmegaF} the values of $\omega_f$ are imposed to construct the full asymptotic solutions.

\subsection{Evaluation of $\mathbb{J}$ (Equation~(\ref{sig:eq:ITilde}))}
	\label{sig:app:ITildeSolution}
In order to evaluate $\mathbb{J}$ in~(\ref{sig:eq:ITilde}) it is broken into $\mathbb{J}_1$ and $\mathbb{J}_2$ such that
\begin{subequations}
\begin{equation}
	\mathbb{J}=\mathbb{J}_1+\mathbb{J}_2,
\end{equation}
\begin{equation}
\label{sig:eq:app:ITilde1}
\mathbb{J}_1=
\int\limits_{-\infty}^{\infty}
-\frac{A_fe^{i(\omega_f)t}}{B^2}
\frac{e^{i(kV)t}e^{iB(k-k_2)(k-k_3)t}}
{(k-k_0)(k-k_1)(k-k_2)(k-k_3)}dk,
\end{equation}
\begin{equation}
\label{sig:eq:app:ITilde2}
\mathbb{J}_2=
\int\limits_{-\infty}^{\infty}
\frac{A_fe^{i(\omega_f)t}}{B^2}\frac{e^{i(kV)t}}
{(k-k_0)(k-k_1)(k-k_2)(k-k_3)}dk.
\end{equation}
\end{subequations}

\noindent
Equation~(\ref{sig:eq:app:ITilde1}) is evaluated with a ``bow-tie'' contour\footnote{The contour is so called because it is shaped like a bow-tie with two lobes extending out symmetrically from a central point.} centered on
\begin{equation}
\label{sig:eq:app:ITilde1Saddle}
k_s=\frac{(c-V)}{2B},
\end{equation}
\noindent
which extends into the first and third quadrants; Supplemental Material Section~\ref{sig:sup:BowTieContour} provides the methodology and Fig.~\ref{sig:fig:app:BowtieContour} (provided in Supplemental Material Section~\ref{sig:sup:BowTieContour}) shows the ``bow-tie'' contour shape.
Equation~(\ref{sig:eq:app:ITilde2}) is evaluated with a half-plane contour in either the upper or lower half-plane, depending on the sign of $V$ (see Supplemental Material Section~\ref{sig:sup:HalfPlaneContour}).
The solution for $\mathbb{J}_1$ is summarized below (from Supplemental Material Sections~\ref{sig:sup:BowTieContour})
\begin{subequations}
	\label{sig:eq:app:ITilde1FormTotal}
\begin{multline}
	\label{sig:eq:app:ITildeForm}
\mathbb{J}_1\sim
\frac{-A_fe^{i\omega_ft}}{B^2}
\Bigg[
\bigg(\Big(
P_A(k_0;k_s)
-P_B(k_0;k_s)+
\Big)F(k_0)\bigg)+
\bigg(\Big(
P_A(k_1;k_s)-
P_B(k_1;k_s)
\Big)F(k_1)\bigg)\\+
\bigg(\Big(
P_A(k_2;k_s)-
P_B(k_2;k_s)
\Big)F(k_2)\bigg)+
\bigg(\Big(
P_A(k_3;k_s)-
P_B(k_3;k_s)
\Big)F(k_3)\bigg)\\+
\frac{
2e^{\left(\frac{i\pi}{4}\right)}
e^{-i\left(\frac{(V-c)^2}{4B}
+\omega_f\right)t}
}
{(k_s-k_0)
(k_s-k_1)
(k_s-k_2)
(k_s-k_3)}
\sqrt{\frac{\pi}{4Bt}}
\Bigg]
\quad
\text{as}
\quad t\to\infty,
\end{multline}
\begin{equation}
k_s=\frac{c-V}{2B},
\quad
F(k_j)=
\lim_{k\to k_j}=
\left[
\frac{(k-k_j)e^{i
(B(k-k_2)(k-k_3)+kV
)t}
}{
(k-k_0)(k-k_1)(k-k_2)(k-k_3)}\right],
\end{equation}
\begin{equation}
\label{sig:eq:app:ITilde:PDef}
P_L(k_j;k_s)=
\begin{cases}
	2i\pi &\text{if pole $k_j$ is enclosed by contour $L$}\\
	i\pi &\text{if pole $k_j$ is intersected by contour $L$}\\
	0 &\text{if pole $k_j$ is not enclosed by contour $L$}
\end{cases}.
\end{equation}
\end{subequations}
\noindent
The notation in~(\ref{sig:eq:app:ITildeForm}) is defined in terms of the $P_L$ function~(\ref{sig:eq:app:ITilde:PDef}) for two contours labeled $A$ and $B$ (so, $L=A$ or $L=B$ in~(\ref{sig:eq:app:ITilde1FormTotal})); see Fig.~\ref{sig:fig:app:BowtieContour} of Supplemental Material Section~\ref{sig:sup:BowTieContour} for a depiction of these contours. The function $P_L$ accounts for if a given pole $k_j$ is contained within a given contour $L$.
By the convention used in this paper, the  contour $A$ is said to be the portion of the bow-tie contour that extends to the right of the saddle point $k_s$ and contour $B$ is said to be the portion of the bow-tie contour that extends to the left of the $k_s$. (see Fig.~\ref{sig:fig:app:BowtieContour} in Supplemental Material Section~\ref{sig:sup:BowTieContour}). 
The solution for $\mathbb{J}_2$ (Supplemental Material Section~\ref{sig:sup:HalfPlaneContour}) is given as 
\begin{subequations}
\label{sig:eq:app:ITilde2FormTotal}
\begin{equation}
\mathbb{J}_2\Big|_{V\geq 0}=
\frac{i\pi A_f}{B^2}e^{i\omega_ft}
\bigg(
2G(k_1)+G(k_2)+G(k_3)
\bigg),
\end{equation}
\begin{equation}
\mathbb{J}_2\Big|_{V\leq 0}=
\frac{i\pi A_f}{B^2}e^{i\omega_ft}
\Big(
-2G(k_0)-G(k_2)-G(k_3)
\Big),
\end{equation}
\begin{equation}
G(z_j)=\lim_{k\to k_j}
\left(
(k-k_j)
\frac{e^{ikVt}}{(k-k_0)(k-k_1)(k-k_2)(k-k_3)}
\right),
\end{equation}
\end{subequations}
where $k_0$ through $k_3$ are defined in~(\ref{sig:eq:PoleLocs}).
Note that the evaluation of the two expressions for $\mathbb{J}_2$ at $V=0$ are equivalent. The form of~(\ref{sig:eq:app:ITildeForm}) does not assume the position of the poles, so it is general for any case where $\omega_f\neq\omega_c$.

\subsection{Evaluation of $\mathbb{L}$ (Equation~(\ref{sig:eq:IHat}))}
	\label{sig:app:IHatSolution}
The integral $\mathbb{L}$ in (\ref{sig:eq:IHat}) is first differentiated with respect to $t$ and broken up into four parts such that
\begin{equation}
\frac{d\mathbb{L}}{dt} = 
	i\frac{A_f}{2B^2}
	\left(
	(V-c)
	\left(
	\mathbb{L}_1+\mathbb{L}_2
	\right)
	+B
	\left(
	\mathbb{L}_3+\mathbb{L}_4
	\right)
	\right),
\end{equation}
\begin{subequations}
	\label{sig:sup:eq:IHatSubequations}
\begin{equation}
	\label{sig:sup:eq:IHat1}
\mathbb{L}_1=
\int\limits_{-\infty}^{\infty}
\frac{e^{i(Bk^2+k(V-c))t}}
{k(k-k_0)(k-k_1)}
dk,
\end{equation}
\begin{equation}
	\label{sig:sup:eq:IHat2}
\mathbb{L}_2=
\int\limits_{-\infty}^{\infty}
\frac{-e^{i(-Bk^2+k(V-c))t}}
{k(k-k_0)(k-k_1)}
dk,
\end{equation}
\begin{equation}
	\label{sig:sup:eq:IHat3}
\mathbb{L}_3=
\int\limits_{-\infty}^{\infty}
\frac{e^{i(Bk^2+k(V-c))t}}
{(k-k_0)(k-k_1)}
dk,
\end{equation}
\begin{equation}
	\label{sig:sup:eq:IHat4}
\mathbb{L}_4=
\int\limits_{-\infty}^{\infty}
\frac{e^{i(-Bk^2+k(V-c))t}}
{(k-k_0)(k-k_1)}
dk.
\end{equation}
\end{subequations}
\noindent 
The four sub-integrals in~(\ref{sig:sup:eq:IHatSubequations}) are evaluated using bow-tie contours described in Fig.~\ref{sig:fig:app:BowtieContour} of Supplemental Material Section~\ref{sig:sup:BowTieContour}.
The contours for equations~(\ref{sig:sup:eq:IHat1}) and~(\ref{sig:sup:eq:IHat3}) are centered at
\begin{equation}
k_{s_1} = \frac{(c-V)}{2B}
\end{equation}
\noindent
and extend into the first and third quadrants. The contours for equations~(\ref{sig:sup:eq:IHat2}) and~(\ref{sig:sup:eq:IHat4}) are centered at
\begin{equation}
k_{s_2} = \frac{(V-c)}{2B}
\end{equation}
\noindent
and extend into the second and fourth quadrants.

Once the contours are evaluated (again, see Supplemental Material Section~\ref{sig:sup:BowTieContour}), the results are combined to yield
\begin{subequations}
\begin{multline}
	\label{sig:eq:app:IHatForm}
\frac{d\mathbb{L}}{dt}\sim
\frac{iA_f}{2B^2}\int\limits_\psi^t\Bigg[
\\
\left[
	\frac{P_A(k_0;k_{s_1})-P_B(k_0;k_{s_1})}{(k_0-k_1)}
\right]
\left(
	B+\frac{(V-c)}{k0}
\right)
\left(
	e^{i(Bk_0^2+k_0(V-c))t}
\right)
\\+
\left[
	\frac{P_A(k_1;k_{s_1})-P_B(k_2;k_{s_1})}{(k_1-k_0)}
\right]
\left(
	B+\frac{(V-c)}{k1}
\right)
\left(
	e^{i(Bk_1^2+k_1(V-c))t}
\right)
\\-
\frac{e^{i\frac{\pi}{4}}\sqrt{\pi B}}
{(k_{s_1}-k_0)(k_{s_1}-k_1)}
\left(
	\frac{e^{ik_{s_1}t(V-c)}}{\sqrt{t}}
\right)
-
\frac{e^{-i\frac{\pi}{4}}\sqrt{\pi B}}
{(k_{s_2}-k_0)(k_{s_2}-k_1)}
\left(
	\frac{e^{ik_{s_2}t(V-c)}}{\sqrt{t}}
\right)
\\+
\left[
	\frac{P_A(k_0;k_{s_2})-P_B(k_0;k_{s_2})}{(k_0-k_1)}
\right]
\left(
	-B+\frac{(V-c)}{k0}
\right)
\left(
	e^{i(-Bk_0^2+k_0(V-c))t}
\right)
\\+
\left[
	\frac{P_A(k_1;k_{s_2})-P_B(k_1;k_{s_2})}{(k_1-k_01)}
\right]
\left(
	-B+\frac{(V-c)}{k1}
\right)
\left(
	e^{i(-Bk_1^2+k_1(V-c))t}
\right)
\\+
\bigg(
P_A(0;k_{s_1})-P_B(0;k_{s_1})
\bigg)
\bigg(
P_A(0;k_{s_2})-P_B(0;k_{s_2})
\bigg)
\frac{(V-c)}{(k_0)(k_1)}
\Bigg]dt + \mathcal{C},
\end{multline}
\begin{equation}
	\label{sig:eq:app:IHatPsiDef}
\psi=
\begin{cases}
	\infty & \text{ if } V\neq c\\
	0 & \text{ if } V = c
\end{cases},
\end{equation}
\end{subequations}

\noindent
The notation in~(\ref{sig:eq:app:IHatForm}) is defined in terms of the same $P_L$ function~(\ref{sig:eq:app:IHat:PDef}) as is used in~(\ref{sig:eq:app:ITilde1FormTotal}). The function $P_L$ accounts for if a given pole $k_j$ is contained within a given contour $L$.
By the convention used in this paper, the contour $A$ is said to be the portion of the bow-tie contour that extends to the right of the saddle point $k_s$ and contour $B$ is said to be the contour portion of the bow-tie that extends to the left of the $k_s$ (see Fig.~\ref{sig:fig:app:BowtieContour} in Supplemental Material Section~\ref{sig:sup:BowTieContour}).
Importantly, the $P_L$ functions have no dependence on $t$, and they can be treated as constants for the purposes of integration. The lower limit of integration $\psi$~(\ref{sig:eq:app:IHatPsiDef}) is chosen such that the integration constant $C$ evaluates to zero which is consistent with the FSS~(Appendix~\ref{sig:app:FSS}) at long times. By integrating the first terms, we obtain
\begin{subequations}
\label{sig:eq:app:IHatFormTotal}
\begin{multline}
\mathbb{L}\sim
\frac{A_f}{2B^2}
\Bigg[
\\
-\bigg(
	P_A(k_0;k_{s_1})-P_B(k_0;k_{s_1})
\bigg]
\left(
	\frac{e^{i\left(Bk_0^2+k_0(V-c)\right)t}}
	{k_0^2(k_1-k_0)}
\right)
\\
+\bigg(
	P_A(k_1;k_{s_1})-P_B(k_1;k_{s_1})
\bigg)
\left(
\frac{e^{i\left(Bk_1^2+k_1(V-c)\right)t}}
{k_1^2(k_1-k_0)}
\right)
\\
-\bigg(
	P_A(k_0;k_{s_2})-P_B(k_0;k_{s_2})
\bigg)
\left(
\frac{e^{i\left(-Bk_0^2+k_0(V-c)\right)t}}
{k_0^2(k_1-k_0)}
\right)
\\
+\bigg(
	P_A(k_1;k_{s_2})-P_B(k_1;k_{s_2})
\bigg)
\left(
\frac{e^{i\left(-Bk_1^2+k_1(V-c)\right)t}}
{k_1^2(k_1-k_0)}
\right)
\\+
\frac{-ie^{i\frac{\pi}{4}}\sqrt{B\pi}}
{(k_{s_1}-k_0)(k_{s_1}-k_1)}
\int\limits_\infty^t
\Bigg(
\frac{e^{i\frac{k_{s_1}(V-c)}{2}\eta}}{\sqrt{\eta}}
\Bigg)d\eta
\\+
\frac{-ie^{-i\frac{\pi}{4}}\sqrt{B\pi}}
{(k_{s_1}+k_0)(k_{s_1}+k_1)}
\int\limits_\infty^t
\Bigg(
\frac{e^{-i\frac{k_{s_1}(V-c)}{2}\eta}}{\sqrt{\eta}}
\Bigg)d\eta
+ \mathcal{C}\Bigg]
\\ \qquad \text{as} \qquad t\to\infty,
\end{multline}

\begin{equation}
	\label{sig:eq:app:IHat:PDef}
P_L(k_j;k_{s_m})=
\begin{cases}
	2i\pi &\text{if pole $k_j$ is enclosed by contour $L$ for $\mathbb{L}_m$}\\
	i\pi &\text{if pole $k_j$ is intersected by contour $L$}\\
	0 &\text{if pole $k_j$ is not enclosed by contour $L$}
\end{cases},
\end{equation}
\end{subequations}
\noindent
where $\psi$ defined in~(\ref{sig:eq:app:IHatPsiDef}), and the solution to the remaining integrals is given in~(\ref{sig:eq:app:IHatFormTotal}).

\section{Long-time asymptotic behavior of the Fourier integral solution}
	\label{sig:app:AsympFormsForOmegaF}
The following sections outline the asymptotic solutions to the Fourier integral solution to~(\ref{sig:eq:h_PDE}) for the cases of $\omega_f<\omega_c$ and $\omega_f>\omega_c$.	These forms are constructed by applying the pole structure associated with the $\omega_f$ value to the forms of $\mathbb{J}$ and $\mathbb{L}$ given in Appendix~\ref{sig:app:AsympFormsGeneral}.
	
\subsection{$\omega_f<\omega_c$}
\label{sig:app:LTC}
This section develops the long-time asymptotic solution for equation~(\ref{sig:eq:h_PDE}) for the pole structure in Fig.~\ref{sig:fig:LTCPoles}. The long-time solution for the $\omega_f<\omega_c$ case is found by applying the pole structure to the solutions for $\mathbb{J}$ (see Appendix~\ref{sig:app:ITildeSolution}) and $\mathbb{L}$ (see Appendix~\ref{sig:app:IHatSolution}). 
Evaluating~(\ref{sig:eq:app:ITilde1FormTotal}),~(\ref{sig:eq:app:ITilde2FormTotal}) and~(\ref{sig:eq:app:IHatFormTotal}) yields
\begin{multline}
\label{sig:eq:app:LTCSolution}
Real[
h(x,t)_{\omega_f<\omega_c}
] \sim 
Real[
 \frac{A_f e^{i\omega_ft}}{B^2}
\left(T_1 + T_2+T_3\right)
-i A_f
\delta_j \sqrt{t} 
]
\\
\text{as } t\to\infty
,\quad \delta_j=
\begin{cases} 
      1, & V=c \\
      0, & V\neq c 
   \end{cases},
\end{multline}
where $T_1$, $T_2$, and $T_3$ are defined as follows
\begin{subequations}
\label{sig:eq:app:LTC_T1}
\begin{multline}
T_1\sim -
\Bigg[
\bigg(P(k_0,k_{s_1})F(k_0)\bigg)+
\bigg(P(k_1,k_{s_1})F(k_1)\bigg)+
\bigg(P(k_2,k_{s_1})F(k_2)\bigg)\\+
\bigg(P(k_3,k_{s_1})F(k_3)\bigg)+
\frac{
2e^{\left(\frac{i\pi}{4}\right)}
e^{-i\left(\frac{(V-c)^2}{4B}
+\omega_f\right)t}
}
{(k_{s_1}-k_0)
(k_{s_1}-k_1)
(k_{s_1}-k_2)
(k_{s_1}-k_3)}
\sqrt{\frac{\pi}{4Bt}}
\Bigg]
\qquad \text{as} \qquad t\to\infty,
\end{multline}
\begin{equation}
F(k_j)=
\lim_{k\to k_j}=
\left[
\frac{(k-k_j)e^{i
(B(k-k_2)(k-k_3)+kV
)t}
}{
(k-k_0)(k-k_1)(k-k_2)(k-k_3)}\right],
\end{equation}
\end{subequations}
\begin{subequations}
\label{sig:eq:app:LTC_T2}
\begin{equation}
T_2=
\text{sgn}(V)i\pi
\Big(
G(k_0)+G(k_1)+G(k_2)+G(k_3)
\Big),
\end{equation}
\begin{equation}
G(z_j)=\lim_{k\to k_j}
\left(
(k-k_j)
\frac{e^{ikVt}}{(k-k_0)(k-k_1)(k-k_2)(k-k_3)}
\right),
\end{equation}
\end{subequations}
\begin{multline}
\label{sig:eq:app:LTC_T3}
T_3\sim
\frac{1}{2}
\Bigg[
-P(k_0,k_{s_1})
\left(
\frac{e^{i\left(Bk_0^2+k_0(V-c)\right)t}}
{k_0^2(k_1-k_0)}
\right)
+
P(k_1,k_{s_1})
\left(
\frac{e^{i\left(Bk_1^2+k_1(V-c)\right)t}}
{k_1^2(k_1-k_0)}
\right)
\\
+P(k_0,k_{s_2})
\left(
\frac{e^{i\left(-Bk_0^2+k_0(V-c)\right)t}}
{k_0^2(k_1-k_0)}
\right)
-
P(k_1,k_{s_2})
\left(
\frac{e^{i\left(-Bk_1^2+k_1(V-c)\right)t}}
{k_1^2(k_1-k_0)}
\right)
\\+
\frac{-ie^{i\frac{\pi}{4}}\sqrt{B\pi}}
{(k_{s_1}-k_0)(k_{s_1}-k_1)}
\int\limits_\infty^t
\Bigg(
\frac{e^{i\frac{k_{s_1}(V-c)}{2}\eta}}{\sqrt{\eta}}
\Bigg)d\eta
+
\frac{-ie^{-i\frac{\pi}{4}}\sqrt{B\pi}}
{(k_{s_1}+k_0)(k_{s_1}+k_1)}
\int\limits_\infty^t
\Bigg(
\frac{e^{-i\frac{k_{s_1}(V-c)}{2}\eta}}{\sqrt{\eta}}
\Bigg)d\eta
+ \mathcal{C}\Bigg]
\\ \qquad \text{as} \qquad t\to\infty
\qquad \text{and} \qquad V\neq c,
\end{multline}
\begin{equation}
\label{sig:eq:app:SaddlePointsDef}
k_{s_1}=\frac{(c-V)}{2B},
\quad
k_{s_2}=\frac{(V-c)}{2B},
\end{equation}
\begin{equation}\label{sig:eq:app:PFunDef}
P(k_j,k_{s_L})=
	\begin{cases}
      i\pi & \text{  if  } k_{s_L}<k_j \\
      \text{undefined} & \text{  if  } k_{s_L}=k_j \\
      -i\pi, & \text{  if  } k_{s_L}>k_j 
	\end{cases}.
\end{equation}

In equation~(\ref{sig:eq:app:LTC_T3}), $\mathcal{C}$ is an unknown constant of integration. The limits of the integrals are chosen for $V\neq c$ such that $\mathcal{C}$ evaluates to zero as motivated by comparison with the FSS~(Appendix~\ref{sig:app:FSS}). The integrals can be evaluated by defining $\zeta\equiv\nicefrac{(V-c)^2}{4B}$, $E_1\equiv-1$, $E_2\equiv 1$ and evaluating the new integral through the asymptotic expansion of the Fresnel integral as
\begin{multline}
\label{sig:sup:eq:IHatSubIntegralSolution}
\int\limits_\infty^t
\Bigg(
\frac{e^{E_j\zeta \eta}}{\sqrt{\eta}}
\Bigg)d\eta
=
\int\limits_0^t
\Bigg(
\frac{e^{E_j\zeta \eta}}{\sqrt{\eta}}
\Bigg)d\eta
-
\int\limits_0^\infty
\Bigg(
\frac{e^{E_j\zeta \eta}}{\sqrt{\eta}}
\Bigg)d\eta
\sim  \\
\sqrt{\frac{2\pi}{\zeta}}
\left[
	\left(
		\frac{1}{2}
		+\frac{\sin(\zeta t)}{\sqrt{2\pi \zeta t}}
	\right)
+iE_j
	\left(
		\frac{1}{2}
		-\frac{\cos(\zeta t)}{\sqrt{2\pi \zeta t}}
	\right)
\right]
-(1+iE_j)
\sqrt{\frac{\pi}{2\zeta}}
\\ \qquad \text{as} \qquad t\to\infty
\qquad \text{and} \qquad V\neq c.
\end{multline}
\noindent
For the case of equation~(\ref{sig:eq:app:LTC_T3}) when $V=c$, the limits of integration must be chosen to be $0\to t$ instead of $\infty\to t$ to ensure a bounded integral, leaving the evaluation of each integral as
\begin{equation}
\int\limits_0^t\frac{1}{\sqrt{\eta}}d\eta=
2\sqrt{t}+\mathcal{C}_1,
\end{equation}

where $\mathcal{C}_1$ evaluates to zero compared to the FSS~(Appendix~\ref{sig:app:FSS}). The emergent algebraic growth only arises in the imaginary part of $T_3$~(\ref{sig:eq:app:LTC_T3}); as such, if $A_f$ is chosen to be purely real, the growth only occurs in the imaginary part of $h(x,t)$ and is not observable in a physical solution.

\subsection{$\omega_f>\omega_c$}
\label{sig:app:GTC}
This section develops the long-time asymptotic solution for equation~(\ref{sig:eq:h_PDE}) for the pole structure in Fig.~\ref{sig:fig:GTCPoles}. The long-time solution for the $\omega_f>\omega_c$ case is found by applying the pole structure to the solutions for $\mathbb{J}$ (see Appendix~\ref{sig:app:ITildeSolution}) and $\mathbb{L}$ (see Appendix~\ref{sig:app:IHatSolution}). 
Evaluating~(\ref{sig:eq:app:ITilde1FormTotal}),~(\ref{sig:eq:app:ITilde2FormTotal}) and~(\ref{sig:eq:app:IHatFormTotal}) yields
\begin{equation}
	\label{sig:eq:app:GTCSolution}
h(x,t)_{\omega_f>\omega_c} \sim
\left(
\frac{A_f e^{i\omega_ft}}{B^2}
\left(T_1 + T_2\right)\right)
-i A_f
\delta_j \sqrt{t}
\quad
\text{as } t\to\infty,
\quad
\delta_j=
\begin{cases} 
      1, & V=c \\
      0, & V\neq c 
   \end{cases}.
\end{equation}

\noindent
Note that, in equation~(\ref{sig:eq:app:GTCSolution}), the growing term is purely imaginary when $A_f$ is taken to be real and even then only occurs along the $V=c$ velocity.
\begin{subequations}
\label{sig:eq:GTC_T1}
\begin{equation}
T_1\sim -
\Bigg[
P(k_2,k_s)F(k_2)+
P(k_3,k_s)F(k_3)
\Bigg]
\qquad \text{as} \qquad t\to\infty,
\end{equation}
\begin{equation}
F(k_j)=
\lim_{k\to k_j}=
\left[
\frac{(k-k_j)e^{i
(B(k-k_2)(k-k_3)+kV
)t}
}{
(k-k_0)(k-k_1)(k-k_2)(k-k_3)}\right],
\end{equation}
\end{subequations}
\begin{subequations}
\label{sig:eq:GTC_T2}
\begin{equation}
T_2\sim
\text{sgn}(V)i\pi
\Big(
G(k_2)+G(k_3)
\Big)
\qquad \text{as} \qquad t\to\infty,
\end{equation}
\begin{equation}
G(k_j)=\lim_{k\to k_j}
\left(
(k-k_j)
\frac{e^{ikVt}}{(k-k_0)(k-k_1)(k-k_2)(k-k_3)}
\right).
\end{equation}
\end{subequations}

\noindent
The definitions of $k_{s_1}$ and $P(k_j,k_s)$ are given above in equations~(\ref{sig:eq:app:SaddlePointsDef}) and~(\ref{sig:eq:app:PFunDef}), and the poles $k_0$ through $k_3$ are defined in~(\ref{sig:eq:PoleLocs}). As with the $\omega_f<\omega_c$ case, the algebraic growth only occurs in the real part of $h$ if $A_f$ is imaginary or complex.
 
\end{appendix} 
 
\setcounter{section}{0} 
 
 \renewcommand{\thepage}{S\arabic{page}}
\renewcommand{\thesection}{S\arabic{section}}
\renewcommand{\thetable}{S\arabic{table}}
\renewcommand{\thefigure}{S\arabic{figure}}
\renewcommand{\theequation}{S\arabic{equation}}

\section{Justification for bounding amplitude expressions given in Section~\ref{sig:sec:BoundingAmplitudes}}
	\label{sig:app:BoundingAmp}

For a given set of parameters, the solution is arranged into the form of~(\ref{sig:eq:SampleExpSum}) (including the terms which are purely imaginary, for completeness),
\begin{multline}	\label{sig:app:eq:SampleExpSum}
	Real[h(x,t)] \sim Real\left[ \sum_{n=1}^N
	\left(
	\alpha_ne^{i\beta_n(V)t}
	\right)+ o(1) + \text{(Imaginary term)} \right]
	\\ \quad V\in[V_1,V_2]
	 \quad as \quad t\to\infty,
\end{multline}
\noindent
for $N$ different modes where $\alpha_n$ is complex and $\beta_n(V)$ is real, such the that argument of the exponential above is purely imaginary. Only imaginary exponents are considered because real exponents either lead to terms that damp with time (subdominant to oscillations) or terms that grow with time (oscillations are subdominant) as time goes to infinity. The terms are grouped in such a way that $\beta_i\neq\beta_j$, as the amplitude of two in-phase exponentials is merely the sum of the amplitudes. To handle the out of phase terms, we first write
\begin{equation}\label{sig:eq:app:AnDef}
\alpha_n\equiv A_n e^{(i\phi)} 
\quad\text{ such that }\quad
A_n=\big|\alpha_n\big|
\end{equation}
\noindent
where $A_n$ is real and positive and $\phi$ is real. Equation~(\ref{sig:app:eq:SampleExpSum}) then becomes
\begin{equation}
	Real[h(x,t)] \sim Real\left[ \sum_{n=1}^N
	\left(
	A_n e^{(i\beta_n t + i\phi)}
	\right) \right]
	\qquad V\in[V_1,V_2]
	\qquad \text{as} \qquad t\to\infty.
\end{equation}
\noindent
In the summation, each term is a complex number with a fixed modulus and changing argument. Through the triangle inequality, the result of the summation is less than or asymptotic to (denoted by $\lesssim$) the sum of $A_n$:
\begin{equation}
	\Big|
	Real[h(x,t)]
	\Big|  \lesssim \sum_{n=1}^N
	\left(
	A_n 
	\right)
	\qquad V\in[V_1,V_2]
	\qquad \text{as} \qquad t\to\infty.
\end{equation}

From~(\ref{sig:eq:app:AnDef}),
\begin{equation}
\label{sig:eq:app:AmplitudeFormula}
	\Big|
	Real[h(x,t)]
	\Big| \lesssim \sum_{n=1}^N
	\big|
	\alpha_n
	\big|
	\qquad V\in[V_1,V_2]
	\qquad \text{as} \qquad t\to\infty.
\end{equation}
\noindent
Equation~(\ref{sig:eq:app:AmplitudeFormula}) is used directly to find the bounding amplitudes presented in Section~\ref{sig:sec:BoundingAmplitudes}.

\section{Exploration of the relative amplitudes in Figs.~\ref{sig:fig:WireframeLT} and~\ref{sig:fig:WireframeGT}}
	\label{sig:sup:RelativeAmplitudes}
The amplitudes of each region shown in Figs.~\ref{sig:fig:WireframeLT} and~\ref{sig:fig:WireframeGT} were examined to determine the generality of the indicated figures qualitatively. In some cases, inspection of the asymptotic results (see~(\ref{sig:eq:AmpLT:Region1}) and (\ref{sig:eq:AmpLT:Region2})) was all that was needed to establish relationships between the region amplitudes.  In some, the bounding amplitudes were calculated to firmly establish the relationship between amplitudes (labeled as ``Tested'' below). In these cases, the following parameter ranges were explored:
\begin{equation}
A_f,B,c \quad
\in (0,10000),
\quad
\omega_f = W \omega_c,
\end{equation}
\noindent
where $W\in(0,1)$ for the $\omega_f<\omega_c$ case and $W\in(1,10000)$ for the $\omega_f>\omega_c$ case. The results are as follows:

\begin{itemize}
\item For the $\omega_f<\omega_c$ case in Fig.~\ref{sig:fig:WireframeLT}, the following general conclusions are drawn.  The bounding amplitude of:
	\begin{itemize}
	\item region $A$ is smaller than that of region $B$ (By inspection of~(\ref{sig:eq:AmpLT:Region1}) and~(\ref{sig:eq:AmpLT:Region2}))
	\item region $B$ is smaller than that of region $B_2$ (By inspection of~(\ref{sig:eq:AmpLT:Region2}) and~(\ref{sig:eq:AmpLT:Region3}))
	\item region $B_2$ is smaller than that of region $C_2$ (Tested)
	\item region $C_2$ is larger than that of region $C$ (By inspection of~(\ref{sig:eq:AmpLT:Region4}) and~(\ref{sig:eq:AmpLT:Region5}))
	\item region $C$ is larger than that of region $D$ (By inspection of~(\ref{sig:eq:AmpLT:Region5}) and~(\ref{sig:eq:AmpLT:Region6}))
	\item region $B$ is smaller than that of region $C$ (Tested)
	\end{itemize}
\item For the $\omega_c>\omega_f$ case in Fig.~\ref{sig:fig:WireframeGT}, the following general conclusions are drawn.  The bounding amplitude of:
	\begin{itemize}
	\item region $A$ is smaller than that of region $B$ (By inspection of~(\ref{sig:eq:AmpGT:Region1}) and~(\ref{sig:eq:AmpGT:Region2}))
	\item region $B$ is smaller than that of region $C$ (Tested)
	\item region $C$ is larger than that of region $D$ (By inspection of~(\ref{sig:eq:AmpGT:Region3}) and~(\ref{sig:eq:AmpGT:Region4}))
	\end{itemize}
\end{itemize}

\noindent
The amplitude of the \textit{Asymptotically Undisturbed Regions} (regions $A$ and $D$ in Figs.~\ref{sig:fig:WireframeLT} and~\ref{sig:fig:WireframeGT}) is zero, so the long-time amplitude of those regions will always be less than or equal to any other region. Also through inspection, the amplitudes of the \textit{Additional Forced Regions} (regions $B_2$ and $C_2$ in Fig.~\ref{sig:fig:WireframeLT}) are equal to the amplitudes of the respective \textit{Forced Regions} (regions $B$ and $C$ in Fig.~\ref{sig:fig:WireframeLT}) plus some positive number. From this investigation, we conclude that the relative amplitudes of the regions shown in Figs.~\ref{sig:fig:WireframeLT} and~\ref{sig:fig:WireframeGT} are sufficiently representative of all solutions to~(\ref{sig:eq:h_PDE}).

\section{Evaluating a bow-tie contour}
	\label{sig:sup:BowTieContour}
We now examine the contour integral and evaluation of Equation~(\ref{sig:eq:app:ITilde1}), that is representative of the other contour integrals used in this paper.  To begin, we rewrite~(\ref{sig:eq:app:ITilde1}) in terms of $\Phi$ as,
\begin{equation}
	\label{sig:sup:eq:ITildeExample}
\mathbb{J}_1=
-\frac{A_f}{B^2}
\int\limits_{-\infty}^{\infty}
\Bigg[
\Bigg(
e^{i \Phi t}
\Bigg)
\Bigg]
\Bigg/
\Bigg[
(k-k_0)
(k-k_1)
(k-k_2)
(k-k_3)
\Bigg]dk,
\end{equation}	

where
\begin{equation}
\Phi=Bk^2-ck-\omega_f+kV.
\end{equation}

\begin{figure}[ht!]
\centering
\includegraphics[keepaspectratio,width=6in]{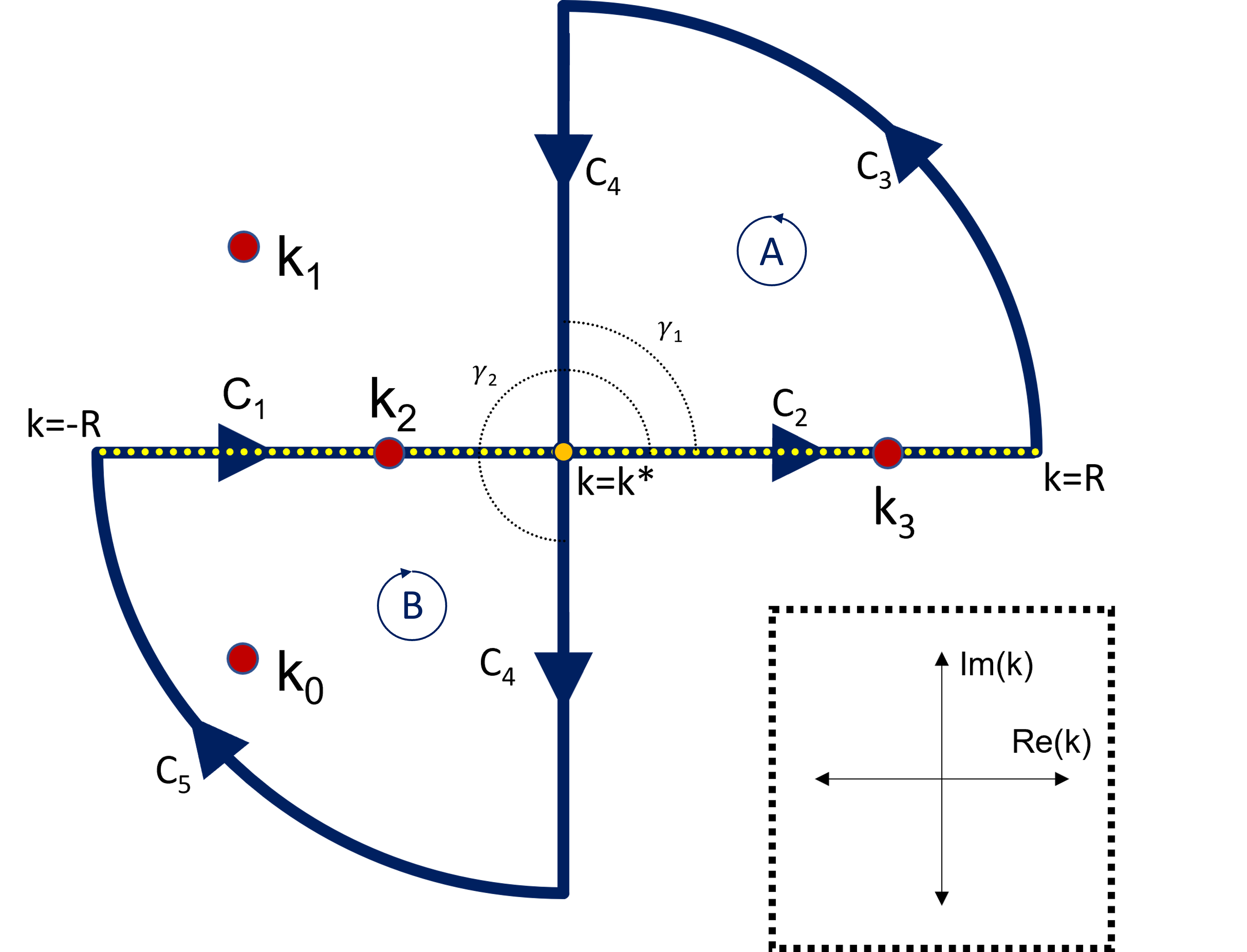}
\caption{
General form of a bow-tie contour. For some central point $k^*$, the angles $\gamma$ are limited to  $(0$, $\nicefrac{\pi}{2})$ for $\gamma_1$ and to $(\pi$, $\nicefrac{3\pi}{2})$ for $\gamma_2$, so that the integrals along the curved sections, $C_3$ and $C_5$ go to zero as $R$ goes to infinity. In this example, the pole $k_0$ is contained within a contour, and the pole $k_1$ is not. The integral along the real axis, the dotted yellow line, is the integral $I_1$ as $R$ goes to infinity. The inset figure shows the orientation of the contour in the complex $k$ plane. Depending on the value of $k^*$, the origin lies somewhere along the dotted yellow line.
For the sake of demonstration, the $\omega_f>\omega_f$ case is presented here using the pole structure in Fig.~\ref{sig:fig:GTCPoles}. In the case of $\omega_f<\omega_c$, all the poles would lie on the dotted yellow line. Each lobe of the bow-tie is a closed subcontour where $A=C_2\cup C_3\cup C_{4\text{ (from $i\infty$ to $k^*$)}}$ and $B=C_5\cup C_1\cup C_{4\text{ (from $k^*$ to $-i\infty$)}}$.
}
\label{sig:fig:app:WideBowtieContour}
\end{figure}

\noindent
By substituting in the definition of $k_2$ and $k_3$ and differentiating $\Phi$ with respect to $k$, we find that there is a second order saddle point at 
\begin{equation*}
k_s=\frac{(c-V)}{2B},
\end{equation*} 
as stated in~(\ref{sig:eq:app:ITilde1Saddle}). A preliminary contour is established in the complex $k$ plane as shown in Fig.~\ref{sig:fig:app:WideBowtieContour}. The limits on the angles $\gamma_1$ and $\gamma_2$ (drawn in the figure) are chosen such that the contour integrals along the curved sections go to zero and the dotted line path in the figure corresponds to the path of the integral~(\ref{sig:eq:app:ITilde1}) as $R\to\infty$.
	
\begin{figure}[ht!]
\centering
\includegraphics[keepaspectratio,width=6in]{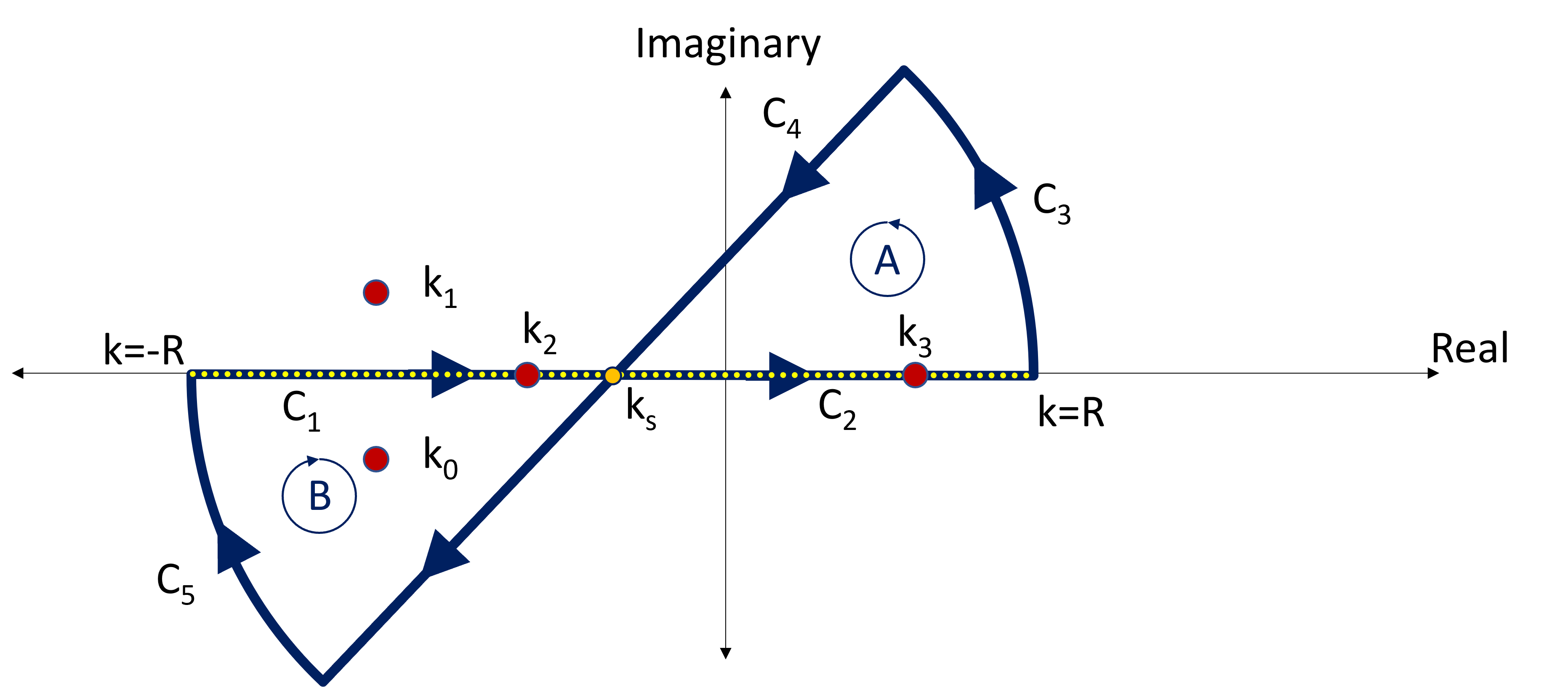}
\caption{
An example bow-tie contour in the complex $k$ plane. As $R$ goes to infinity, contour sections $C_1$ and $C_2$ become the integral of interest, and contour sections $C_3$ and $C_5$ go to zero. The pole $k_0$ is enclosed, and the pole $k_1$ is not enclosed. The integral along the contour $C_4$ can be evaluated asymptotically as time goes to infinity due to the choice of $k^*$ (Fig.~\ref{sig:fig:app:WideBowtieContour}) as the saddle point $k_s$ and the choice of the angles $\gamma$. The position of the saddle point $k_s$ is a function of velocity. Contour is depicted for $V>c$ and $\omega_f>\omega_c$. Varying $V$ translates the contour along the real $k$ axis. In the case of $\omega_f<\omega_c$, all the poles would lie on the dotted yellow line. Each lobe of the bow-tie is a closed subcontour where $A=C_2\cup C_3\cup C_{4\text{ (from $(1+i)\infty$ to $k^*$)}}$ and $B=C_5\cup C_1\cup C_{4\text{ (from $k^*$ to $-(1+i)\infty$)}}$
}
\label{sig:fig:app:BowtieContour}
\end{figure}

Through consideration of~(\ref{sig:sup:eq:ITildeExample}), the contour angles, $\gamma_1$ and $\gamma_2$, are restricted to lie in the intervals $\left[0,\nicefrac{\pi}{2}\right]$ and $\left[\pi,\nicefrac{3\pi}{2}\right]$, as indicated in Fig.~\ref{sig:fig:app:WideBowtieContour}.
When the contour angles are chosen to be $\nicefrac{\pi}{4}$ and $\nicefrac{5\pi}{4}$, the integrals along the diagonal can be evaluated asymptotically as time goes to infinity using the method of steepest descent (see Fig.~\ref{sig:fig:app:BowtieContour}).

\noindent
As indicated in~(\ref{sig:eq:app:ITilde1Saddle}), the saddle point is a function of $V$ and is always real-valued. Thus, the contour in Fig.~\ref{sig:fig:app:BowtieContour} will be shifted along the real axis for different velocities. 
In what follows, no distinction will be made between the locus of saddles and the real $k$ axis as they are equivalent for this problem.

\begin{figure}[ht!]
\centering
\includegraphics[keepaspectratio,width=6in]{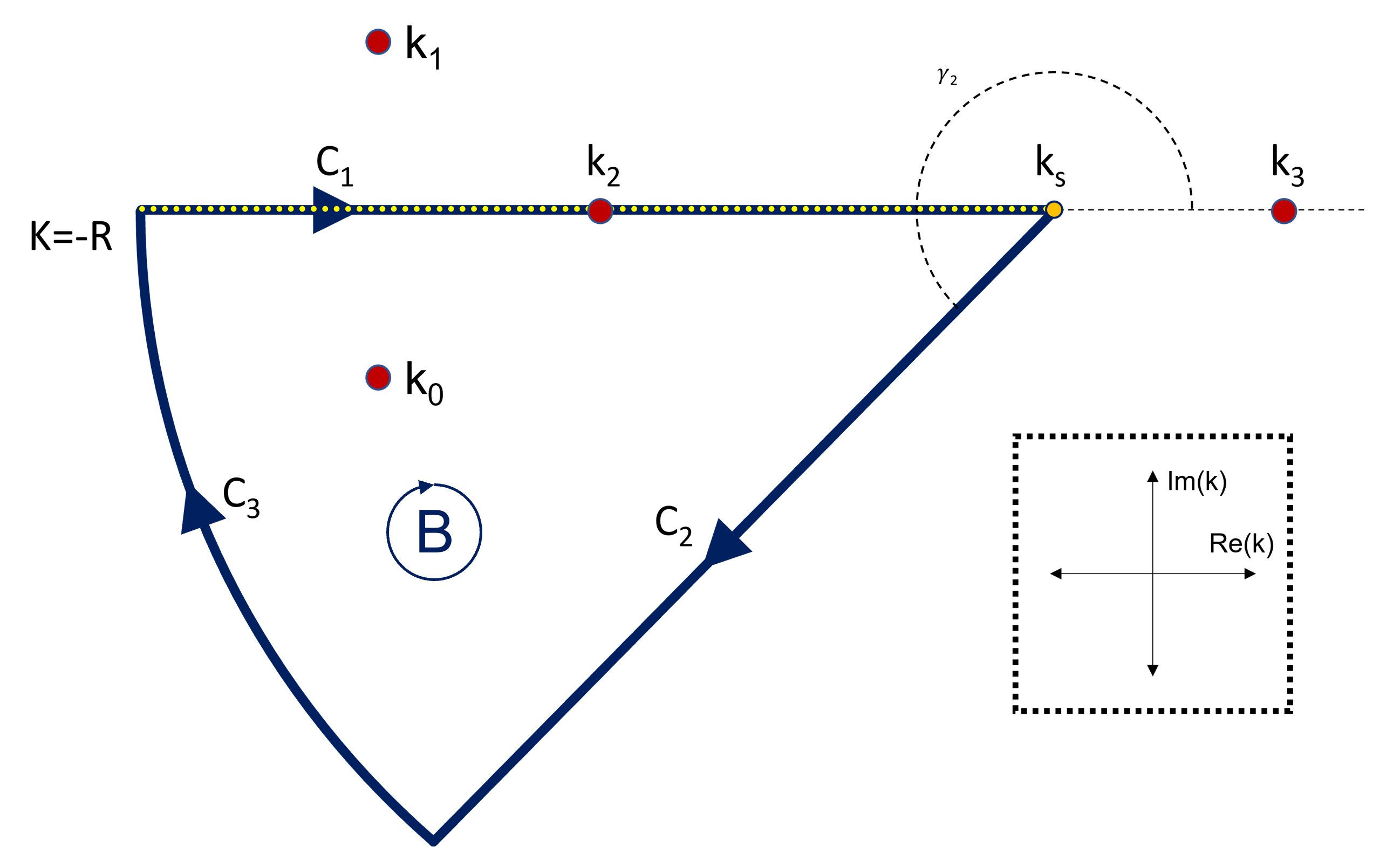}
\caption{
Example integration contour $B$ from Fig.~\ref{sig:fig:app:BowtieContour}. As before, the angle $\gamma_2$ is chosen such that the integral along $C_3$ goes to zero as $R$ goes to infinity, the integral along $C_2$ can be determined asymptotically as time goes to infinity, and the integral along the real axis ($C_1$, dotted yellow line) is the integral of interest as $R$ goes to infinity. For the sake of example, the pole $k_0$ is contained in the contour and the pole $k_2$ is intersected by the contour.
}
\label{sig:fig:app:LeftLoopContour}
\end{figure}

The complete contour in Fig.~\ref{sig:fig:app:BowtieContour} is broken up into two closed sub-contours, $A$ and $B$ which are separately evaluated through Cauchy's residue theorem (see Fig.~\ref{sig:fig:app:BowtieContour}). As an example, we focus here on contour $B$ of Fig.~\ref{sig:fig:app:BowtieContour}, shown in Fig.~\ref{sig:fig:app:LeftLoopContour}. 
Because the real axis segment of Fig.~\ref{sig:fig:app:BowtieContour} is the integral of interest from $k=-\infty$ to $k=\infty$, it follows that the real axis segment of Fig.~\ref{sig:fig:app:LeftLoopContour} is the integrand of interest integrated from $k=-\infty$ to $k_s$.

\noindent
Around the closed contour $B$ in Fig.~\ref{sig:fig:app:LeftLoopContour} the integral can be expressed with $\zeta$ as a placeholder of the integrand as
\begin{equation}
\int_{C_1}\zeta dk
+\int_{C_2}\zeta dk
+\int_{C_3}\zeta dk
=2\pi i \text{Res}(\zeta,k_0)
+\pi i \text{Res}(\zeta,k_2),
\end{equation}

\noindent
where Res$(f,k_j)$ is the residue of the integrand $\zeta$ at the pole $k_j$. The contained pole $k_0$ has the full contribution, but the intersected pole $k_2$ only has half the contribution. Also, this expression is only valid when $k_0$ is contained (as it is in Fig.~\ref{sig:fig:app:LeftLoopContour}). For a larger velocity, the saddle point may be shifted far enough to the left such that $k_0$ is no longer contained.
Through the choice of $\gamma_2$, the integral along $C_3$ goes to zero as $R\to\infty$.
Therefore, we can say that 
\begin{equation}
\int_{C_1}\zeta dk
=2\pi i \text{Res}(\zeta,k_0)
+\pi i \text{Res}(\zeta,k_2)
-\int_{C_2}\zeta dk.
\end{equation}

Evaluating the residue of~(\ref{sig:sup:eq:ITildeExample}) at the saddle points yields
\begin{equation}
\text{Res}(\zeta ;k_j)=
\lim_{k\to k_j}
\left(
-\frac{A_f}{B^2}
\left(
\frac{(k-k_j)e^{i \Phi t}}
{(k-k_0)(k-k_1)(k-k_2)(k-k_3)}
\right)
\right).
\end{equation}

\noindent
The contour along the diagonal contour, $C_2$, is determined by converting the integral into polar form as $k = k_s + r e^{i\gamma}$, $dk = e^{i\gamma}dr$. This transformation yields
\begin{equation}
\int_{C_2}\zeta dk=\int\limits_0^{\infty}
e^{i\gamma}\zeta(k_s+r e^{i\gamma})dr,
\end{equation}
\begin{multline}
\int\limits_{C_2}\zeta dk
=e^{i\left(\gamma_2-
\left(\frac{(V-c)^2t}{4B}
\right)
-\omega_ft\right)}
\\ \times
\int\limits_{0}^{\infty}
\frac{e^{iBr^2 e^{2i\gamma} t}dr}
{(re^{i\gamma_2}+k_s-k_0)
(re^{i\gamma_2}+k_s-k_1)
(re^{i\gamma_2}+k_s-k_2)
(re^{i\gamma_2}+k_s-k_3)}.
\end{multline}

Next, $\gamma$ is chosen in order to put the exponent into the form of $-r^2$, $\gamma=\nicefrac{5\pi}{4}$. The integral becomes:
\begin{equation}
\label{sig:eq:app:BowTieBeforeMoSD}
\int\limits_{C_2}\zeta dk=
e^{\frac{5i\pi}{4}}
e^{-i\left(\frac{(V-c)^2}{4B}
+\omega_f\right)t}
\int\limits_{0}^{\infty}
\frac{e^{-Br^2 t}dr}
{(re^{i\frac{5\pi}{4}}+k_s-k_0)
(re^{i\frac{5\pi}{4}}+k_s-k_1)
(re^{i\frac{5\pi}{4}}+k_s-k_2)
(re^{i\frac{5\pi}{4}}+k_s-k_3)}.
\end{equation}
The integral in~(\ref{sig:eq:app:BowTieBeforeMoSD}) can be split up into two separate integrals, one from $r=0$ to $r=\epsilon$ and one from $r=\epsilon$ to $r=\infty$ where $\epsilon << 1$ .
Through integration by parts, the latter integral is asymptotically subdominant to the former. Over the small range $0$ to $\epsilon$ the integral can be replaced with the integral linearized about $r=0$. In order to evaluate the integral, $\epsilon$ is taken to infinity without adding any asymptotically dominant terms in accordance with the method of steepest descent. Doing so yields
\begin{equation}
\int\limits_{C_2}\zeta dk\sim
e^{\frac{5i\pi}{4}}
e^{-i\left(\frac{(V-c)^2}{4B}
+\omega_f\right)t}
\int\limits_{0}^{\infty}
\frac{e^{-Br^2 t}dr}
{(k_s-k_0)
(k_s-k_1)
(k_s-k_2)
(k_s-k_3)}
\quad
\text{as}
\quad t\to\infty,
\end{equation}
\begin{equation}
\int\limits_{C_2}\zeta dk
\sim
\frac{-e^{\frac{i\pi}{4}}
e^{-i\left(\frac{(V-c)^2}{4B}
+\omega_f\right)t}}
{(k_s-k_0)
(k_s-k_1)
(k_s-k_2)
(k_s-k_3)}
\int\limits_{0}^{\infty}
e^{-Br^2 t}dr
\quad
\text{as}
\quad t\to\infty,
\end{equation}
\begin{equation}
\int\limits_{C_2}\zeta dk
\sim
\frac{-e^{\frac{i\pi}{4}}
e^{-i\left(\frac{(V-c)^2}{4B}
+\omega_f\right)t}}
{(k_s-k_0)
(k_s-k_1)
(k_s-k_2)
(k_s-k_3)}
\sqrt{\frac{\pi}{4Bt}}
\quad
\text{as}
\quad t\to\infty.
\end{equation}

\noindent
For the arrangement of the poles and saddle point shown in Fig.~\ref{sig:fig:app:LeftLoopContour}, the section of interest can be expressed as 
\begin{equation}
\int_{C_1}\zeta dk
=2\pi i \text{Res}(\zeta,k_0)
+\pi i \text{Res}(\zeta,k_2)
-\frac{-e^{\frac{i\pi}{4}}
e^{-i\left(\frac{(V-c)^2}{4B}
+\omega_f\right)t}}
{(k_s-k_0)
(k_s-k_1)
(k_s-k_2)
(k_s-k_3)}
\sqrt{\frac{\pi}{4Bt}}
\quad
\text{as}
\quad t\to\infty,
\end{equation}
\begin{equation}
\text{Res}(\zeta ;k_j)=
\lim_{k\to k_j}
\left(
-\frac{A_f}{B^2}
\left(
\frac{(k-k_j)e^{i \Phi t}}
{(k-k_0)(k-k_1)(k-k_2)(k-k_3)}
\right)
\right).
\end{equation}

\noindent
Note again that the pole structure and the velocity that is being studied will determine which of the residue terms contribute to the solution.

\section{Evaluating a half-plane contour}
	\label{sig:sup:HalfPlaneContour}

Here we examine~(\ref{sig:eq:app:ITilde2}), which we rewrite in terms of $\Phi$ as,
\begin{equation}
\mathbb{J}_2=
\int\limits_{-\infty}^{\infty}
\Bigg[
\frac{A_f}{B^2}e^{i\Phi t}
\Bigg]
\Bigg/
\Bigg[
(k-k_0)
(k-k_1)
(k-k_2)
(k-k_3)
\Bigg]dk,
\end{equation}
\begin{equation}
	\label{sig:sup:eq:PhiHalfPlane}
\Phi=\omega_f+kV.
\end{equation}

\noindent
Next we establish a half-plane contour in either the upper half-plane or lower-half plane depending on the sign of $V$.

\begin{figure}[ht!]
\centering
\includegraphics[keepaspectratio,width=6in]{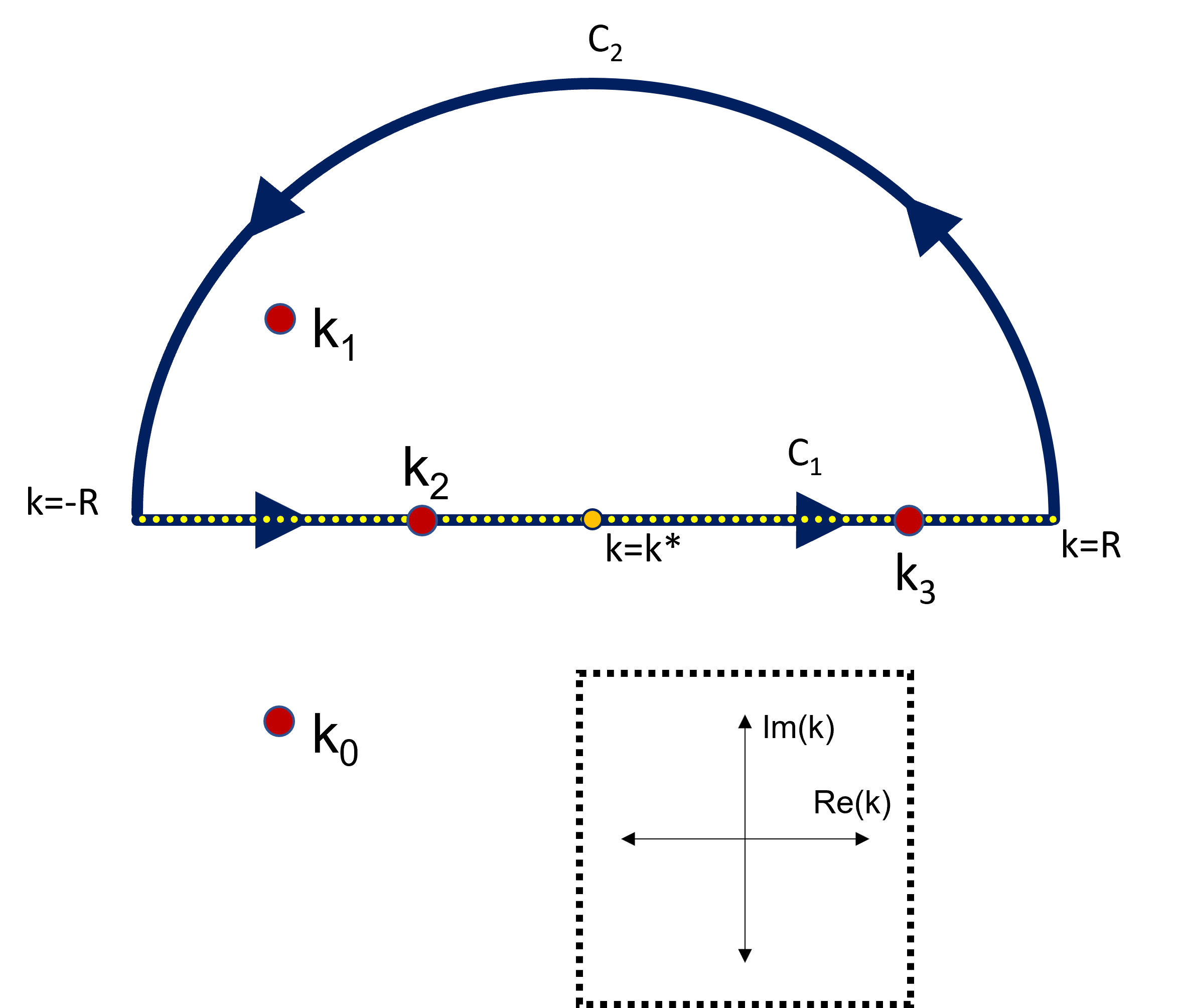}
\caption{Sample half-plane contour for~(\ref{sig:eq:app:ITilde2}) for $V>0$ and $\omega_f>\omega_c$. The poles $k_1$ is contained and the poles $k_2$ and $k_3$ are intersected. See~(\ref{sig:eq:PoleLocs}) for the definitions of the poles. The integral along the curved contour goes to zero as $R$ goes to infinity.
}
\label{sig:fig:app:HalfPlaneContour}
\end{figure}

The integral~(\ref{sig:eq:app:ITilde2}) is expressed as
\begin{multline}
\oint\left(
		\frac{A_f}{B^2}
		\frac{e^{i\Phi t}}
		{(k-k_0)(k-k_1)(k-k_2)(k-k_3)}dk
	\right)
\\
=
\int_{C_1}\left(
		\frac{A_f}{B^2}
		\frac{e^{i\Phi t}}
		{(k-k_0)(k-k_1)(k-k_2)(k-k_3)}dk
	\right)
+
\int_{C_2}\left(
		\frac{A_f}{B^2}
		\frac{e^{i\Phi t}}
		{(k-k_0)(k-k_1)(k-k_2)(k-k_3)}dk
	\right)
\\
=
2\pi i \text{Res}(\zeta,k_1)
+\pi i \text{Res}(\zeta,k_2)
+\pi i \text{Res}(\zeta,k_3),
\end{multline}
\noindent
where $\zeta$ is a placeholder for the integrand of~(\ref{sig:eq:app:ITilde2}). We note that, because $V>0$ and the contour is in the upper half-plane, the integral along $C_2$ goes to zero as $R$ goes to infinity. Additionally, the integral along $C_1$ becomes $\mathbb{J}_2$ under the same limit, yielding
\begin{equation}
	\label{sig:eq:sup:ITilde2}
\mathbb{J}_2=
2\pi i \text{Res}(\zeta,k_1)
+\pi i \text{Res}(\zeta,k_2)
+\pi i \text{Res}(\zeta,k_3),
\end{equation}
\begin{equation}
	\label{sig:eq:sup:Residue}
\text{Res}(\zeta ;k_j)=
\lim_{k\to k_j}
\left(
-\frac{A_f}{B^2}
\left(
\frac{(k-k_j)e^{i (\omega_f+kV) t}}
{(k-k_0)(k-k_1)(k-k_2)(k-k_3)}
\right)
\right).
\end{equation}

\noindent
Note that if $V<0$ then the contour must be established in the lower half-plane, so $k_0$ would be contained instead of $k_1$. If $\omega_f<\omega_c$, all the poles lie on the real axis and the choice of direction for the contour has no effect on which poles are contained. The result~(\ref{sig:eq:sup:ITilde2}) is the desired exact result for $\mathbb{J}_2$ used in the asymptotic results~(\ref{sig:eq:app:LTCSolution}) and~(\ref{sig:eq:app:GTCSolution}).

\end{document}